\documentclass[a4paper,11pt]{article}
\usepackage{jheppub}

\pdfoutput=1
\usepackage{amsmath}
\usepackage{gensymb}
\usepackage{amsfonts}
\usepackage{comment}
\usepackage{amssymb}
\usepackage{mathrsfs}
\usepackage{graphicx}
\usepackage{color}
\usepackage{multirow}
\usepackage{array}
\usepackage{bm}
\usepackage{tabularx}
\usepackage{subcaption}
\usepackage{blindtext}
\usepackage{wasysym}
\usepackage{hyperref}
\usepackage{float}
\hypersetup{colorlinks=true,allcolors=blue}
\usepackage{orcidlink}
\usepackage[normalem]{ulem}
\usepackage{lineno}
\usepackage[T1]{fontenc}
\usepackage{soul}
\usepackage[normalem]{ulem}
\usepackage{rotating}
\usepackage{graphicx}
\usepackage{multirow}
\usepackage{array}
\newcolumntype{L}{>{\centering\arraybackslash}m{1.99cm}}

\newcommand{\dchsq}{\Delta\chi^2}

\newcommand{\nova}{NO$\nu$A~}
\newcommand{\tk}{T2K}

\newcommand{\bi}{\begin{itemize}}
\newcommand{\ei}{\end{itemize}}

\newcommand{\dcp}{\delta_{CP}}

\newcommand{\be}{\begin{equation}}
\newcommand{\ee}{\end{equation}}
\newcommand{\ba}{\begin{array}}
\newcommand{\ea}{\end{array}}
\newcommand{\bea}{\begin{eqnarray}}
\newcommand{\eea}{\end{eqnarray}}

\newcommand{\beq}{\begin{equation}}
\newcommand{\eeq}{\end{equation}}
\newcommand{\beqa}{\begin{eqnarray}}
\newcommand{\eeqa}{\end{eqnarray}}

\newcommand{\tx}{{\theta_{12}}}
\newcommand{\ty}{{\theta_{13}}}
\newcommand{\tz}{{\theta_{23}}}

\newcommand{\dl}{{\Delta_{31}}}
\newcommand{\ds}{{\Delta_{21}}}

\newcommand{\ahat}{\hat{A}}

\newcommand{\dhat}{\hat{\Delta}}

\newcommand{\pme}{P_{\mu e}}

\newcommand{\pmebar}{P_{\bar{\mu} \bar{e}}}

\begin{document}
 
\title{ Enhancing the sensitivity to neutrino oscillation parameters using synergy between T2K, \nova and JUNO} 

\author[a]{Srubabati Goswami,}
\affiliation[a]{Theoretical Physics Division, Physical Research Laboratory, Navrangpura, Ahmedabad 380009, India}
\author[b,c]{Aman Gupta\footnote{\footnotesize Present address: Institute of High Energy Physics, Chinese Academy of Sciences, Beijing 100049, China}\footnote{\footnotesize Spallation Neutron Source Science Center, Dongguan, Guangdong, 523803, China}\orcidlink{0000-0002-7247-2424},}
\affiliation[b]{Theory Division, Saha Institute of Nuclear Physics, 1/AF, Bidhannagar, Kolkata 700064, India}
\affiliation[c]{School of Physical Sciences, Indian Association for the Cultivation of Science,2A $\&$ 2B Raja S.C. Mullick Road, Jadavpur, Kolkata 700032, India}
\author[d,e]{Ushak Rahaman,}
\affiliation[d]{Department of Physics, University of Toronto, Toronto, ON M5S 1A7, Canada}
\affiliation[e]{Tata Institute of Fundamental Research, Homi
Bhabha Road, Colaba, Mumbai 400005, India}
\author[f]{Sushant K. Raut}
\affiliation[f]{Division of Sciences, Krea University, Sri City, India 517646}

\emailAdd{sruba@prl.res.in}
\emailAdd{amann16.iitr@gmail.com}
\emailAdd{ushak.rahaman@cern.ch}
\emailAdd{sushant.raut@krea.edu.in}

\abstract{We study the impact of combining the present NO$\nu$A and T2K data with simulated data from the JUNO experiment on the determination of the leptonic CP phase and the neutrino mass hierarchy. The current NO$\nu$A data exhibit a hierarchy--$\dcp$ degeneracy, admitting both normal hierarchy (NH) with $\dcp \in [0,180^\circ]$, and inverted hierarchy (IH) with $\dcp \in [180^\circ,360^\circ]$ solutions at comparable significance, while T2K prefers $\dcp\simeq 270^\circ$ for both hierarchies, leading to a $2\sigma$ tension between the two experiments for normal hierarchy. Using detailed \textsc{GLoBES} simulations, we show that future JUNO data with excellent hierarchy sensitivity, can lift the hierarchy--$\dcp$ degeneracy in NO$\nu$A and strengthen the hierarchy reach of T2K in spite of having no $\dcp$ sensitivity. Allowing the hierarchy to be a free parameter in the fit, if the true ordering is IH, JUNO aligns the NO$\nu$A and T2K allowed regions and resolves their present tension; if NH is true, the tension continues to persist. \textit{We also show that JUNO’s precise measurement of $|\dl|$ leads to improved constraints on $\sin^2\theta_{23}$ and $\dcp$ for normal mass hierarchy in NO$\nu$A even though JUNO itself is insensitive to these parameters}. Finally, updated solar-parameter measurements from JUNO’s first data release further enhance the combined precision. Our results demonstrate that JUNO plays a crucial synergistic role in the global neutrino-oscillation programme, enabling a more robust determination of the mass ordering and improving the sensitivity to the CP phase when combined with long-baseline data.
}


\maketitle

\section{Introduction}

Neutrino masses and mixing were first proposed as a solution to the solar and atmospheric neutrino anomalies, in which the experimentally detected numbers of neutrinos of {\bf{ specific}} flavours were not in accordance with the theoretical expectations. Eventually, data from atmospheric neutrino experiments such as Super-Kamiokande, solar neutrino experiments such as SNO, and reactor neutrino experiments such as Double-CHOOZ, Daya Bay, and RENO confirmed the neutrino mixing hypothesis, and the anomalous results could be explained through the phenomenon of neutrino oscillations~\cite{Super-Kamiokande:1998kpq,Hisano:1998fj}, i.e. the conversion of neutrinos from one flavour to another during propagation. The discovery of neutrino oscillations (and hence non-zero neutrino masses) can be considered to be the first signature of the existence of physics beyond the canonical Standard Model (BSM).

In the standard oscillation scenario, the mixing between the three neutrino flavours is parametrized by three mixing angles -- $\theta_{12}$, $\theta_{13}$, and $\theta_{23}$, and one  Dirac-CP phase, $\dcp$. The oscillation probabilities depend on these four mixing parameters, the two independent mass-squared differences $\Delta_{21}$ and $\Delta_{31}$, the energy of the neutrino, and the distance traversed by it. In addition, neutrino oscillation probabilities in the presence of background matter are modified by the Mikheyev-Smirnov-Wolfenstein potential $V=\sqrt{2} G_F n_e$ \cite{PhysRevD.17.2369,Zaglauer:1988gz,Blennow:2013rca} where the density of background electrons ($n_e$) can be easily expressed in terms of the density of the medium. 

Determination of the neutrino oscillation parameters can play a significant role in discriminating between BSM models that have been built to explain the existence of the tiny non-zero neutrino masses. They can also point to additional symmetries in the leptonic sector, and possibly the origin of the matter-antimatter asymmetry of the Universe. 

Numerous experiments over the past few decades have measured these parameters to varying degrees of precision. The current best-fit values from global analysis of all neutrino oscillation experiments are presented in Table~\ref{tab:t1}.
This table shows that the parameters $\theta_{12}$, $\theta_{13}$, $\Delta_{21}$, and $|\Delta_{31}|$ have been measured with percent-level precision. Thus, the current unknowns are the sign of the larger mass-squared difference, $\textrm{sgn}(\Delta_{31})$ which is known as the neutrino mass hierarchy, the maximal/non-maximal nature of $\theta_{23}$ and its octant (if non-maximal), and the value of the phase $\dcp$. The T2K and NOvA long-baseline superbeam experiments are currently collecting data to determine these unknown parameters by measuring the $\nu_\mu \to \nu_e$ oscillation probability, $P_{\mu e}$, and the $\bar{\nu}_\mu\to\bar{\nu}_e$ oscillation probability, $P_{\bar{\mu}\bar{e}}$. Since these probabilities depends on all the above unknowns, their measurement is hampered by the presence of parameter degeneracies \cite{Barger:2001yr,Kajita:2006bt}.  
After the precise measurement of the mixing angle $\theta_{13}$, the remaining degeneracies can be understood well in terms of a generalized hierarchy$-\theta_{23}-\delta_{CP}$ degeneracy \cite{Ghosh:2015ena}.
Data from experiments with different energies and distances, and hence different functional dependence on the parameters, can help to alleviate this problem.   

Data from T2K and \nova have limited sensitivity due to the presence of hierarchy - $\delta_{CP}$  and octant - $\delta_{CP}$ degeneracies. In fact, there is currently a tension in the preferred values of the parameters inferred from data at these two experiments~\cite{T2K:2023smv,NOvA:2021nfi,Rahaman_2022,Mikola:2024jnj}. Many studies have attempted to explain this tension by invoking non-standard oscillation scenarios~\cite{Denton:2020uda,PhysRevLett.126.051802,Rahaman:2021zzm,Chatterjee:2024kbn}. The T2K and \nova collaborations have undertaken a joint analysis of both sets of data to try to pin down the source of the tension. In a recent paper~\cite{Chatterjee:2024kbn}, it is shown that in light of new data by T2K and \nova, the tension in the determination of the standard CP-phase extracted by
the two experiments in the normal neutrino mass ordering persists and has a statistical significance of $2\sigma$. 

In this work, we study the implications of the JUNO experiment on our knowledge of the oscillation parameters in light of the T2K-\nova tension.
JUNO can measure the neutrino mass hierarchy using the $\overline{\nu}_e \to \overline{\nu}_e$ oscillation channel which is independent of $\dcp$ and $\theta_{23}$. The effect of simulated data from JUNO on the combined T2K-\nova determination of the mass hierarchy has been studied in Ref.~\cite{Cao:2020ans}. 
 We investigate if the hierarchy sensitivity of JUNO can lift the hierarchy-$\dcp$ degeneracy in combination with current and future data from \nova and T2K, leading to an enhanced sensitivity to discover CP violation in the leptonic sector. 

The layout of this article is as follows. In Sec.~\ref{experiment}, we discuss the details of the NO$\nu$A, T2K and JUNO experiments. The theoretical discussions on neutrino oscillation and survival probabilities and parameter degeneracies have been done in section~\ref{probability}. Section~\ref{analysis} is about the details of the analysis used in this paper and the results of the analysis have been discussed in section~\ref{result}. The sensitivities of the future \nova and T2K, combined with the data from future JUNO experiment have been discussed in section~\ref{future}. The first results from the JUNO collaboration have been considered in the simulation of the future JUNO data, and the corresponding results have been presented in chapter~\ref{juno_data}. The final conclusions have been drawn in section~\ref{conclusion}.

\section{Experimental specifications}
\label{experiment}
\begin{center}
    \it {\textbf{1. T2K}}
\end{center}
The Tokai-To-Kamioka (T2K) is an accelerator-based long-baseline (LBL) neutrino oscillation experiment which uses $\nu_\mu$ beam with a peak energy of $0.6$ GeV from the J-PARC accelerator at Tokai to measure the three-flavor neutrino mixing parameters. The T2K far detector, Super-Kamiokande (SK), consists of a $50$ kton water Cherenkov detector and is located under the Kamioka mine in Japan at distance of $295$ km away from the source. This detector is placed $2.5\degree$ off-axis with respect to the beam direction. In this analysis, we consider the total exposure of $1.97\times 10^{21}$ protons on target (POT) in $\nu-{\rm mode}$ and $1.63\times 10^{21}$ POT in $\bar{\nu}-{\rm mode}$ which corresponds to the total data samples collected by T2K from January 2010 to February 2020. T2K also proposes to extend the run until 2026 and will collect data with exposure of $20\times 10^{21}$ POT~\cite{T2K:2016siu}.

\begin{center}
    \it {\textbf{2. \nova}}
\end{center}
\nova is also a LBL experiment which consists of a $14$ kt totally active liquid  scintillator 
detector, placed at a distance of $810$ km away from the Fermilab's NuMI beam which serves as an intense source of muon neutrinos. The far detector of \nova, akin to T2K, is located off-axis i.e. at an angle of $0.8\degree$ to the direction of the neutrino beam. With such a long baseline, \nova will have mass hierarchy sensitivity through the standard matter effect. The flux of neutrinos from the NuMI beam peaks at 2 GeV which is close to the first oscillation maximum. For the analysis of \nova experiment, we consider the data taken from $2014$ to $2024$ and presented in the recent data release \cite{Wolcott:2024}. This corresponds to a total exposure of $2.660\times 10^{21}$ POT in $\nu-{\rm mode}$ and $1.250\times 10^{21}$ POT in $\bar{\nu}-{\rm mode}$.   

\begin{center}
    \it {\textbf{3. JUNO}}
\end{center}

The Jiangmen Underground Neutrino Observatory (JUNO) experiment \cite{JUNO:2015zny}  is a multi-national neutrino experiment based in China. JUNO started taking data from August, 2025 and the JUNO collaboration has recently published their first results \cite{JUNO:2025gmd}. JUNO observes reactor antineutrinos from several nuclear power plants located at Yangjiang and Taishan. It consists of a 20 kton fiducial mass liquid scintillator detector situated at an average baseline of approximately $53$ km from the reactors. This detector is projected with the ability to reconstruct the incoming neutrino energy with an unprecedented resolution $\Delta E/E \sim 0.03/\sqrt{E_{\text{vis}}(\text{MeV})}$ \cite{JUNO:2015zny}, where $E_{\rm vis}$ is the visible neutrino energy.

The principal purpose of JUNO is to measure the neutrino mass ordering. In addition, it will also be able to measure $\theta_{12}$, $\Delta_{21}$ and $|\Delta_{31}|$ quite precisely~\cite{Petcov:2001sy, Bandyopadhyay:2003du, Choubey:2003qx}. The distant reactors at Daya Bay and Huizhou will also have small contributions of neutrino flux at JUNO, but in this work, we ignore these reactor cores as their contribution to signal events is very small. Here, we have taken into account the neutrino sources only at the Yangjiang and Taishan nuclear power plants (with their respective thermal powers and baselines) as mentioned in Table 2 of Ref.~\cite{JUNO:2015zny}. The details of backgrounds and systematic errors are adapted from Refs.~\cite{JUNO:2015zny, JUNO:2021vlw}. The main backgrounds come from geo-neutrino events at low energies. We have considered $5\%$ systematic errors for signal and $20\%$ systematic errors for backgrounds. We use $2\%$ energy calibration error for both signal and background. For the analysis presented in this work, we consider a combined signal and background events of around $140,000$. The signals at JUNO are the inverse beta-decay (IBD) events, $\bar{\nu}_{e} + p \rightarrow e^{+} + n$. The bulk of the signal will lie in the energy range $\sim\left[1.8, 8\right]$ MeV. The antineutrino fluxes and IBD cross-sections are relatively well-known \cite{JUNO:2015zny}.


\section{Probabilities and Degeneracies} 
\label{probability}

\begin{figure}[htbp]
\begin{center}
\includegraphics[width=0.8\textwidth,height=0.4\textheight]{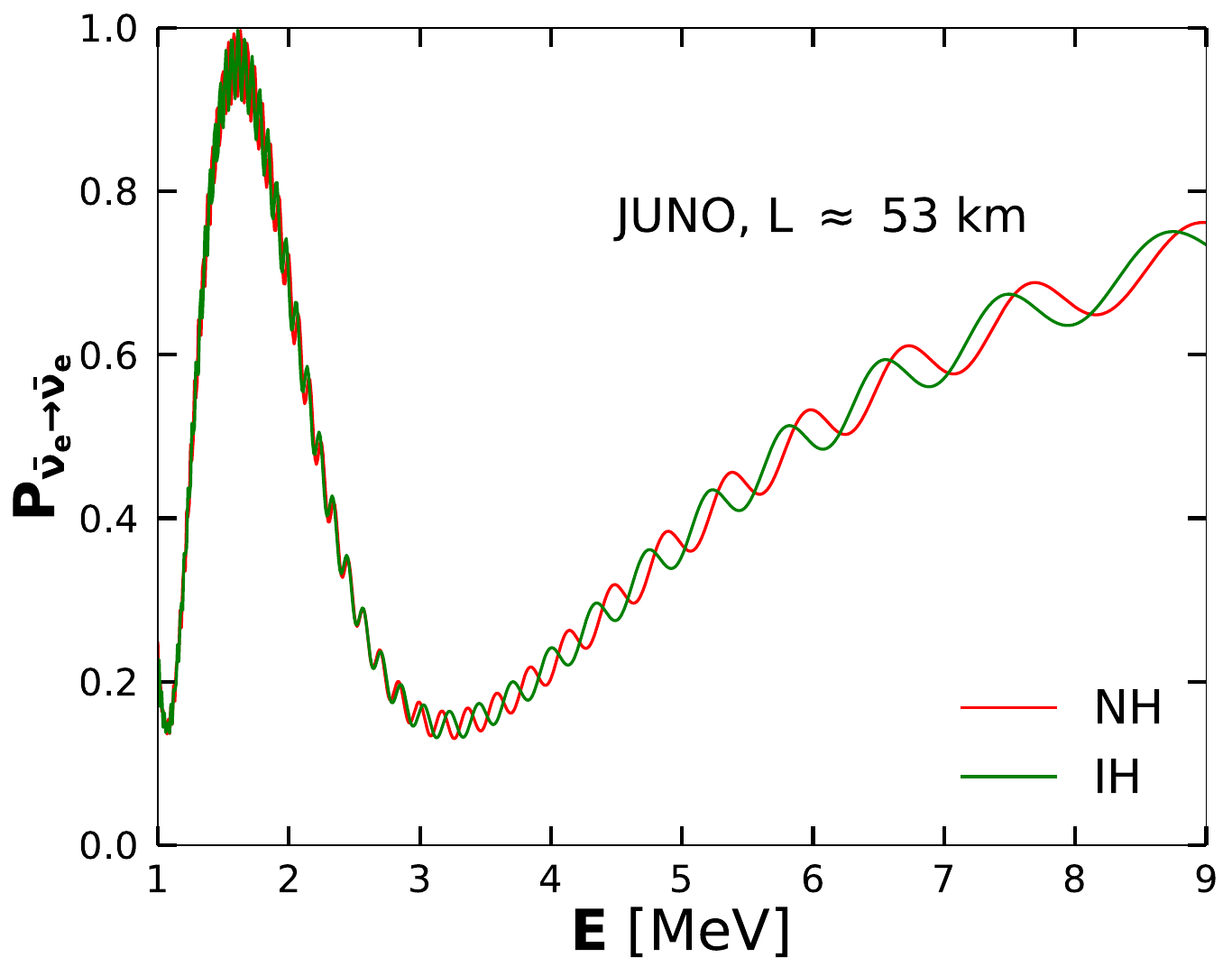}
\vspace*{0.7cm}
 
	\caption{The electron antineutrino survival probability as a function of antineutrino energy for the JUNO experiment. The red (green) colour corresponds to NH (IH) of neutrino masses. Here, the values of oscillation parameters are adopted from NuFIT 6.0 as presented in Table~\ref{tab:t1}.}
	\label{fig:Juno_prob}
    \end{center}
\end{figure}

With the average baseline distance of 53 km, the matter effect is negligible at JUNO. The $\bar{\nu}_e$ survival probability in vacuum can be written as \cite{JUNO:2022mxj}
\begin{eqnarray}
    P_{\bar{e}\bar{e}}=&&1- \sin^22\tx c_{13}^{4}\sin^2 \hat{\Delta}_{21}-\frac{1}{2} \sin^22\ty(\sin^2 \hat{\Delta}_{31}+\sin^2\hat{\Delta}_{32})\nonumber \\
    &&-\frac{1}{2} \cos 2\tx \sin^22\ty \sin \hat{\Delta}_{21}\sin (\hat{\Delta}_{31}+\hat{\Delta}_{32}),
    \label{prob-juno}
\end{eqnarray}
where $\hat{\Delta}_{ij}=\Delta_{ij}L/(4E)$. In Eq.~\ref{prob-juno}, the second and third terms are dominated by solar and atmospheric parameters, respectively. The sensitivity to the neutrino mass hierarchy comes from the fourth term. The modification of the survival probability in Eq.~\ref{prob-juno}, due to matter effect will be small. This modification has been studied in Refs.~\cite{JUNO:2022mxj, Li:2016txk, Capozzi:2013psa, Khan:2019doq}. In our numerical calculations and analyses throughout the paper, we have considered matter effect due to a constant matter density for JUNO. From Eq.~\ref{prob-juno}, it is clear that JUNO will have no sensitivity to $\dcp$ or $\tz$. Also, the hierarchy sensitivity of JUNO will be free of parameter degeneracy. In Fig.~\ref{fig:Juno_prob}, we can see that at the first probability maximum around 2 MeV, the $\bar{\nu}_e$ survival probabilities for NH and IH overlap. However, at higher energy, the survival probabilities of $\bar{\nu}_e$ for NH and IH become distinct. The hierarchy sensitivity for JUNO comes from this region. 

\begin{center}
\begin{table}
\begin{tabular}{|c |c| c|} 
 \hline
Oscillation parameters ($3\nu$) & Normal ordering (NO) & Inverse Ordering (IO) \\ [0.5ex] 
 \hline\hline
$\theta_{12} (^{\circ})$ & $33.68^{+0.73}_{-0.70}$ &$33.68^{+0.73}_{-0.70}$ \\
 \hline
$\theta_{23} (^{\circ})$ & $43.3^{+1.0}_{-0.8}$ &  $47.9^{+0.7}_{-0.9}$\\
\hline
$\theta_{13} (^
{\circ})$ & $8.56^{+0.11}_{-0.11}$ &  $8.59^{+0.11}_{-0.11}$\\
\hline
$\delta_{CP} (^{\circ})$ & $212^{+26}_{-41}$ & $274^{+22}_{-25}$ \\
\hline
$\ds$ (eV$^2$) & $7.49^{+0.19}_{-0.19}\times 10^{-5}$ & $7.49^{+0.19}_{-0.19}\times 10^{-5}$\\
\hline
$\Delta_{3l}$ (eV$^2$) & $+2.513^{+0.021}_{-0.019}\times 10^{-3}$ & $-2.484^{+0.020}_{-0.020}\times 10^{-3}$\\
\hline
\end{tabular}
\caption{Best-fit values of the standard three flavour neutrino oscillation parameters for both NH and IH. These values, along with their $1\sigma$ uncertainty intervals, are taken from NuFIT 6.0 (2024)~\cite{NuFIT6.0}, including IceCube and Super-K atmospheric data \cite{Esteban_2024}. In the Table, $l = 1$ for NH ($\dl > 0$) and $l = 2$ for IH ($\Delta_{32} < 0$).}  
\label{tab:t1}
\end{table}
\end{center}


In case of \nova and T2K, the $\nu_\mu \to \nu_e$ oscillation probability for a neutrino of energy $E$ traversing a distance $L$ \cite{Cervera:2000kp} is 
\begin{eqnarray}
  &\pme& \simeq \sin^2 2 \ty \sin^2 \tz\frac{\sin^2\dhat(1-\ahat)}{(1-\ahat)^2}\nonumber\\
  &+&\alpha \cos \ty \sin2\tx \sin 2\ty \sin 2\tz \cos(\dhat+\dcp)\nonumber\\
 &&\frac{\sin\dhat \ahat}{\ahat}
  \frac{\sin \dhat(1-\ahat)}{1-\ahat},
  \label{pme}
   \end{eqnarray}
    where $\alpha=\frac{\ds}{\dl}$, $\dhat=\frac{\dl L}{4E}$, $\ahat=\frac{A}{\dl}$, and the matter effect is represented by $A=2\sqrt{2}G_Fn_eE=0.76\times 10^{-4} \rho ({\rm g/cc})E(\textrm{GeV})\, \, ({\rm in\, \, eV^2})$, where $\rho$ is the matter density. The antineutrino oscillation probability $\pmebar$ can be obtained by changing the sign of $A$ and $\dcp$ in Eq.~\ref{pme}. The oscillation probability formula shows the dependence on mass hierarchy, octant of $\tz$ and $\dcp$.

    From Eq.~\ref{pme}, it is seen that the dominant term in $\pme$ is proportional to $\sin^2 2\ty$, and thus this probability is rather small. Matter effect can enhance (suppress) the probability if $\dl$ is positive (negative). The situation is reversed for $\pmebar$. The dominant term is also proportional to $\sin^2\tz$. For $\sin^22\tz<1$, there can be two possible octant choices for $\tz$. When $\tz$ is in lower octant (LO), i.e. $\sin^2 \tz<0.5$, both $\pme$ and $\pmebar$ is suppressed  relative to the maximal mixing, i.e. $\sin^2 \tz=0.5$. When $\tz$ is in higher octant (HO), i.e. $\sin^2 \tz>0.5$, both $\pme$ and $\pmebar$ is enhanced  relative to the maximal mixing. $\dcp$ sensitivity of the experiments comes from the second term which is proportional to $\alpha\approx 0.03$. When $180^\circ<\dcp<360^\circ$, we say $\dcp$ is in the lower half plane (LHP), and when $0 <\dcp<180^\circ$, we say $\dcp$ is in the upper half plane (UHP). When $\dcp$ is in UHP, $\pme$ ($\pmebar$) is smaller (larger), compared to the CP conserving case. However, when $\dcp$ is in LHP, $\pme$ ($\pmebar$) is larger (smaller), compared to the CP conserving case. Since each of the unknowns can choose two different values, there are eight possible combinations of the unknowns. Any given value of $\pme$ can be reproduced with any combination of the unknowns by choosing the value of $\ty$ appropriately. Thus there is an eight-fold degeneracy \cite{Fogli:1996pv, Burguet-Castell:2001ppm, Minakata:2001qm, Burguet-Castell:2002ald, Mena:2004sa, Prakash:2012az, Meloni:2008bd, Agarwalla:2013ju, Nath:2015kjg, Bora:2016tmb, Barger:2001yr, Kajita:2006bt} in the expression of $\pme$ and $\pmebar$ if $\sin^2 2\ty$ is not known precisely. The recent precision measurement of $\ty$ breaks this eight-fold degeneracy into $(1+3+3+1)$ pattern \cite{Bharti:2018eyj}.

\begin{figure}[H]
    \centering


    \begin{subfigure}[b]{0.45\textwidth}
        \centering
        \includegraphics[width=\textwidth, height=6cm]{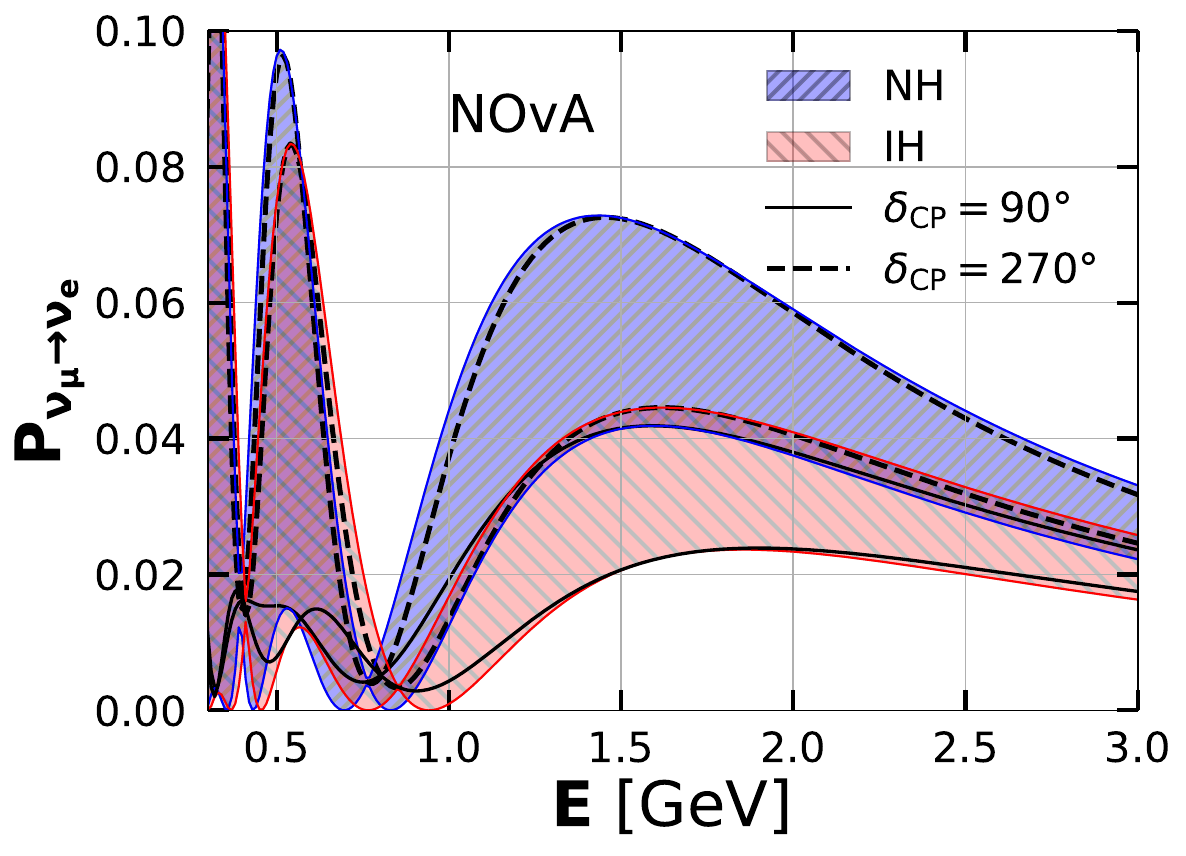}
        \caption{}
    \end{subfigure}
    \hfill
    \begin{subfigure}[b]{0.45\textwidth}
        \centering
        \includegraphics[width=\textwidth, height=6cm]{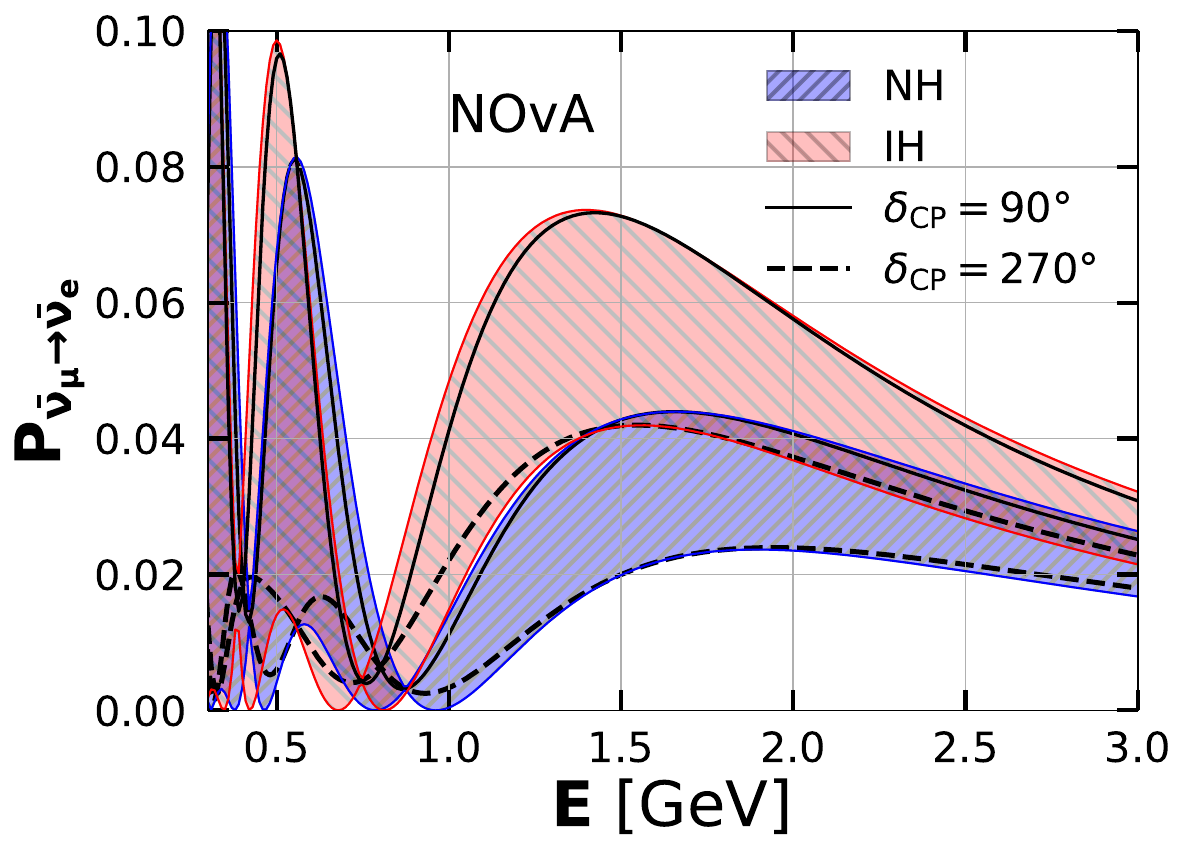}
        \caption{}
    \end{subfigure}
    \begin{subfigure}[b]{0.45\textwidth}
        \centering
        \includegraphics[width=\textwidth, height=6cm]{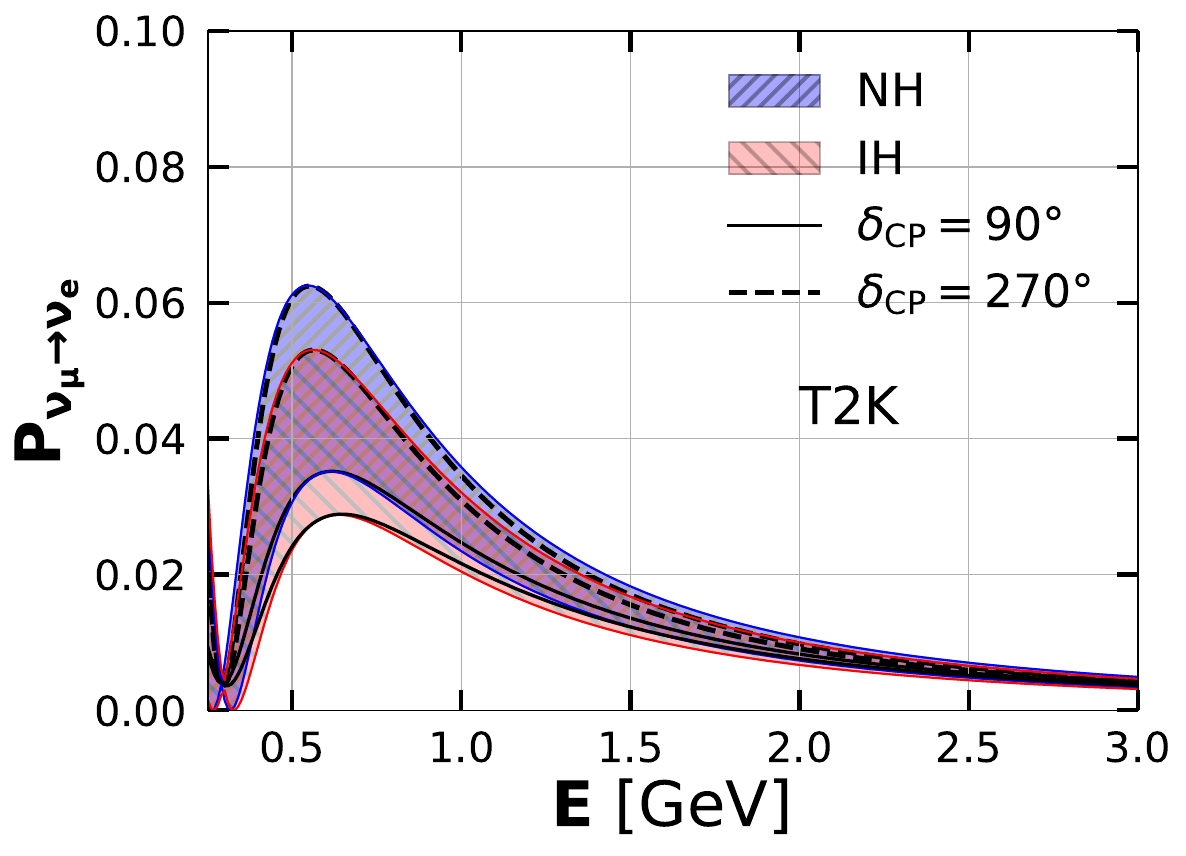}
        \caption{}
    \end{subfigure}
    \hfill
    \begin{subfigure}[b]{0.45\textwidth}
        \centering
        \includegraphics[width=\textwidth, height=6cm]{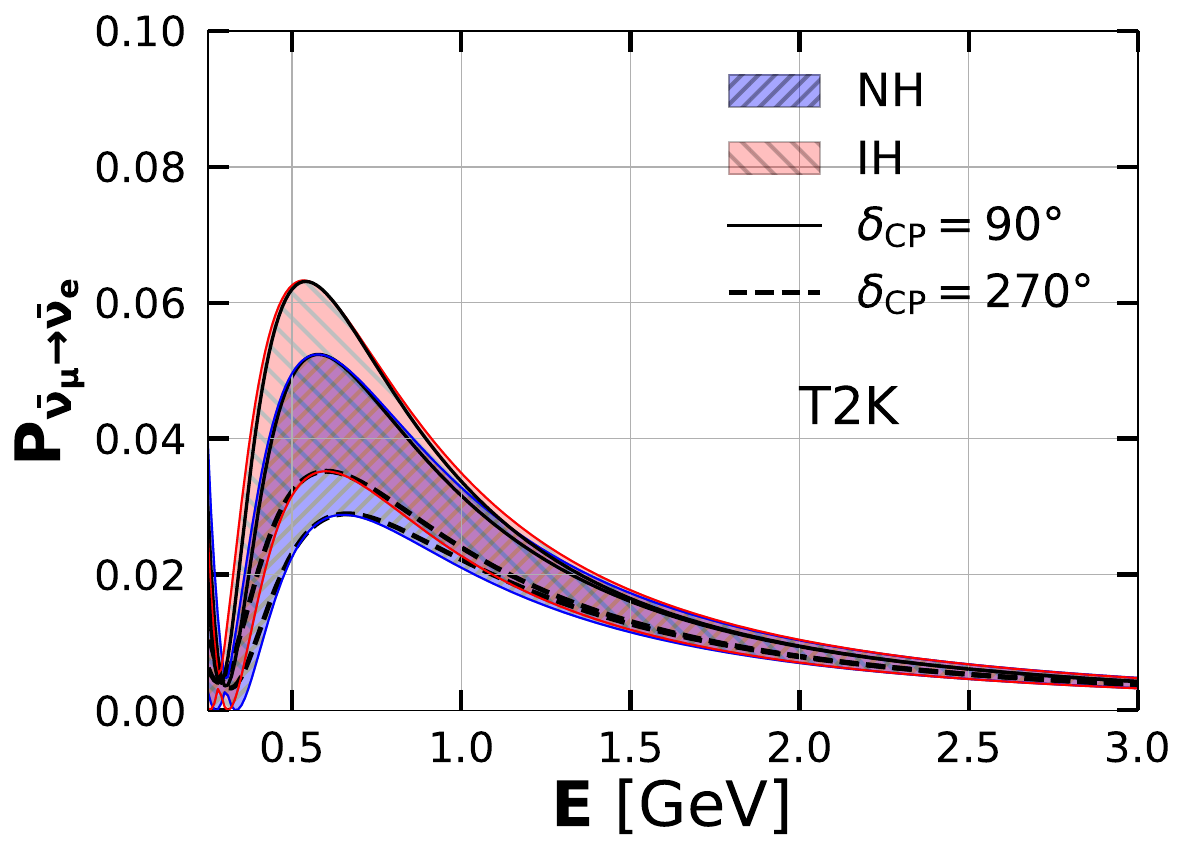}
        \caption{}
    \end{subfigure}

    \vspace{0.5cm} 
    \caption{$\pme$ (left panel) and $\pmebar$ (right panel) oscillation probabilities as a function of neutrino energy depicting the hierarchy-$\delta_{\rm CP}$ degeneracy for \nova (upper panel) and T2K (lower panel) experiments. The blue (red) band in all the plots represents the variation of $\delta_{\rm CP}$ when the neutrino mass hierarchy is NH (IH). The value of $\theta_{23}$ is taken to be maximal (i.e. $45\degree$), and the remaining oscillation parameters are adopted from NuFIT 6.0 given in Table~\ref{tab:t1}. }
    \label{fig:t2k_nova_probband}
\end{figure}

    For a precisely measured $\ty$, there can be three different parameter degeneracies, namely hierarchy-$\dcp$ degeneracy \cite{Barger:2001yr, Minakata:2003wq, Mena:2004sa}, hierarchy-octant degeneracy, and octant-$\dcp$ degeneracy \cite{Agarwalla:2013ju}. Among these, the hierarchy-octant and octant-$\dcp$ degeneracies can be removed by the combination of neutrino and antineutrino data\cite{Agarwalla:2013ju, Prakash:2013dua}. However, the hierarchy-$\dcp$ degeneracy cannot be removed by the combination of neutrino and antineutrino data, and this degeneracy can affect the results of \nova and T2K experiments.

    From Fig.~\ref{fig:t2k_nova_probband}, we can see that for NO$\nu$A, $\pme$ ($\pmebar$) becomes maximum (minimum) for NH and $\dcp=270^\circ$ and it becomes minimum (maximum) for IH and $\dcp=90^\circ$. Therefore, $\pme$ ($\pmebar$) for these two hierarchy-$\dcp$ combinations are well separated from other hierarchy-$\dcp$ combinations. But, the $\pme$ and $\pmebar$ for NH and $\dcp=+90^\circ$, and IH and $\dcp=270^\circ$ are degenerate with each other, and it is not possible to distinguish them from each other. In general, NH (IH) and $\dcp$ in LHP (UHP) is the favourable hierarchy-$\dcp$ combination to determine these two unknowns in NO$\nu$A. On the other hand, NH (IH) and $\dcp$ in UHP (LHP) is the unfavourable hierarchy-$\dcp$ combination to determine these two unknowns in NO$\nu$A. For these unfavourable combinations, the change in the first term of Eq.~\ref{pme} due to matter effect, is cancelled by the change due to the unfavourable value of $\dcp$, leading to the hierarchy-$\dcp$ degeneracy. Since the hierarchy-$\dcp$ degeneracy between the two unfavourable hierarchy-$\dcp$ combinations exists in both $\pme$ and $\pmebar$, this degeneracy cannot be removed by antineutrino run.
    
    The energy of T2K is only one-third of the energy of NO$\nu$A, and hence the matter effect for T2K is correspondingly smaller. Therefore, T2K has very little hierarchy sensitivity. From Fig.~\ref{fig:t2k_nova_probband}, we can see that $\pme$ and $\pmebar$ of T2K for $\dcp=270^\circ$ are well separated from those for $\dcp=90^\circ$. However, for a particular $\dcp$ value, $\pme$ and $\pmebar$ for NH and IH are not well separated. Hence, T2K has a better CP sensitivity than its hierarchy sensitivity. The cancellation of change due to matter effect in the expression of $\pme$ in Eq.~\ref{pme} occurs for different values of $\dcp$ for \nova and T2K. Hence, their combination can lead to small hierarchy sensitivity even for the unfavourable combinations of hierarchy and $\dcp$ \cite{Mena:2004sa, Prakash:2012az, Agarwalla:2012bv}.

\section{Analysis details}
\label{analysis}
Here we first analyze the \nova 2024 \cite{Wolcott:2024} and T2K 2020~\cite{T2K:2021xwb, T2K:2025yoy} data individually and then perform the joint fit of T2K and \nova data, obtaining the best-fit values of the oscillation parameters for both the cases of individual and joint analyses. We want to emphasize that the latest data release from T2K was in 2020. All the results from T2K after that, including the one in Ref.~\cite{T2K:2025yoy}, used the same amount of POTs as in Ref.~\cite{T2K:2021xwb}. Even the joint analysis of \nova and T2K \cite{Mikola:2024jnj} used the same data. On the other hand, \nova released their latest data in 2024 \cite{Wolcott:2024}. Hence, we used \nova 2024 and T2K 2020 data in our analyses. For this purpose, we keep solar parameters namely $\sin^2{\theta_{12}}$ and $\Delta_{21}$ at their best-fit values $0.310$ and $7.39\times 10^{-5}$ eV$^2$, respectively \cite{Esteban:2018azc}\footnote{Since, we have analysed the T2K data published in 2020, we have used the global fit prior to that data.}. For $\theta_{13}$ and $\theta_{23}$, we vary $\sin^2{\theta_{13}}$ in its $3\sigma$ range around its best-fit value $0.02237$ while $\sin^2{\theta_{23}}$ has been varied from $[0.35:0.65]$. Similarly, we vary $|\Delta_{3l}|$ ($l =1$ for NH and $l=2$ for IH) in its $3\sigma$ range around its best-fit values given in Ref.~\cite{Esteban:2018azc}. The Dirac CP phase $\dcp$ has been varied in its full range $[0:2\pi]$. It is worth noting that the analysis presented in Sec.~\ref{result} involves the latest available experimental data from \nova and T2K. Since the most recent T2K data release included in our analysis dates back to 2020, we adopt the best-fit values of the fixed parameters $\tx$ and $\ds$, as well as the allowed ranges of the free parameters, from a global fit~\cite{Esteban:2018azc} that predates this data release. This choice avoids potential bias arising from using global-fit results that already incorporate the T2K data under consideration.

We determine the theoretical event rates and compute the $\chi^2$ value by comparing the data with these theoretical rates using the GLoBES~\cite{Huber:2004ka,Huber:2007ji} package. The experimental data are sourced from Ref.~\cite{Wolcott:2024, T2K:2021xwb}. To compute the theoretical event rates, we fix the signal and background efficiencies to align with the Monte Carlo simulations provided by the collaborations~\cite{T2K:2021xwb, Wolcott:2024}. Energy smearing for the generated theoretical events is automatically applied within GLoBES~\cite{Huber:2004ka,Huber:2007ji} using a Gaussian smearing function on a bin-by-bin basis (for detailed analysis, see ~\cite{Rahaman:2021zzm, Yu:2024nkc}). 

A Poissonian $\chi^2$ between the theory and the experiment is calculated as:
\begin{eqnarray}
\chi^2 &=& 2\sum_i \left\{
 N_i^{\rm th} - N_i^{\rm exp} + N_i^{\rm exp} 
\ln\left[ \frac{N_i^{\rm exp}}{ N_i^{\rm th}} \right]
\right\} + 2 \sum_j  N_j^{\rm th},  \,\, \nonumber \\
\label{poisionian}
\end{eqnarray}
where $i$ stands for the bins for which $N_i^{\rm exp}\neq 0$ and  $j$ stands for the bins for which $N_j^{\rm exp} = 0$. Finally the minimum is subtracted from each $\chi^2$ to calculate $\Delta\chi^2$s. We have also used a prior on $\ty$, with central value $=0.02237$ and uncertainty $ = 2.9\%$ of the central value.

\begin{table}[h!]
\hspace{-1.0cm}
\renewcommand{\arraystretch}{1.3}
\begin{tabular}{|c|c|c|c|c|c|c|c|c|c|}
\hline
\multirow{2}{*}{\textbf{Osc. Params.}} & \multicolumn{2}{c|}{\textbf{NuFIT 4.1}} & \multicolumn{2}{c|}{\textbf{T2K}} & \multicolumn{2}{c|}{\textbf{NO$\nu$A}} & \multicolumn{2}{c|}{\textbf{T2K+NO$\nu$A}} \\
\cline{2-9}
& NH & IH & NH & IH & NH & IH & NH & IH \\
\hline
$\sin^2\theta_{12}$         &   0.310   & 0.310     & 0.310     & 0.310     &  0.310    & 0.310     &   0.310   & 0.310     \\
                   (fixed)  &      &      &      &      &      &      &      &      \\

\hline
$\sin^2\theta_{23}$         & 0.563     &  0.565    &  0.560    &  0.560    &    0.570  &  0.570    &  0.570    &  0.570    \\
                  (free range $[0.43:0.62]$)   &      &      &      &      &      &      &      &      \\
\hline
$\sin^2\theta_{13}$        &  0.02237    &  0.02259    &   0.02237   &  0.02237    &  0.02237    & 0.02237     & 0.02237     &  0.02237    \\
           (prior with error $2.9\%$)          &      &      &      &      &      &      &      &      \\
\hline
$\delta_{\text{CP}}~(\degree)$ &  221    &  282    & 250     &     270 &  150   & 280     &  180    & 270     \\
             (free range $[0:360^\circ]$)        &      &      &      &      &      &      &      &      \\
\hline
$\ds~(\times 10^{5})$ eV$^2$     &  7.39    & 7.39     & 7.39     & 7.39     &   7.39     &   7.39   &   7.39   &7.39      \\
               (fixed)      &      &      &      &      &      &      &      &      \\
\hline
$\dl~(\times 10^{3})$ eV$^2$    & 2.528     & -2.436      &    2.540   & -2.436    &  2.504     &   -2.418   & 2.528    &  -2.418    \\
                (free range $[2.451:2.578]$)     &      &      &      &     &      &      &      &      \\
\hline
\end{tabular}
\caption{Best fit values of standard 3-flavours neutrino oscillation parameters for normal hierarchy (NH) and inverted hierarchy (IH) of neutrino masses used in this analysis. For T2K, \nova and T2K+\nova best fit values we fit the actual data of respective experiments and then obtain the values of the oscillation parameters. We have also mentioned if the parameter has been treated as a fixed or free parameter, ranges of the free parameters, and if any prior has been added on the parameter in the parentheses.}
\label{Table:bestfit}
\end{table}

\begin{table}[h!]
\centering
\renewcommand{\arraystretch}{1.3}
\begin{tabular}{|c|c|c|c|c|c|}
\hline
\multirow{2}{*}{\textbf{Uncertainty Source}} & \multicolumn{2}{c|}{\textbf{T2K}} & \multicolumn{2}{c|}{\textbf{NO$\nu$A}} & \textbf{JUNO} \\
\cline{2-6}
& \textbf{App.} & \textbf{Disapp.} & \textbf{App.} & \textbf{Disapp.} & \textbf{App.}\\
\hline
Signal uncertainty & 5\% & 5\% & 5\% & 5\% & 5\% \\
\hline
Background uncertainty & 5\% & 5\% & 5\% & 5\% & 20\% \\
\hline
Energy calibration error & 5\% & 5\% & 5\% & 0.01\% & 2\% \\
\hline
\end{tabular}
\caption{Systematic uncertainties considered for T2K, NO$\nu$A, and JUNO in this phenomenological study. Here, App. = appearance channel, Disapp. = disappearance channel.}
\label{tab:syst_uncertainty}
\end{table}


We also combine the JUNO simulation with the T2K and \nova data. The corresponding systematic uncertainties for each experiment are listed in Table~\ref{tab:syst_uncertainty}. Note that the actual data from T2K and \nova contain random fluctuations, whereas the simulated JUNO data do not. Therefore, to be consistent, we simulate JUNO data with random fluctuations. In doing so, we use the ``true values'' of the oscillation parameters mentioned in Table~\ref{Table:bestfit} to be their central values, considering one column at a time. Simulations are performed with both NH and IH as the true mass hierarchy. The fluctuations in the simulated data are included in the following way. Using GLoBES, we first calculate the JUNO expected events in the $i$th energy bin, say $N_i^{\rm e}$, with a total of 200 energy bins. Considering this $N_i^{\rm e}$ as the mean, we compute 100 Poissonian random numbers. This procedure is repeated for all 200 energy bins. Thus, for each energy bin, we have 100 possible random events corresponding to the JUNO simulation for a given ``true values'' of the oscillation parameters and mass hierarchy. By collecting these possible event numbers for each bin cautiously, we obtain 100 independent data sets for the JUNO simulations which include the random Poissonian fluctuations expected in experiments. An average $\bar{\chi}^2$ is obtained from these 100 sets of $\chi^2$ values.  The final $\dchsq$ is obtained by subtracting the minimum $\bar{\chi}^2$ from all the $\bar{\chi}^2$ values. 
We find that 100 independent Poissonian-fluctuated data sets are sufficient to match the simulation results provided by the collaboration in Ref.~\cite{JUNO:2015zny}.

In case of JUNO, we have considered different beam powers for the neutrino beams from different reactor cores, as given in Table 1-2 of ref.~\cite{JUNO:2015zny}. However, we have not considered backgrounds and systematics from the individual 
reactor core. We have considered an overall average systematic and background error. The consideration of backgrounds and systematics from individual reactor cores will lead to only a subleading modification for the ordering $\dchsq$.

For the analyses of \tk~, \nova~ or \tk+\nova~ along with JUNO simulated data, we first compute the $\bar{\chi}^2$ from 100 JUNO like experiments as described above. Let us call this as $\bar{\chi}^2_{\rm JUNO}$, and this is a function of the test values of the oscillation parameters for both NH and IH. This process is repeated first assuming NH as the true mass hierarchy and then IH. Thus, we have 4 sets of $\bar{\chi}^2_{\rm JUNO}$ for the true-test combinations of mass hierarchy, i.e., $\bar{\chi}^2_{\rm JUNO}{\rm (NH-NH)}$, $\bar{\chi}^2_{\rm JUNO}{\rm (NH-IH)}$, $\bar{\chi}^2_{\rm JUNO}{\rm (IH-IH)}$ and $\bar{\chi}^2_{\rm JUNO}{\rm (IH-NH)}$. We also calculate the $\chi^2$s for \tk~ and \nova as a function of oscillation parameters for both NH and IH as test hierarchy. To analyse the simulated data with the experimental data of \nova we add the $\chi^2$ from \nova with $\bar{\chi}^2_{\rm JUNO}$ for the same test values of oscillation parameters and test hierarchy of the two experiments. By finding the minimum of all the $\chi^2$ using the same procedure mentioned previously, we then calculate the $\Delta \chi^2$, i.e., $\Delta\chi^2_{\rm NOvA+JUNO}$ for both the test hierarchies. In the similar manner, we also calculate $\Delta\chi^2_{\rm T2K+JUNO},\Delta\chi^2_{\rm T2K+NOvA+JUNO}$ for NH and IH test hierarchy cases. 

In the next section, we discuss our results where we have calculated the allowed regions for $\sin^2\theta_{23}-\delta_{\rm CP}$ considering different choices of the true-test combinations. To do so, we marginalise these $\dchsq$s over the test values of oscillation parameters, except test values of $\sin^2\theta_{23}$ and $\delta_{\rm CP}$. We also present our results in the form of CP violation (CPV) sensitivity plots. For these plots, we marginalise the $\dchsq$ over all the test oscillation parameters except $\delta_{\rm CP}$. In case of known hierarchy sensitivities, true and test hierarchies for JUNO simulations have been kept same. For \nova and T2K data analysis, there are no true values of parameters, because we have used the actual experimental data. In the case of unknown hierarchy, marginalisation is been done over test hierarchy as well, for both the case of real experimental data of \nova and T2K as well as the simulated fluctuation-induced data of JUNO. In our analysis, we simulate JUNO experiments with true parameter values fixed at
\begin{itemize}
    \item \nova best-fit values,
    \item T2K best-fit values,
    \item NO$\nu$A+T2K best-fit values,
    \item NuFIT4.1 \cite{Esteban:2018azc} best-fit values.
\end{itemize}
The parameter values for all these four cases are listed in Table \ref{Table:bestfit}.

\begin{figure}[htbp]
\includegraphics[width=0.8\textwidth,height=0.4\textheight]{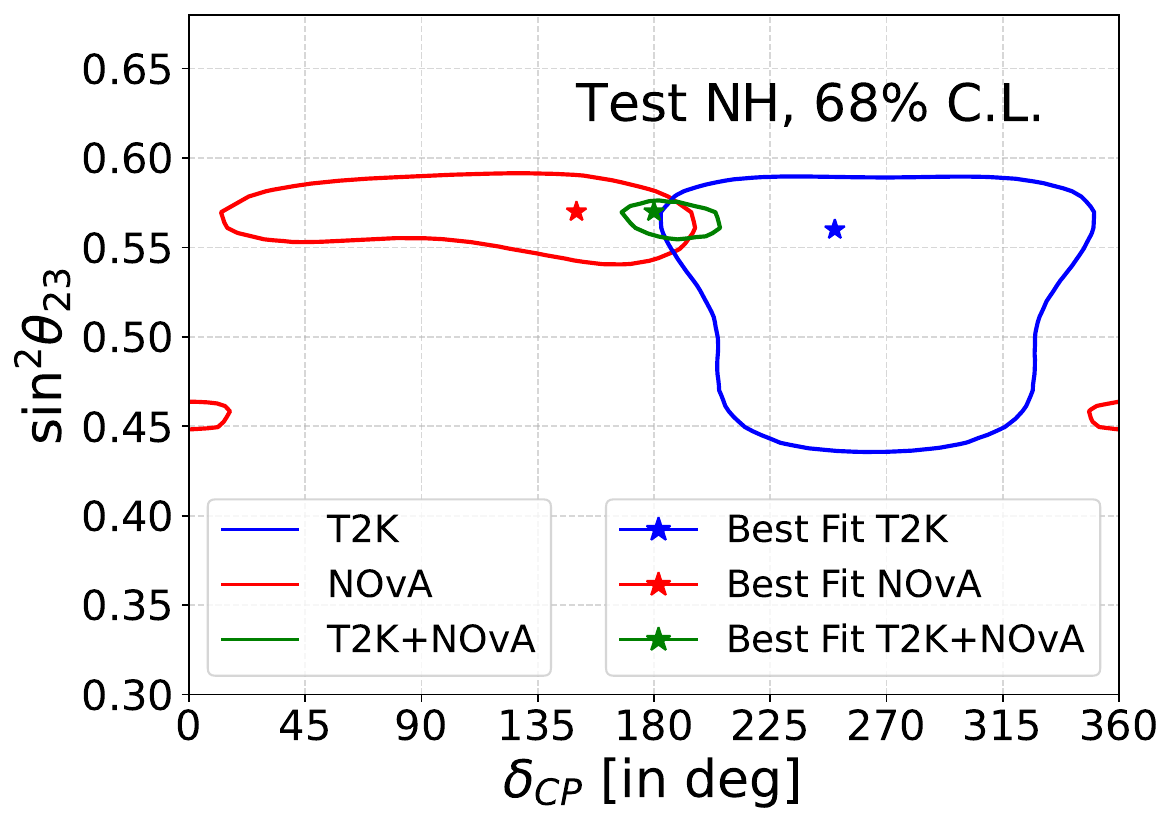}
\vspace*{0.7cm}

\includegraphics[width=0.8\textwidth,height=0.4\textheight]{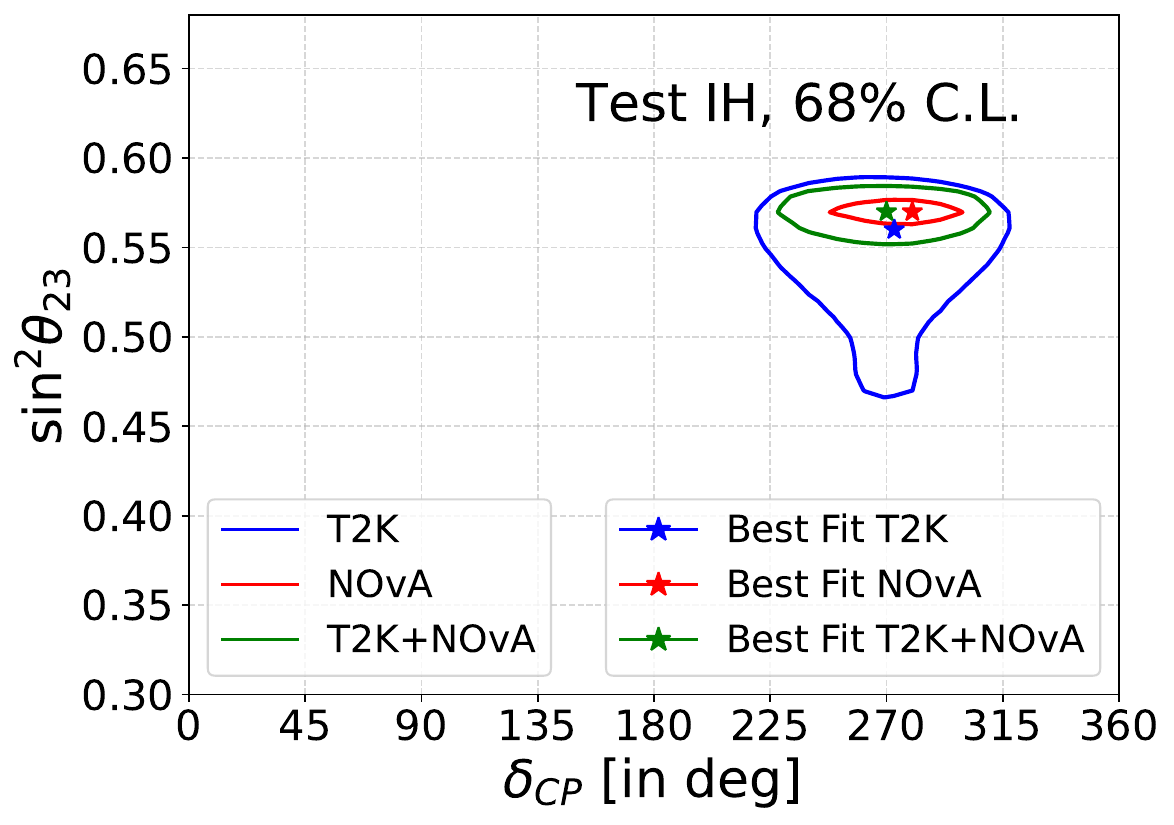}
 
	\caption{Allowed region in the $\sin^2{\theta_{23}}-\delta_{\rm CP}$ plane after complete analysis of \nova 2024 and T2K 2020 data.}
	\label{fig:t2k_nova_match}
\end{figure}
\section{Results}
\label{result}
In this section, we will present the results of our analysis. 
\subsection{Current status of the \nova and T2K data}

Figure~\ref{fig:t2k_nova_match} shows the $1\sigma$ allowed regions obtained independently from the \nova and T2K datasets. Both experiments obtain their respective best-fit points in the normal hierarchy. However, the corresponding $1\sigma$ allowed regions do not overlap, indicating a clear incompatibility between the two datasets if NH is assumed. This reproduces the well-known T2K--\nova tension under normal ordering.

In contrast, when the inverted hierarchy is assumed, the best-fit points of the two experiments become nearly degenerate: we obtain $\Delta\chi^2 = 1.90$ for \nova and $\Delta\chi^2 = 0.34$ for T2K relative to their NH minima. Under IH, the $1\sigma$ contours overlap substantially, and the two datasets are statistically consistent. Consequently, the combined analysis favours IH as the global best-fit.

In the combined fit of both experiments, for NH, only a small CP-conserving region survives at $1\sigma$, with a minimum $\Delta\chi^2 \approx 1.63$ relative to the best-fit obtained at IH for the combined data set of \nova and T2K. It is also worth noting that \nova continues to favour the so-called unfavourable hierarchy--$\dcp$ combinations, namely NH-UHP and IH-LHP, in agreement with earlier analyses~\cite{Himmel:2020, NOvA:2021nfi, Rahaman:2021zzm, Rahaman_2022}.

\subsection{Effects of combining JUNO simulations with the present data from \nova and T2K: hierarchy and $\dcp$--octant structure}\label{subsec: fig5_6_7}

We now study how the addition of simulated JUNO data modifies the constraints obtained from the present \nova and T2K datasets. In the first set of simulations, JUNO is generated assuming the NuFIT~4.1 best-fit values as the true oscillation parameters. The resulting allowed regions in the $\sin^2\theta_{23}-\dcp$ plane are shown in Fig.~\ref{fig:s23_dcp_nufit2019}. The top panels correspond to analyses with the mass hierarchy fixed, whereas the bottom panels marginalise over the test hierarchy. The left (right) panels assume NH (IH) as the true hierarchy of the JUNO simulation.

\subsubsection{Known hierarchy}
When the hierarchy is fixed, adding JUNO to either \nova or T2K does not qualitatively modify the allowed regions. Since JUNO has no intrinsic $\dcp$ sensitivity, the $\dcp$--octant structure continues to be governed mostly by the accelerator data.

\subsubsection{Unknown hierarchy}
A qualitatively different behaviour emerges when the hierarchy is treated as unknown. Without JUNO, the \nova data display a hierarchy--$\dcp$ degeneracy at $1\sigma$: the NH best-fit lies in the UHP of $\dcp$, while an alternative IH solution with $\dcp$ in the LHP exists due to NO$\nu$A's preference for unfavourable hierarchy--$\dcp$ combinations. Consequently, \nova admits two disconnected $1\sigma$ regions in the $\sin^2\theta_{23}-\dcp$ plane.

T2K, by contrast, prefers $\dcp$ in the LHP for both NH and IH, and therefore has only a single $1\sigma$ allowed region under free hierarchy. One of the degenerate \nova solutions thus conflicts with the T2K preference.

JUNO's strong hierarchy sensitivity resolves this degeneracy by selecting the correct mass ordering in all simulated cases:
\begin{itemize}
    \item \textbf{If JUNO is simulated with NH true:} JUNO selects NH. The surviving \nova solution lies in the UHP, while T2K prefers the LHP. Hence, the \nova--T2K tension persists.
    \item \textbf{If JUNO is simulated with IH true:} JUNO selects IH (min. $\dchsq$ for NH test in case of JUNO only simulation is $\approx$ 72.31). Under IH, both \nova and T2K prefer $\dcp$ in the LHP, and the tension between the experiments disappears.
\end{itemize}
Thus, JUNO always removes the hierarchy--$\dcp$ degeneracy inherent to \nova, but the survival or removal of the \nova--T2K tension depends entirely on the underlying hierarchy.

\begin{figure}[htbp]
    \centering

    \begin{subfigure}[b]{0.45\textwidth}
        \centering
        \includegraphics[width=\textwidth, height=6cm]{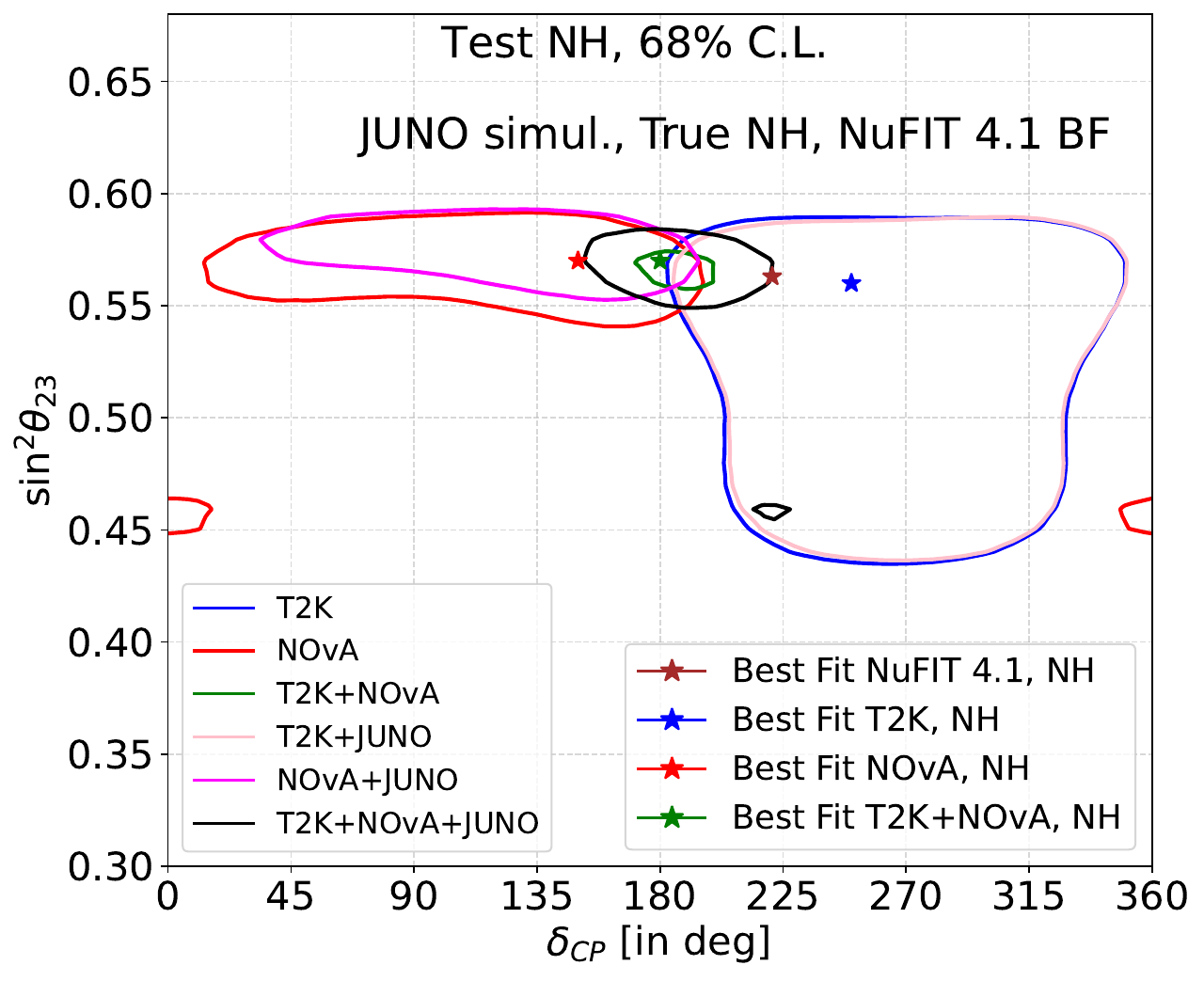}
        \caption{}
    \end{subfigure}
    \hfill
    \begin{subfigure}[b]{0.45\textwidth}
        \centering
        \includegraphics[width=\textwidth, height=6cm]{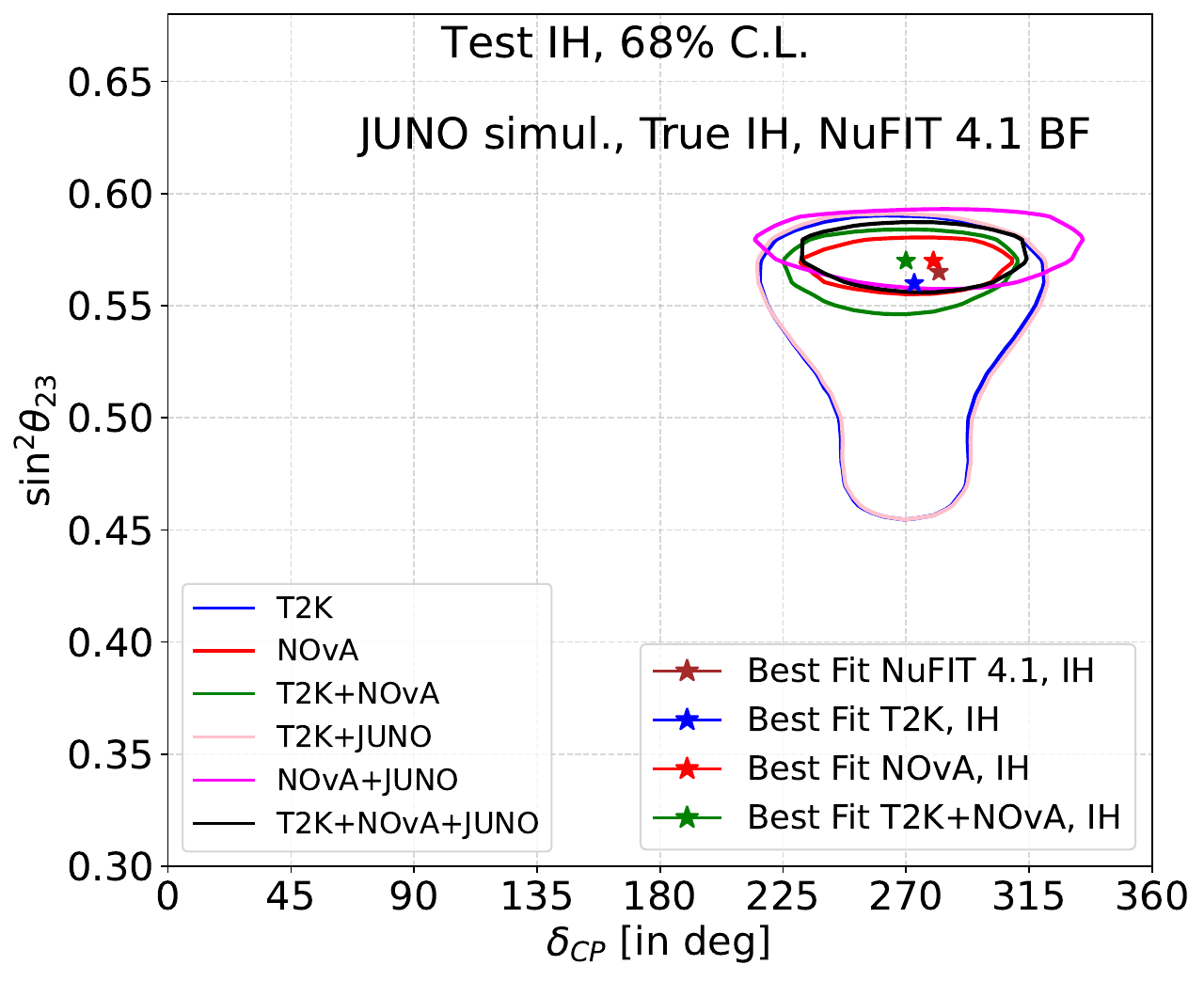}
        \caption{}
    \end{subfigure}

    \vspace{0.5cm} 

    \begin{subfigure}[b]{0.45\textwidth}
        \centering
        \includegraphics[width=\textwidth, height=6cm]{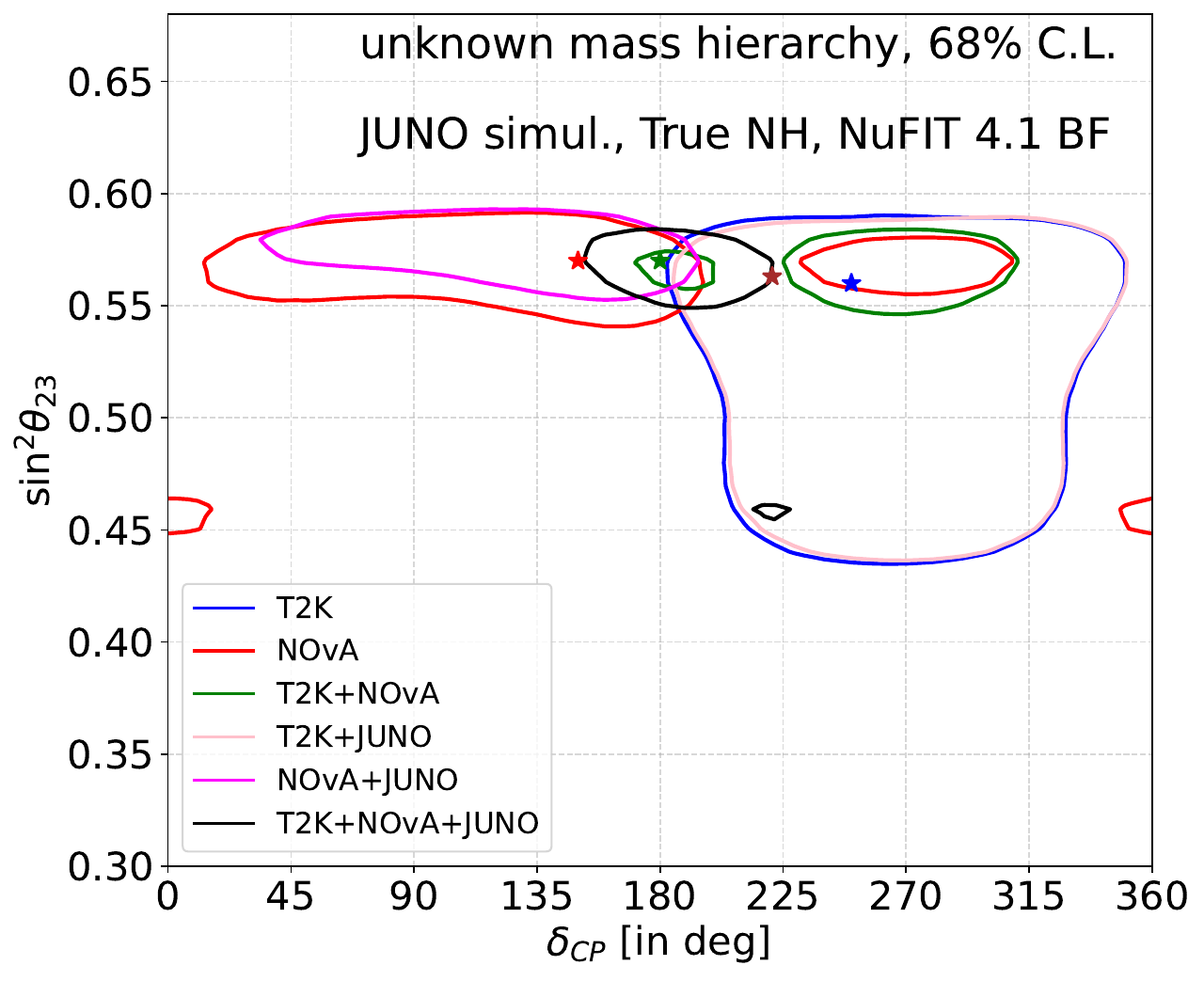}
        \caption{}
    \end{subfigure}
    \hfill
    \begin{subfigure}[b]{0.45\textwidth}
        \centering
        \includegraphics[width=\textwidth, height=6cm]{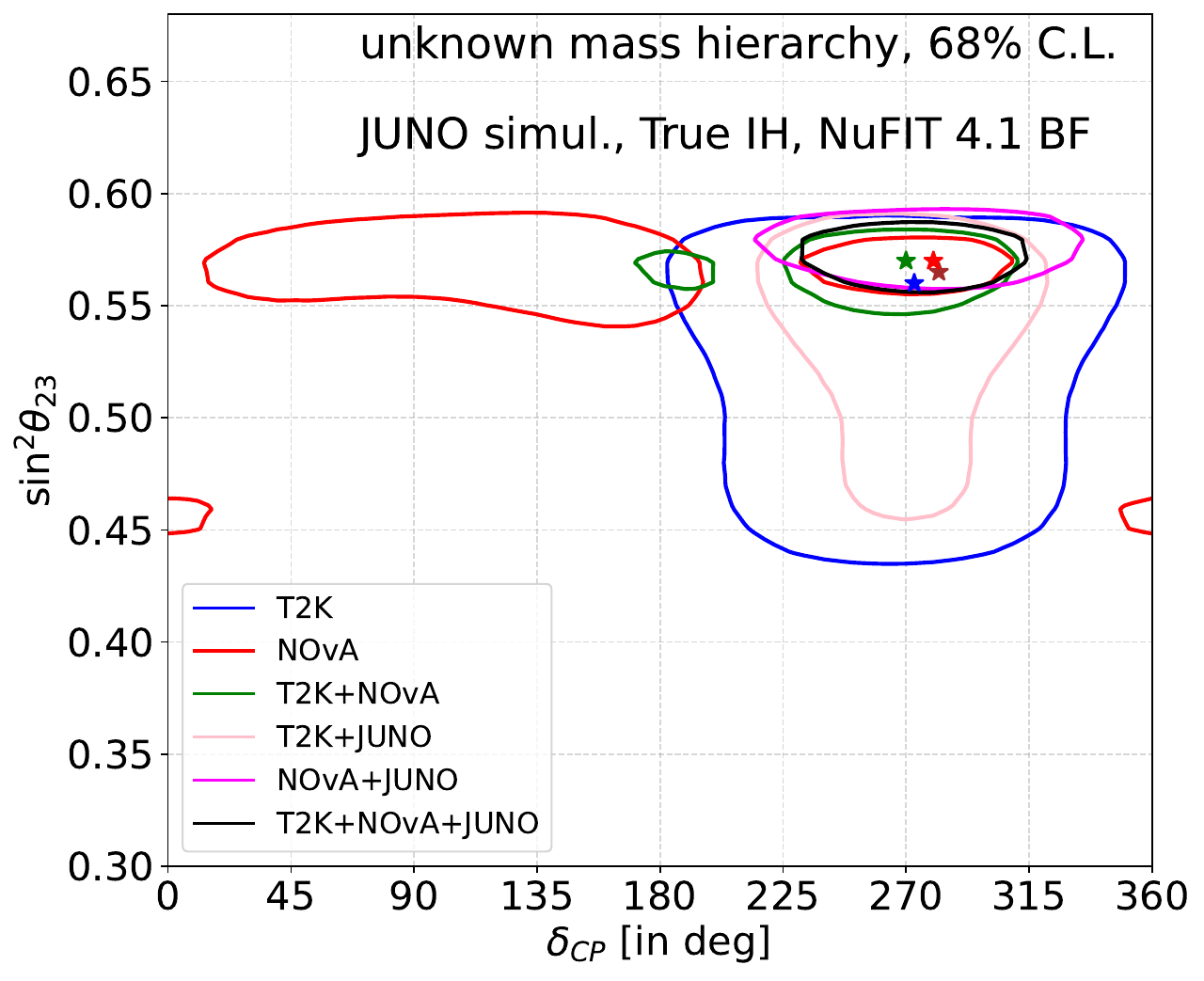}
        \caption{}
    \end{subfigure}

    \caption{Allowed regions in the $\sin^2{\theta_{23}}-\delta_{\rm CP}$ plane including T2K, \nova data and JUNO simulation. The true events of JUNO have been simulated with the best-fit values from Ref.~\cite{Esteban:2018azc} as the true values of oscillation parameters. The top (bottom) panels present the cases for known (unknown) mass hierarchy and the left (right) panels present the cases when the true hierarchy for JUNO is NH (IH).}
    \label{fig:s23_dcp_nufit2019}
\end{figure}

\begin{figure}[htbp]
\includegraphics[width=0.8\textwidth,height=0.4\textheight]{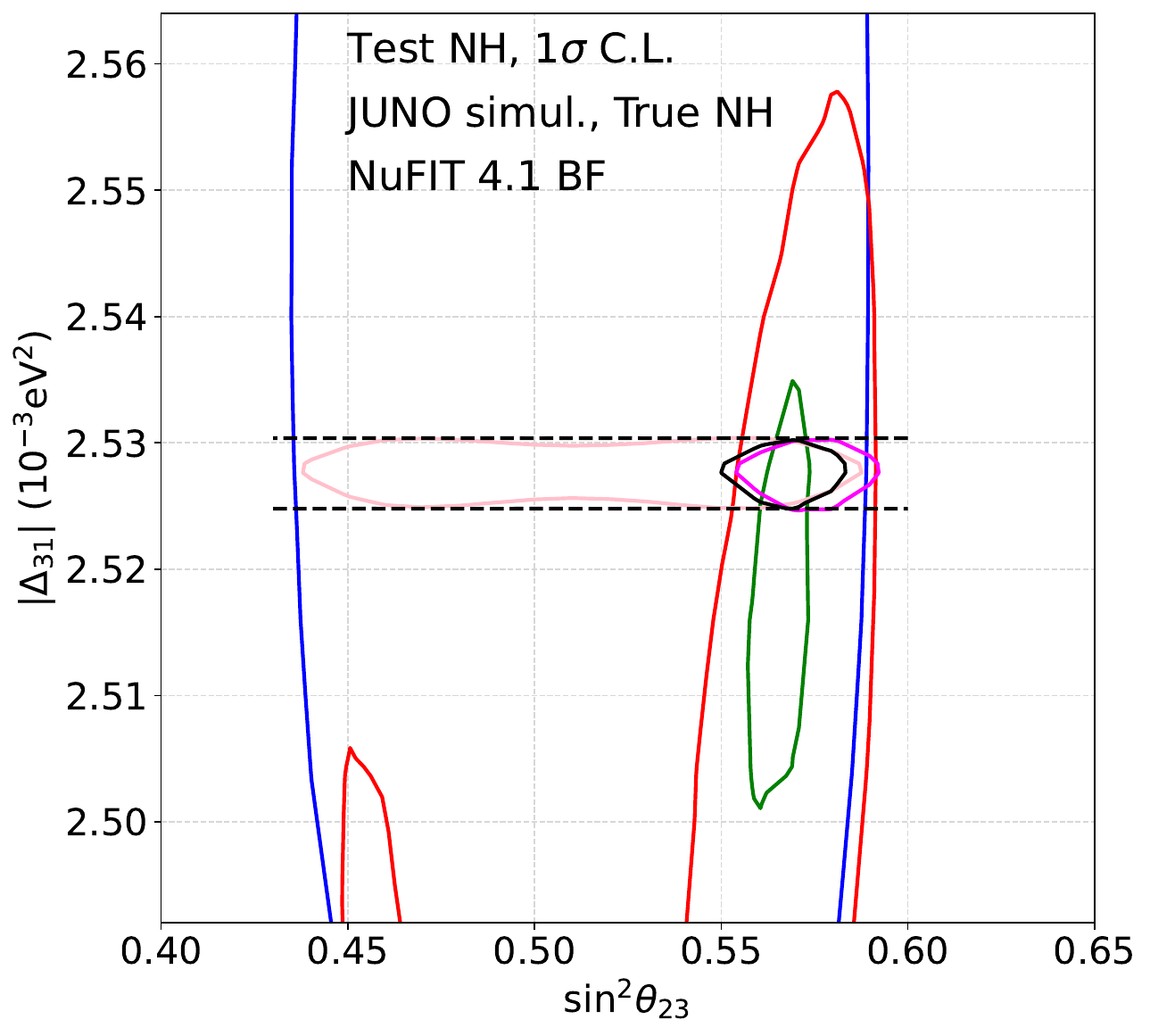}
\vspace*{0.7cm}

\includegraphics[width=0.8\textwidth,height=0.4\textheight]{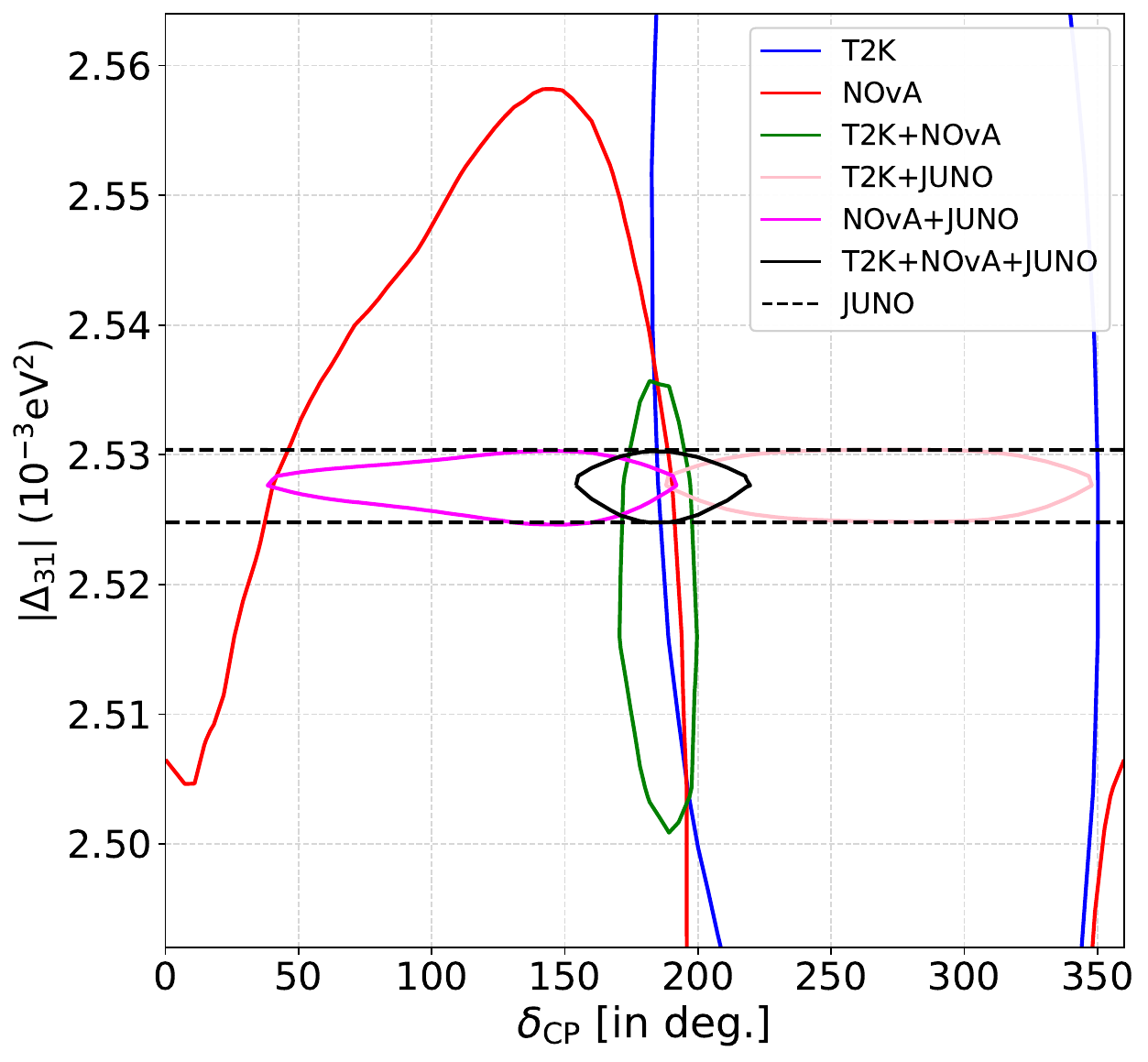}
 
	\caption{$1\,\sigma$ allowed region in the $\Delta_{31}-\sin^2{\theta_{23}}$ ($\dl-\dcp$)  plane after complete analysis of \nova 2024 and T2K 2020 data in the top (bottom) panel. The true events of JUNO have been simulated with the best-fit values from Ref.~\cite{Esteban:2018azc} as the true values of oscillation parameters.}
	\label{fig:th23_dm31}
\end{figure}
\subsubsection{Robustness to the choice of JUNO true values.}
Figures~\ref{fig:s23_dcp_t2k}, \ref{fig:s23_dcp_nova}, and \ref{fig:s23_dcp_t2k_nova} (in Appendix~\ref{app:A}) show the results obtained when JUNO is simulated using the T2K best-fit, \nova best-fit, or the combined best-fit points as the true values. The qualitative behaviour is unchanged in all cases: JUNO fixes the hierarchy, while the $\dcp$ preference remains dictated by the accelerator data. Since the main variation among these true points is in $\dcp$, to which JUNO is insensitive, the overall conclusions are unaffected.

\subsubsection{Impact on $|\dl|$, $\sin^2\theta_{23}$, and $\dcp$ precision}
A notable consequence of adding JUNO is the improvement of the \nova constraints on $\sin^2\tz$ and $\dcp$ when NH is true and the hierarchy is known. JUNO provides excellent precision on $|\Delta_{31}|$ (or $|\Delta_{32}|$), whereas \nova exhibits comparatively poor precision in this parameter for NH (IH)(see Fig.~\ref{fig:th23_dm31}). Their combination therefore yields a sharper determination of $|\Delta_{31}|$, which in turn slightly tightens the $\sin^2\theta_{23}-\dcp$ contours.

For T2K, although JUNO improves the measurement of $|\dl|$, the $1\sigma$ regions in $\sin^2\theta_{23}$ and $\dcp$ remain essentially unchanged, as these are largely uncorrelated with $|\dl|$ within the T2K dataset.

Finally, the absence of closed contours in Fig.~\ref{fig:th23_dm31} reflects the fact that the present \nova and T2K datasets allow values of $|\dl|$ outside the $3\sigma$ range of NuFIT~4.1~\cite{Esteban:2018azc}.

\begin{figure}[htbp]
    \centering

    \begin{subfigure}[b]{0.45\textwidth}
        \centering
        \includegraphics[width=\textwidth, height=6cm]{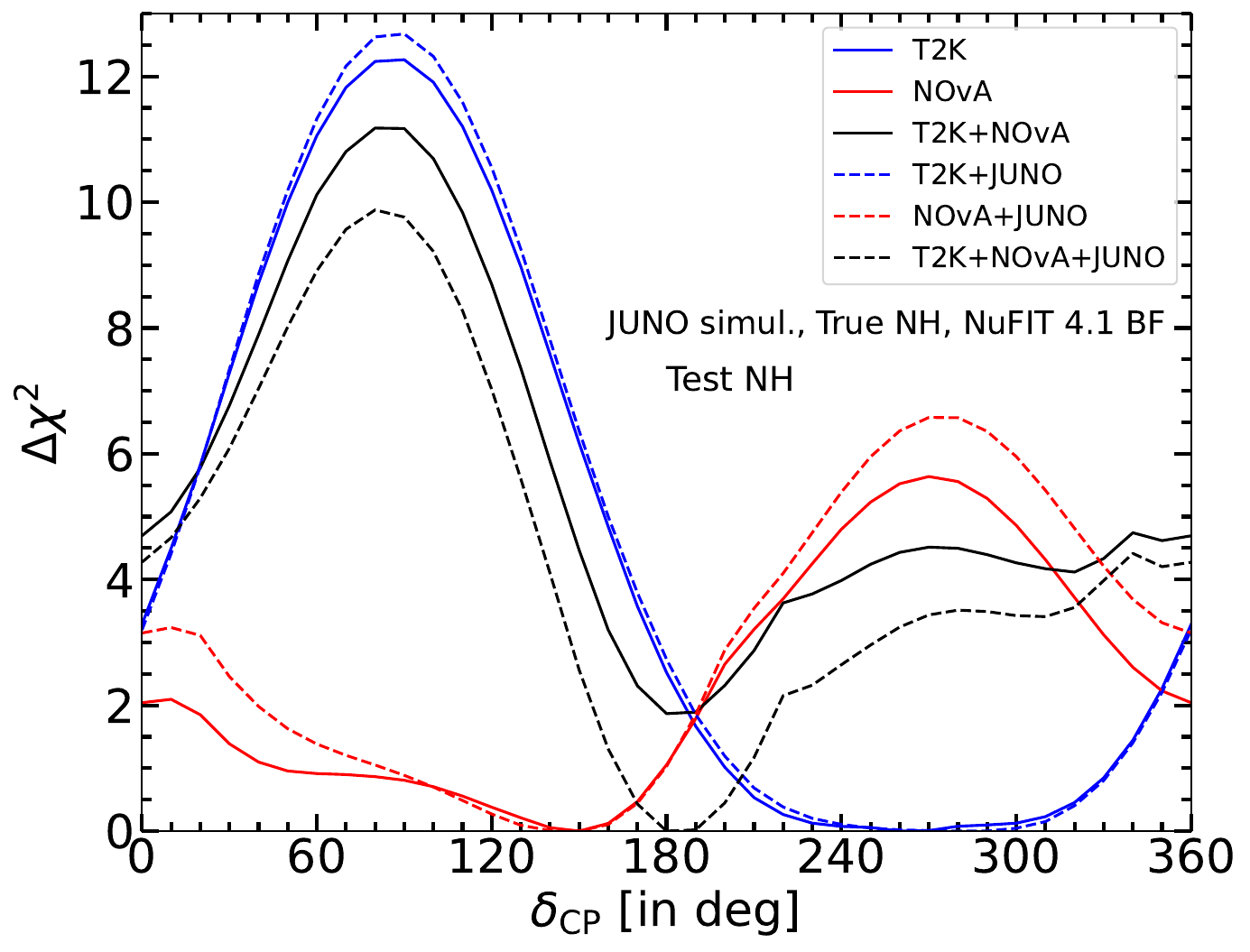}
        \caption{}
    \end{subfigure}
    \hfill
    \begin{subfigure}[b]{0.45\textwidth}
        \centering
        \includegraphics[width=\textwidth, height=6cm]{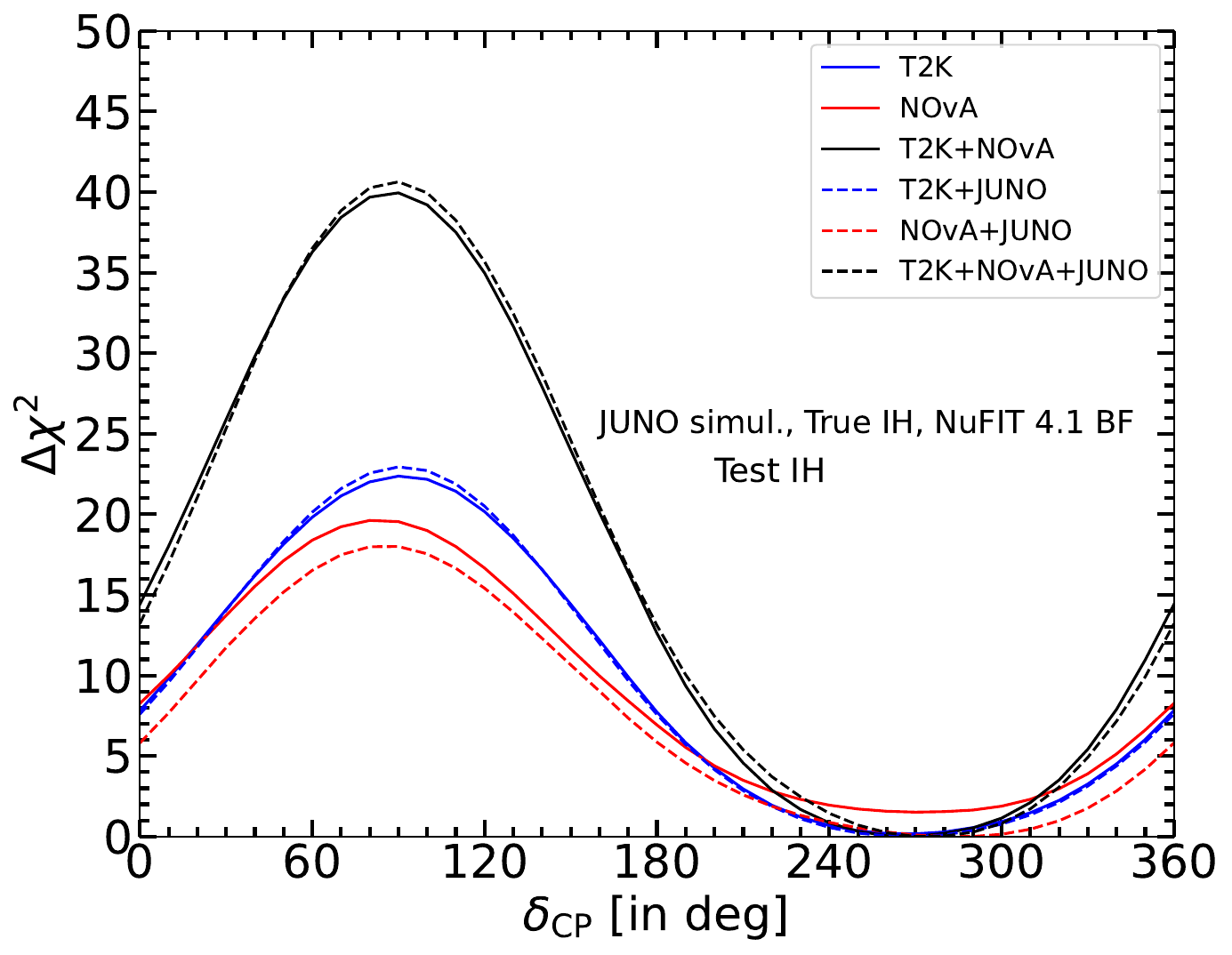}
        \caption{}
    \end{subfigure}

    \vspace{0.5cm} 

    \begin{subfigure}[b]{0.45\textwidth}
        \centering
        \includegraphics[width=\textwidth, height=6cm]{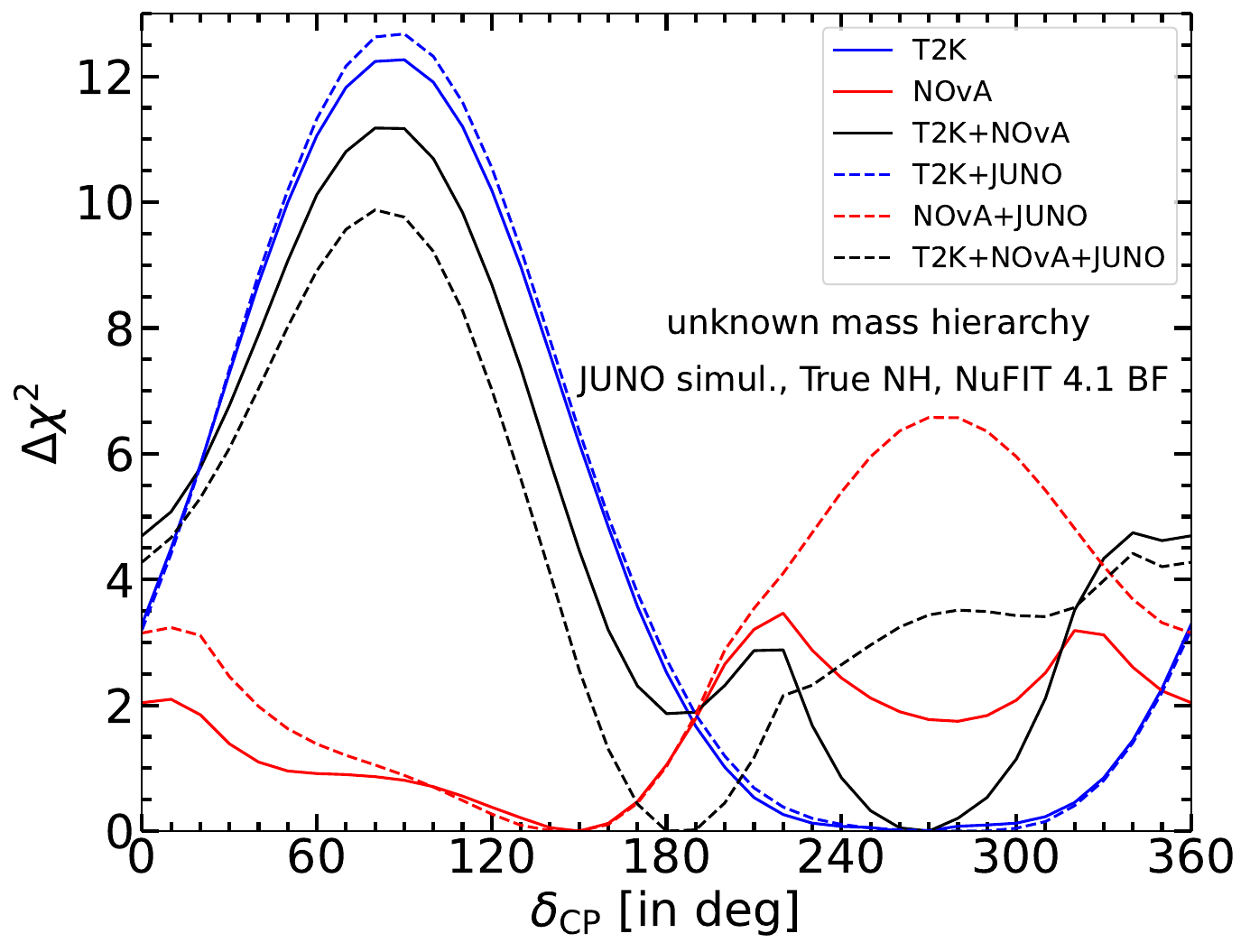}
        \caption{}
    \end{subfigure}
    \hfill
    \begin{subfigure}[b]{0.45\textwidth}
        \centering
        \includegraphics[width=\textwidth, height=6cm]{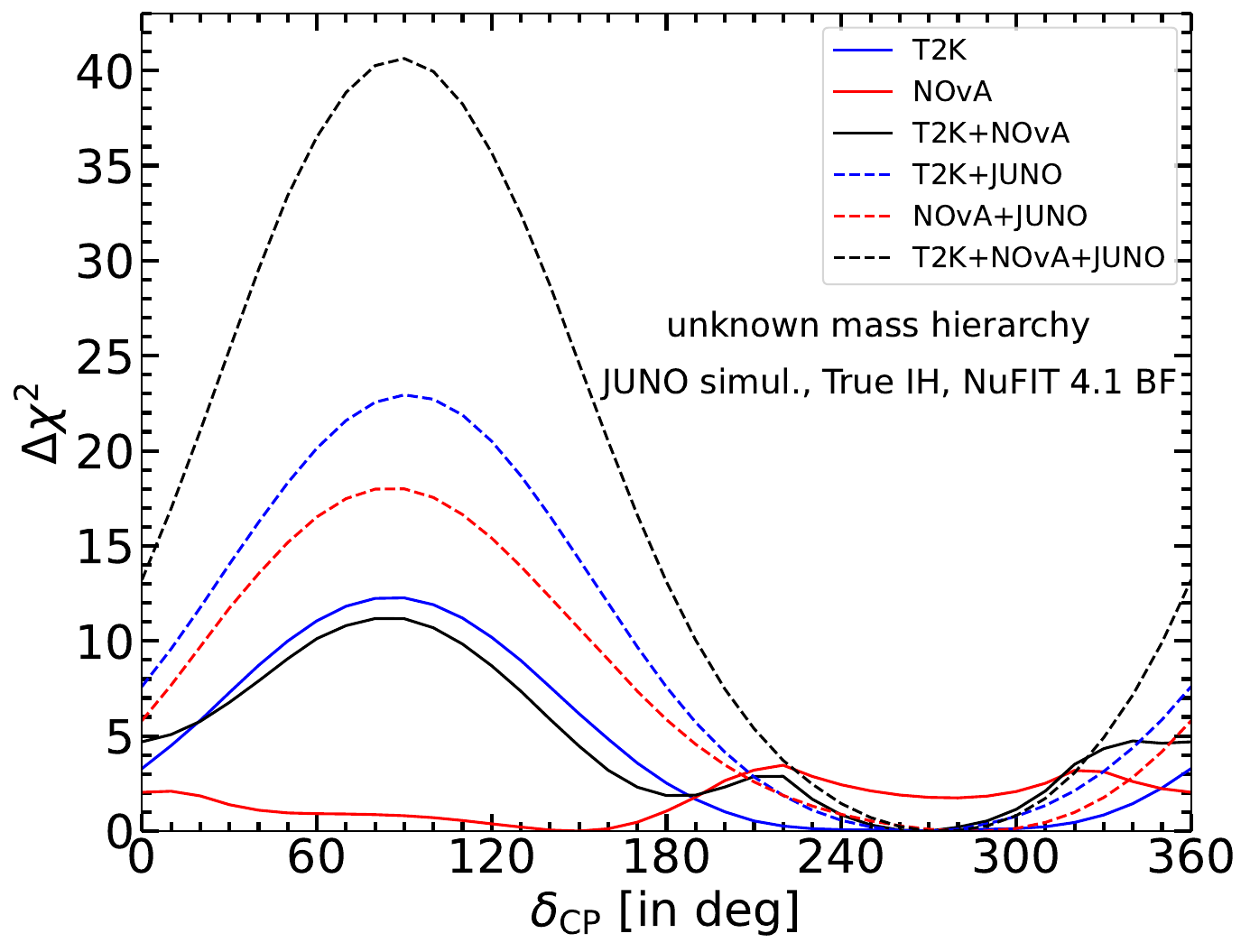}
        \caption{}
    \end{subfigure}

    \caption{CPV sensitivity plots for T2K and \nova data, and JUNO simulations with global best-fit as true parameter values \cite{Esteban:2018azc}. The top (bottom) panels present the cases for known (unknown) mass hierarchy and the left (right) panels present the cases when the true hierarchy for JUNO is NH (IH).}
    \label{fig:res_nufit2019}
\end{figure}

\subsection{Effects of JUNO simulation on \nova and T2K data: CP sensitivity}

The impact of adding JUNO simulations to the present \nova and T2K data on the CP sensitivity is shown in Fig.~\ref{fig:res_nufit2019}. As in the previous subsection, the behaviour depends crucially on whether the mass hierarchy is fixed or marginalised.

\subsubsection{Known hierarchy}
When the hierarchy is fixed, JUNO does not induce any qualitative change in the CP sensitivity. Since JUNO has almost no intrinsic sensitivity to $\dcp$, the CP sensitivity of the combined dataset is determined entirely by the accelerator data. The combined analysis of \nova and T2K dataset, as already discussed, prefers IH over NH as the best-fit solution. If the JUNO simulation assumes NH (IH) as the true hierarchy, then the combined NO$\nu$A+T2K+JUNO fit also selects NH (IH), but the shape of the $\Delta\chi^2(\dcp)$ curves remains essentially unchanged from the NO$\nu$A+T2K fit. In case of IH being the true hierarchy for JUNO simulation, the min. $\dchsq$ for NH test in case of NO$\nu$A+T2K+JUNO data set is $\approx$ 74.68.

\subsubsection{Unknown hierarchy}
A qualitatively different behaviour emerges when the hierarchy is marginalised. As discussed earlier, the \nova data alone exhibit a hierarchy--$\dcp$ degeneracy: NH with $\dcp$ in the UHP is the best-fit, but an alternative IH solution with $\dcp$ in the LHP is nearly degenerate. Consequently, \nova cannot exclude the wrong half-plane of $\dcp$ at better than about $1.5\sigma$.

When JUNO is included, its strong hierarchy sensitivity resolves this ambiguity:
\begin{itemize}
    \item If JUNO is simulated with \textbf{NH true}, the combined fit of \nova + JUNO selects NH, and only the NH--UHP branch survives. In this case, $\dcp$ in the LHP can be excluded at more than $2.5\sigma$.
    \item If JUNO is simulated with \textbf{IH true}, the combined fit of \nova + JUNO selects IH (min. $\dchsq$ for NH test for the \nova + JUNO data set is $\approx$ 70.46), and only the IH--LHP branch survives. Here $\dcp$ in the UHP can be excluded at nearly $4\sigma$.
\end{itemize}
Thus, JUNO helps to determine which of the two degenerate solutions of \nova survives.

\subsubsection{Combined \nova + T2K + JUNO}
For the combined accelerator dataset, the best-fit point occurs at $\dcp = 270^\circ$ under IH. A secondary minimum appears near $\dcp = 180^\circ$, corresponding to the NH best-fit with a $\dchsq=1.63$. When JUNO is added:
\begin{itemize}
    \item If JUNO is simulated with \textbf{NH true}, the global minimum shifts to $\dcp = 180^\circ$ and the other minimum disappears.
    \item If JUNO is simulated with \textbf{IH true}, the global minimum remains at $\dcp = 270^\circ$ and the $\dcp = 180^\circ$ minimum disappears.
\end{itemize}
In all cases, JUNO determines the neutrino mass ordering, while the accelerator data continue to drive the $\dcp$ preference.

\subsubsection{Robustness to true parameter choices}
We have repeated this study using JUNO simulations generated with the T2K best-fit, \nova best-fit, and the combined NO$\nu$A+T2K best-fit points as true values. In all cases, the qualitative conclusions remain unchanged. Since JUNO has negligible sensitivity to $\dcp$, variations in the true value of $\dcp$ do not affect our results. The Figures for JUNO simulations with the other three true parameter values have been presented in Appendix~\ref{app:B}.

\subsection{Summary}

The results of the above subsections can be summarised as follows. When the mass hierarchy is treated as a variable over which $\Delta\chi^2$ is marginalised:

\begin{itemize}

    \item \textbf{\nova exhibits a hierarchy--$\dcp$ degeneracy.}
    NH with $\dcp$ in the UHP and IH with $\dcp$ in the LHP form two nearly degenerate $1\sigma$ solutions.

    \item \textbf{T2K does not exhibit such a degeneracy.}
    Although T2K cannot rule out either hierarchy, it consistently prefers $\dcp$ in the LHP for both NH and IH.

    \item \textbf{A tension exists between the two experiments.}
    One of the degenerate \nova solutions in the $\sin^2\theta_{23}-\dcp$ plane is incompatible with the corresponding T2K allowed region.

    \item \textbf{JUNO resolves the hierarchy--$\dcp$ degeneracy of \nova.}
    Its strong hierarchy sensitivity and $\dcp$ insensitivity remove the two-branch structure in \nova for both NH and IH true hierarchies.

    \item \textbf{If NH is the true hierarchy:}
    JUNO forces the combined NO$\nu$A+JUNO fit to NH, leaving only the \nova solution that is in tension with T2K; the \nova solution consistent with T2K is removed.

    \item \textbf{If IH is the true hierarchy:}
    JUNO forces the combined NO$\nu$A+JUNO fit to IH (min. $\dchsq$ for NH test is $\approx$ 70.46), leaving only the \nova solution that is compatible with T2K; the tension between the two experiments is lifted.

    \item \textbf{T2K benefits from JUNO through improved hierarchy sensitivity.}
    Although JUNO has no $\dcp$ sensitivity, its hierarchy measurement strengthens the combined accelerator hierarchy preference.

    \item \textbf{Dependence on JUNO true parameter values.}
    Using JUNO simulations generated with
    \begin{enumerate}
        \item \textbf{the global best-fit values of Ref.~\cite{Esteban_2024}}: The best-fit point for NH is included in the $1\,\sigma$ allowed region of NO$\nu$A+T2K+JUNO fit,
        \item \textbf{the \nova best-fit for NH}: The best-fit point is outside of the $1\,\sigma$ allowed region of NO$\nu$A+T2K+JUNO fit,
        \item \textbf{the T2K best-fit for NH}: The best-fit point is close to the $1\,\sigma$ allowed region of NO$\nu$A+T2K+JUNO fit,
        \item \textbf{the NH best-fit of the combined NO$\nu$A+T2K analysis}: The best-fit point is included in the $1\,\sigma$ allowed region of NO$\nu$A+T2K+JUNO fit.
    \end{enumerate}
\end{itemize}

Apart from these, when mass hierarchy is known:

\textbf{Precision improvements:}
    JUNO significantly improves the $|\dl|$ (or $|\Delta_{32}|$) precision of both \nova and T2K for NH (IH).  
    For \nova, this improvement propagates to modest tightening of the $\sin^2\theta_{23}$ and $\dcp$ constraints at $1\sigma$ for NH, while for T2K the improvement remains confined to $|\dl|$.

\section{Future Sensitivities}
\label{future}

In this section, we present the future CP--violation discovery, octant and mass--hierarchy sensitivities expected from \nova, T2K, and JUNO. We assume exposures of $13.305\times 10^{21}$ ($6.25\times 10^{21}$) POT for the $\nu$ ($\bar\nu$) run of \nova, $9.85\times 10^{21}$ ($8.15\times 10^{21}$) POT for the $\nu$ ($\bar\nu$) run of T2K, and 6 years of JUNO data (corresponding to $\sim 10^5$ IBD events). The true value of $\dcp$ is varied over its full range $[-180^\circ,180^\circ]$, while the remaining oscillation parameters are fixed at the current global best--fit values from Ref.~\cite{NuFIT6.0,Esteban_2024}. In this section, we analyze the future sensitivities of NO$\nu$A, T2K, and JUNO. Accordingly, we adopt the latest global-fit results~\cite{NuFIT6.0,Esteban_2024} as the baseline for the fixed parameters as well as the allowed ranges of the free parameters.

\begin{figure}[htbp]
    \centering
    \begin{subfigure}[b]{0.45\textwidth}
        \centering
        \includegraphics[width=\textwidth, height=6cm]{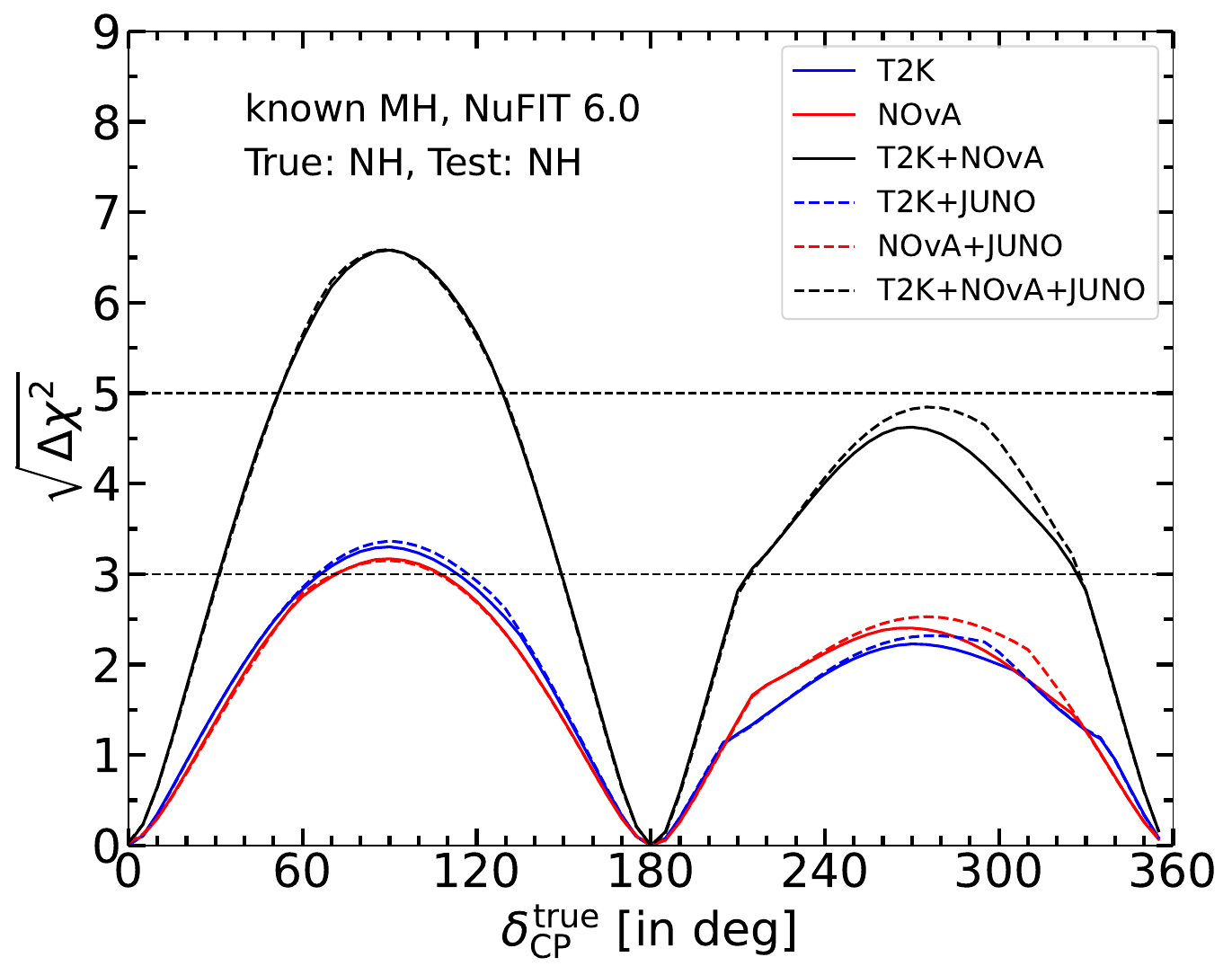}
        \caption{}
    \end{subfigure}
    \hfill
    \begin{subfigure}[b]{0.45\textwidth}
        \centering
        \includegraphics[width=\textwidth, height=6cm]{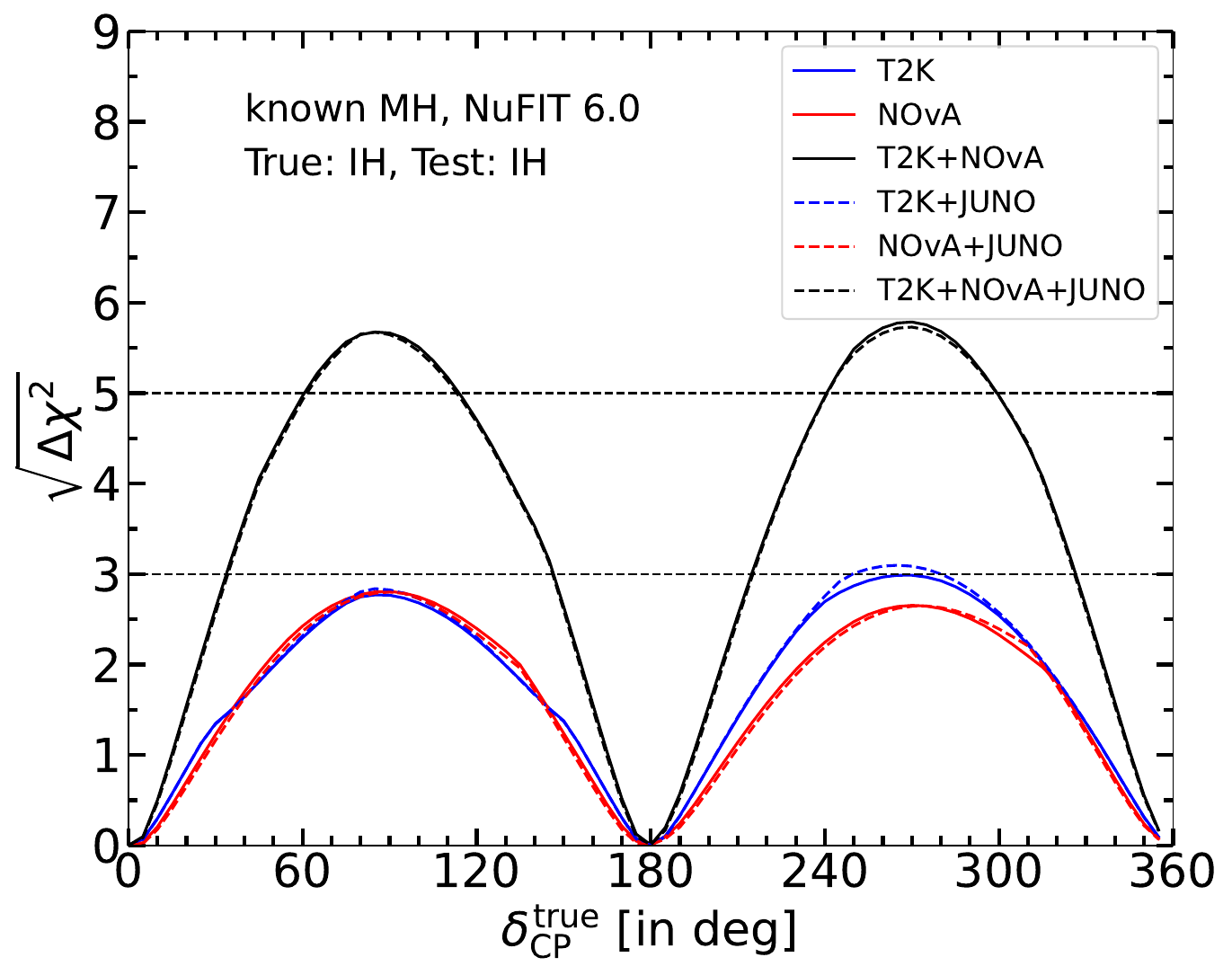}
        \caption{}
    \end{subfigure}

    \vspace{0.5cm} 

    \begin{subfigure}[b]{0.45\textwidth}
        \centering
        \includegraphics[width=\textwidth, height=6cm]{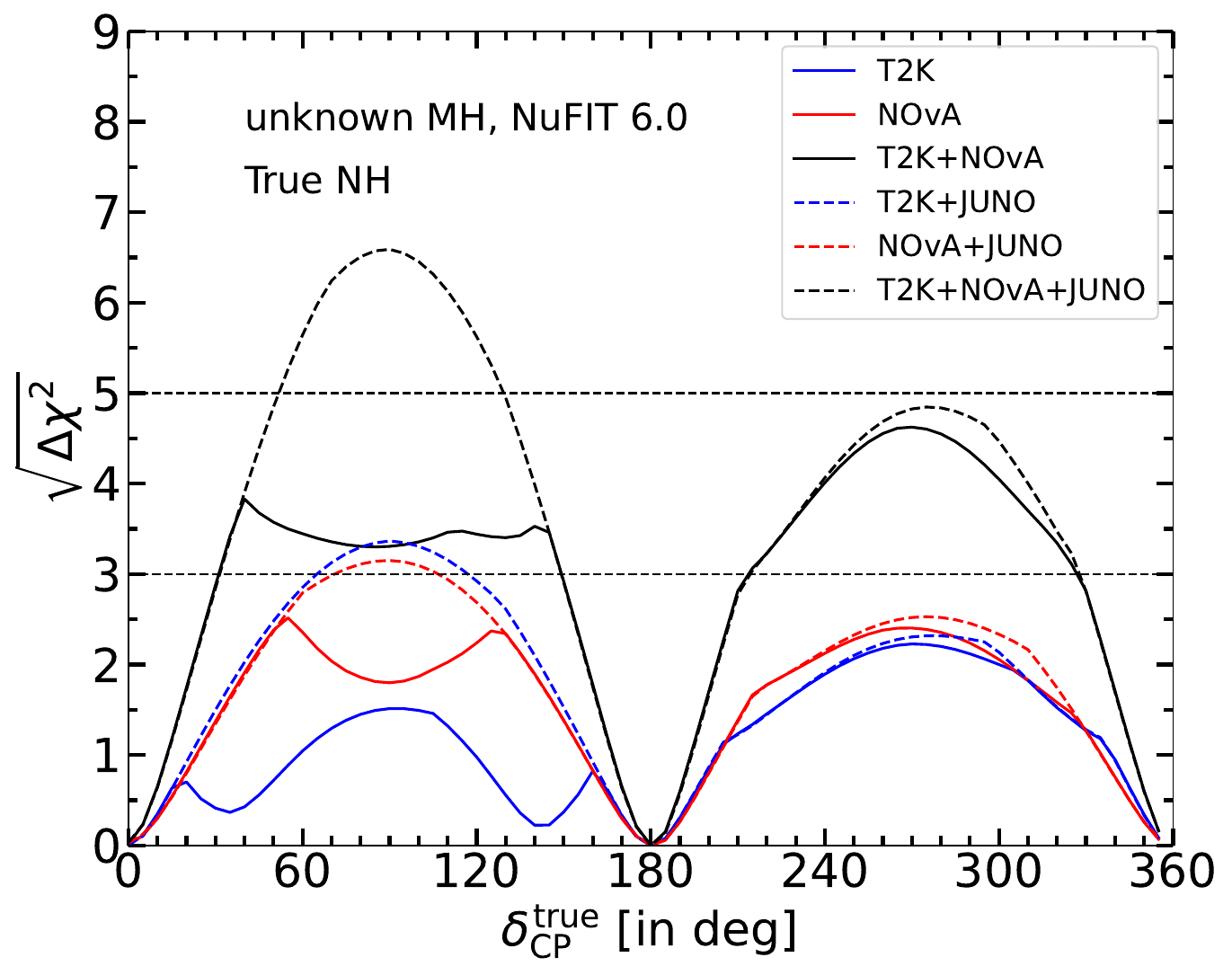}
        \caption{}
    \end{subfigure}
    \hfill
    \begin{subfigure}[b]{0.45\textwidth}
        \centering
        \includegraphics[width=\textwidth, height=6cm]{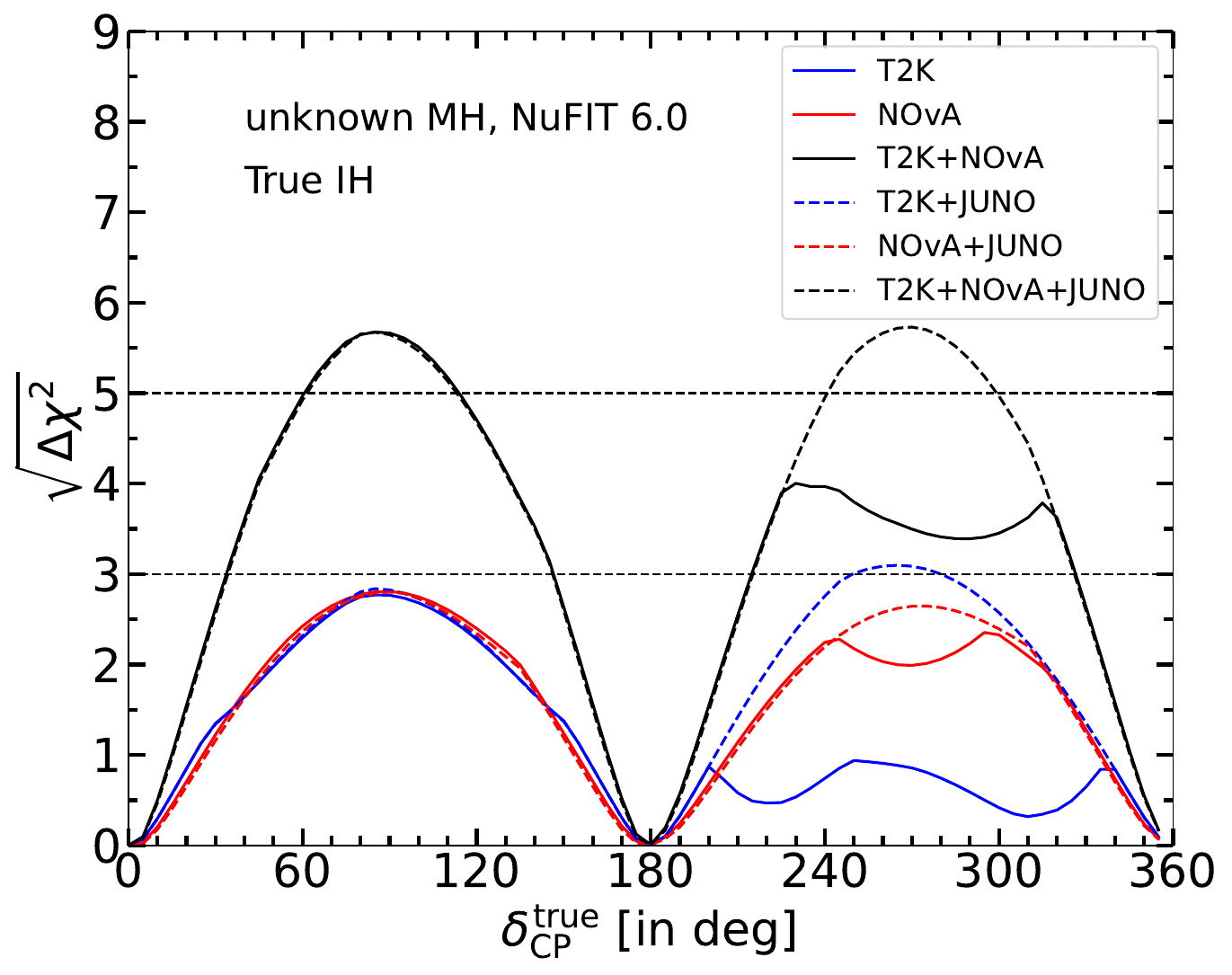}
        \caption{}
    \end{subfigure}

    \caption{Future sensitivity of CPV of T2K, \nova and the combination for T2K, \nova and JUNO.}
    \label{fig:cpv_future}
\end{figure}

In Fig.~\ref{fig:cpv_future}, we show the CP--violation discovery sensitivity. For these plots, the test values of $\dcp$ are restricted to the CP--conserving choices $0$ and $180^\circ$. The $\chi^2$ between the true and test spectra is computed with \textsc{GLoBES}, and for simulated data the resulting $\Delta\chi^2$ is equivalent to $\chi^2$. The $\Delta\chi^2$ values are marginalised over all test oscillation parameters. In the case of a known hierarchy, the true and test hierarchies are fixed to be identical. In the case of an unknown hierarchy, both NH and IH are included as test hypotheses and subsequently marginalised.

For known hierarchy, both \nova and T2K individually can exclude CP--conserving values of $\dcp$ at the $3\sigma$ level for true $\dcp\approx 90^\circ$, regardless of the true hierarchy. For true $\dcp=270^\circ$, \nova and T2K reach $2.5\sigma$ and $3\sigma$ sensitivity respectively when IH is true, whereas none of them reaches $3\sigma$ for NH true. The combined NO$\nu$A+T2K future dataset excludes CP conservation at the $5\sigma$ level for both hierarchies for true $\dcp=90^\circ$. For true $\dcp=270^\circ$, NO$\nu$A+T2K future dataset excludes CP conservation at the $5\sigma$ ($4.5\sigma$) level for IH (NH) being the true hierarchy. The inclusion of future JUNO data does not lead to qualitative changes.

For an unknown hierarchy, the hierarchy--$\dcp$ degeneracy produces dips in the CP--violation sensitivity for the unfavourable hierarchy--$\dcp$ combinations: NH with $\dcp=90^\circ$ and IH with $\dcp=270^\circ$. In these cases, $P_{\mu e}$ and $P_{\bar\mu\bar e}$ at the true point resemble those at the opposite test hierarchy with CP--conserving $\dcp$, resulting in reduced discrimination power. JUNO’s strong hierarchy sensitivity removes these degeneracies by eliminating the wrong test hierarchy, restoring the CP--violation sensitivity to the same level obtained in the known--hierarchy case, even for the unfavourable combinations.

\begin{figure}[htbp]
    \centering
    \begin{subfigure}[b]{0.45\textwidth}
        \centering
        \includegraphics[width=\textwidth, height=6cm]{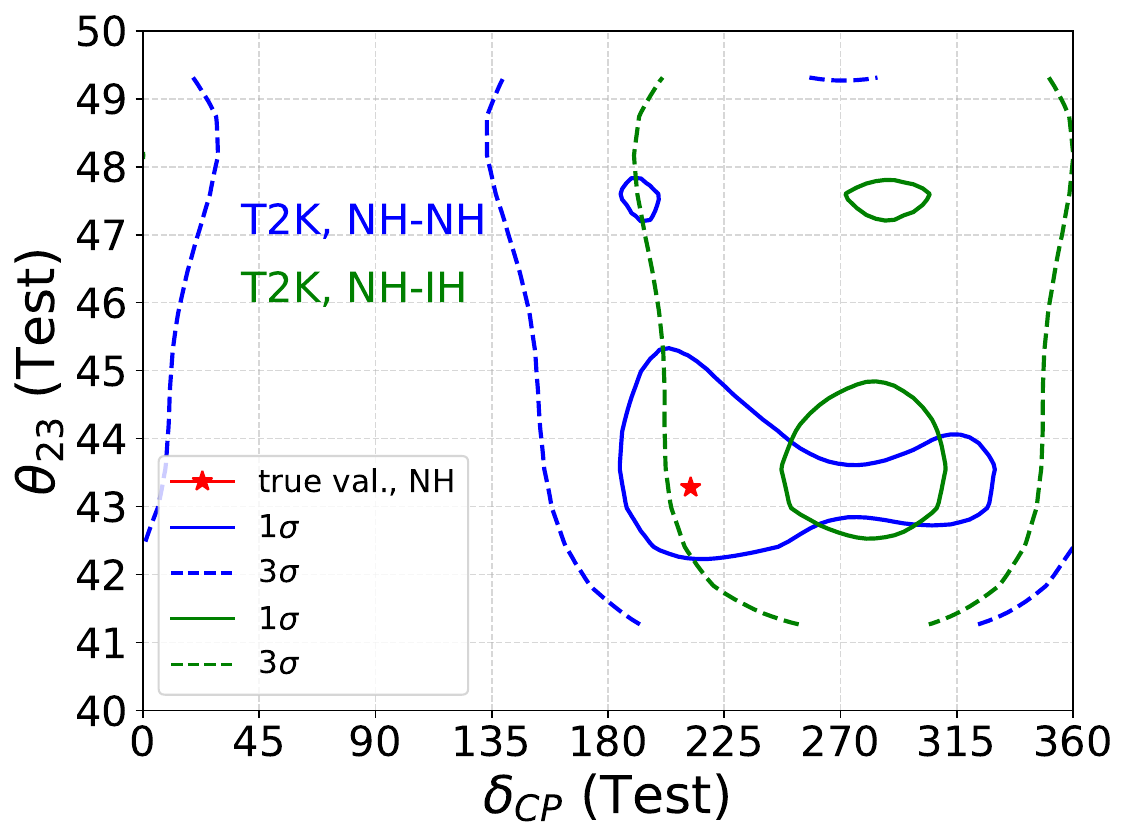}
        \caption{}
    \end{subfigure}
    \hfill
    \begin{subfigure}[b]{0.45\textwidth}
        \centering
        \includegraphics[width=\textwidth, height=6cm]{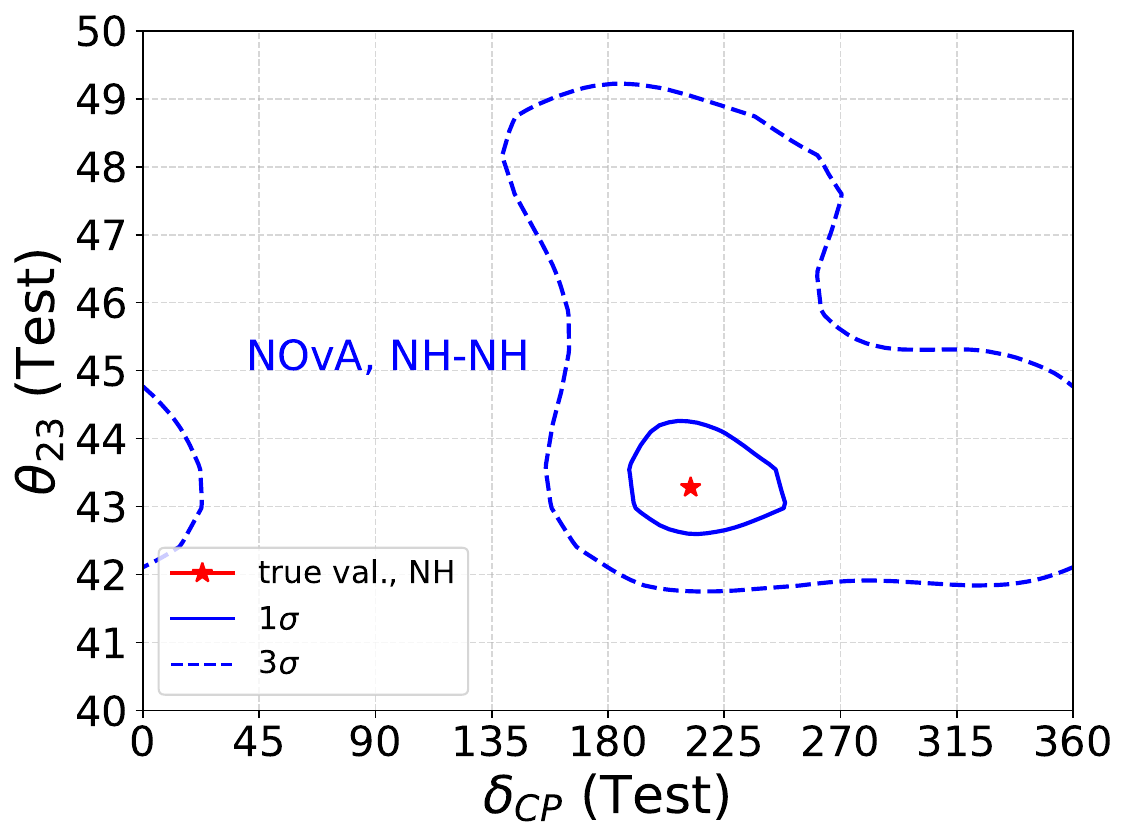}
        \caption{}
    \end{subfigure}

    \vspace{0.5cm} 

    \begin{subfigure}[b]{0.45\textwidth}
        \centering
        \includegraphics[width=\textwidth, height=6cm]{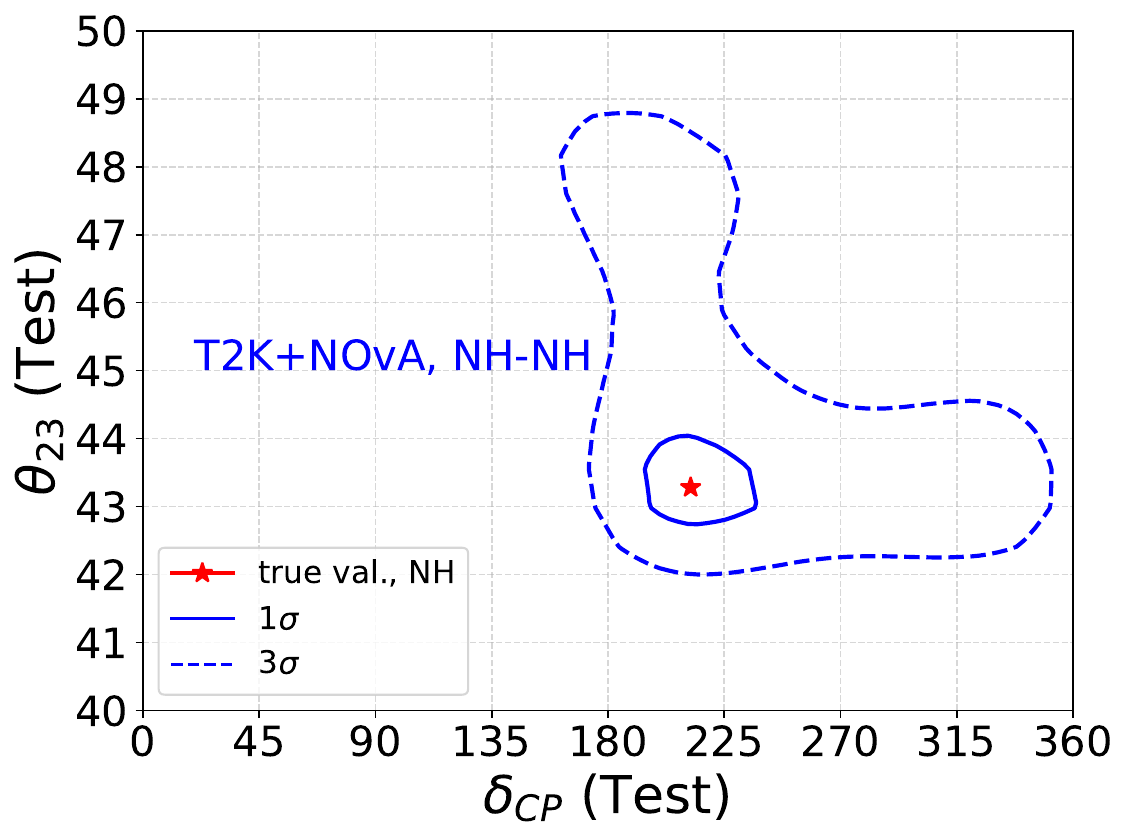}
        \caption{}
    \end{subfigure}
    \hfill
    \begin{subfigure}[b]{0.45\textwidth}
        \centering
        \includegraphics[width=\textwidth, height=6cm]{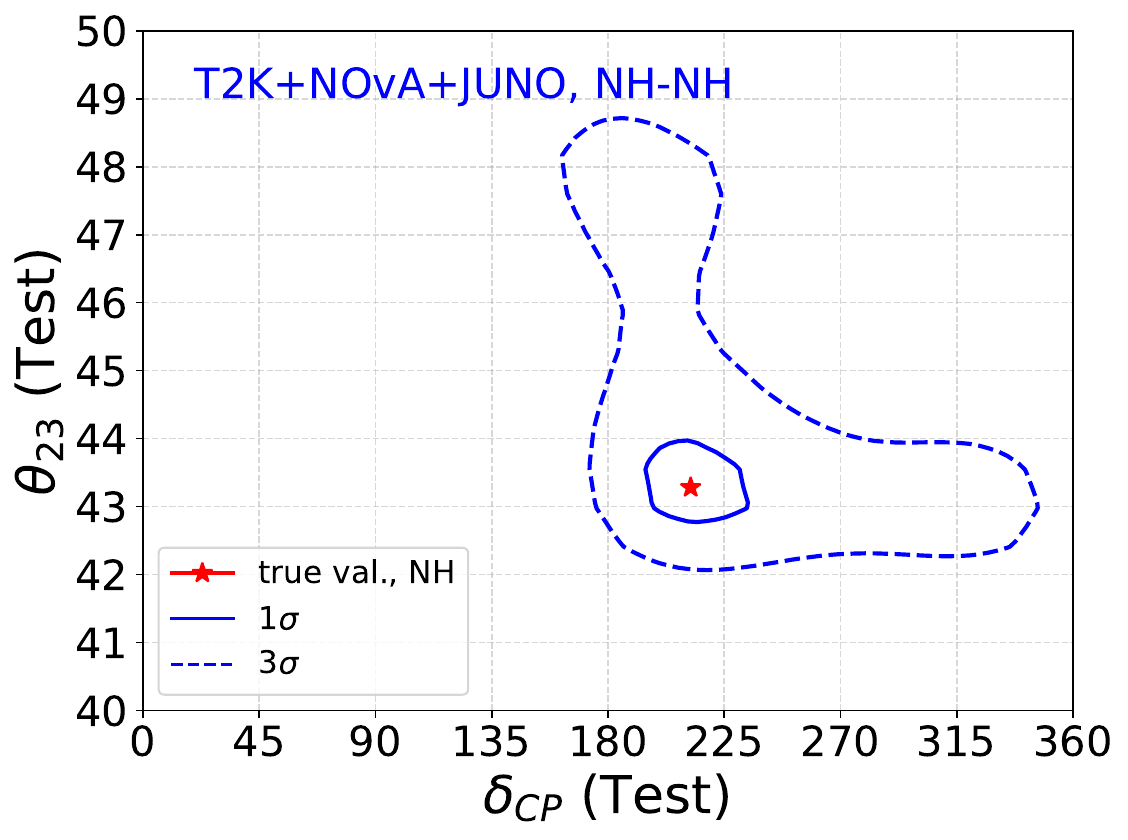}
        \caption{}
    \end{subfigure}

    \caption{Allowed regions in the test $\delta_{\rm CP}-\theta_{23}$ plane for the true values of $\theta_{23} = 43.3\degree$ and $\delta_{\rm CP} = 212\degree ({\rm or~} -148\degree )$ considering NH as true mass hierarchy. These true values and best-fit values of other oscillation parameters are adopted from NuFIT 6.0 (2024). In all the above plots, blue colour corresponds to the case where NH is assumed in both true and test (i.e. NH-NH), and green colour represents the NH-IH case. The absence of green colour contours means that there are no allowed regions for the NH-IH case within the considered significance limit.}
    \label{fig:testth23_testdcp_NHtrue}
\end{figure}

\begin{figure}[htbp]
    \centering
    \begin{subfigure}[b]{0.45\textwidth}
        \centering
        \includegraphics[width=\textwidth, height=6cm]{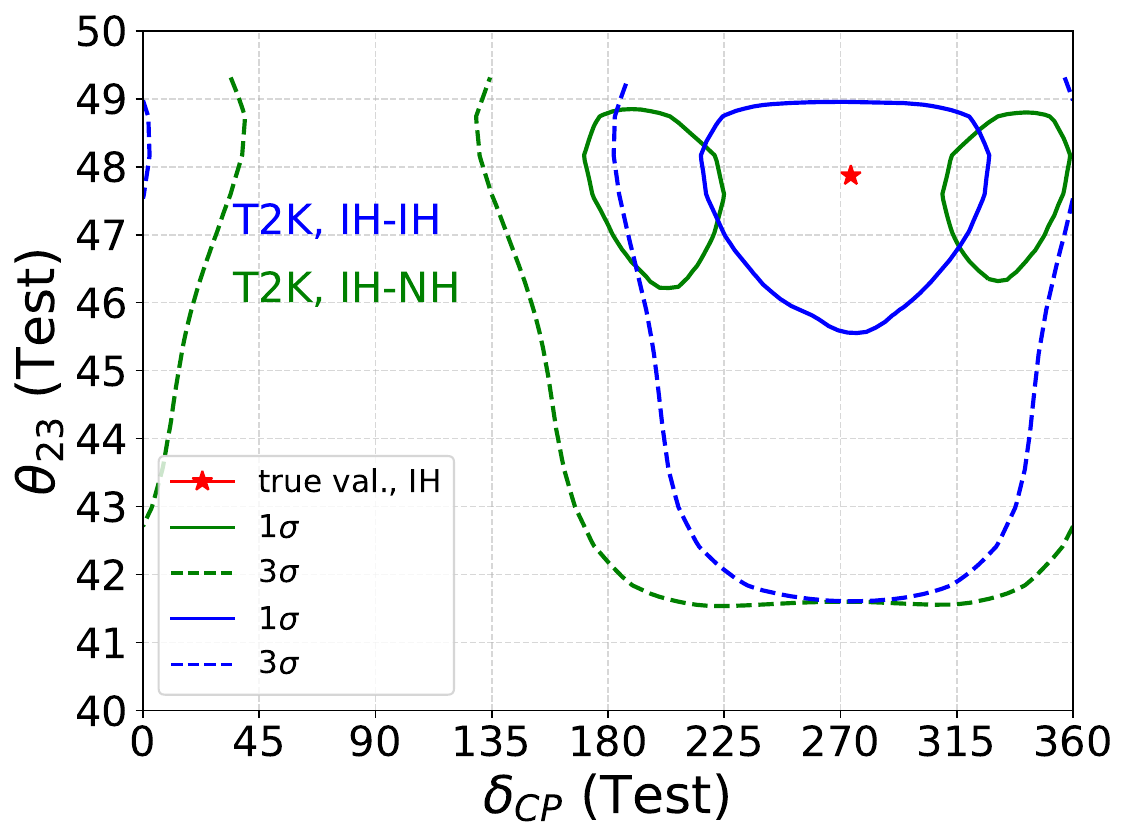}
        \caption{}
    \end{subfigure}
    \hfill
    \begin{subfigure}[b]{0.45\textwidth}
        \centering
        \includegraphics[width=\textwidth, height=6cm]{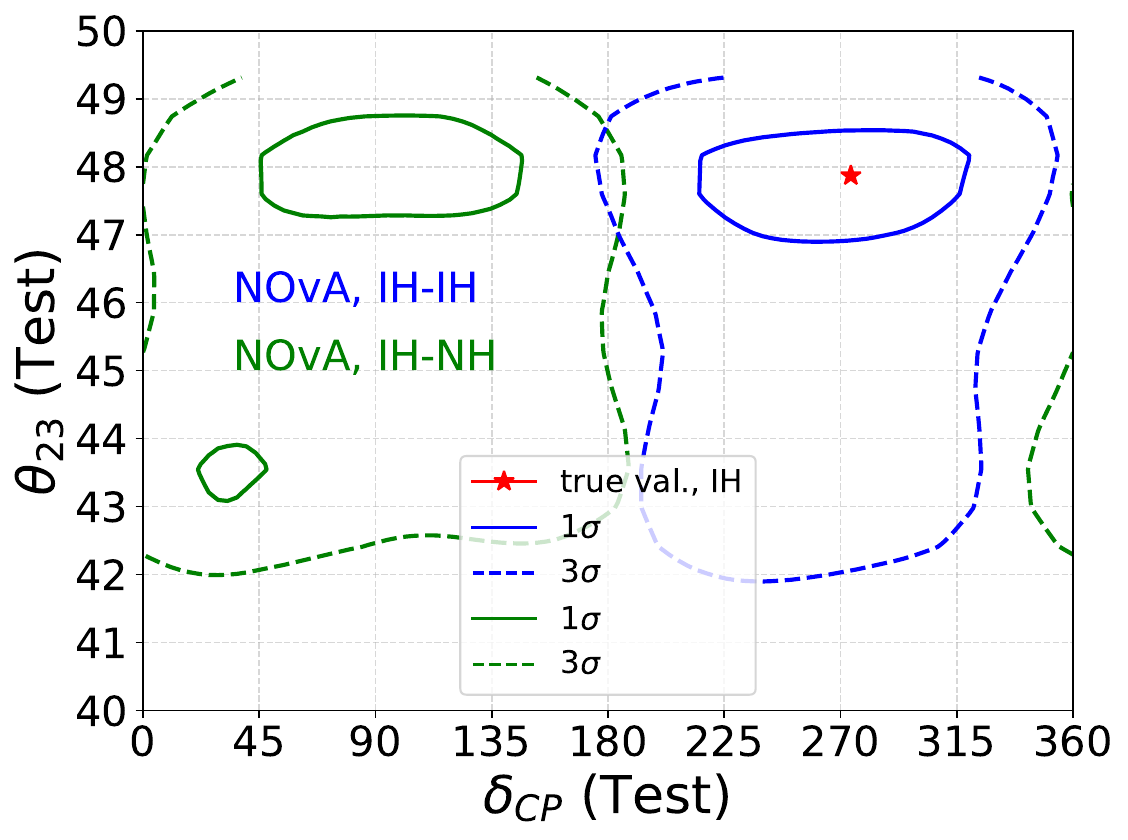}
        \caption{}
    \end{subfigure}

    \vspace{0.5cm} 

    \begin{subfigure}[b]{0.45\textwidth}
        \centering
        \includegraphics[width=\textwidth, height=6cm]{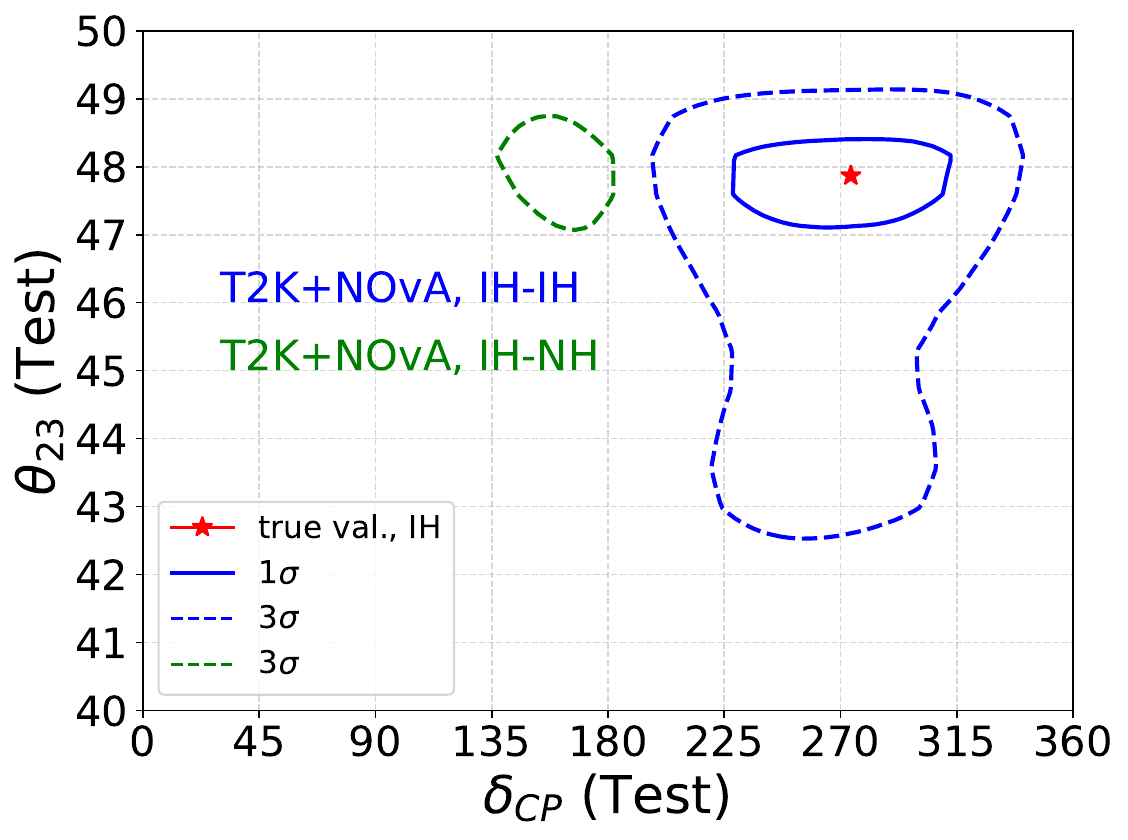}
        \caption{}
    \end{subfigure}
    \hfill
    \begin{subfigure}[b]{0.45\textwidth}
        \centering
        \includegraphics[width=\textwidth, height=6cm]{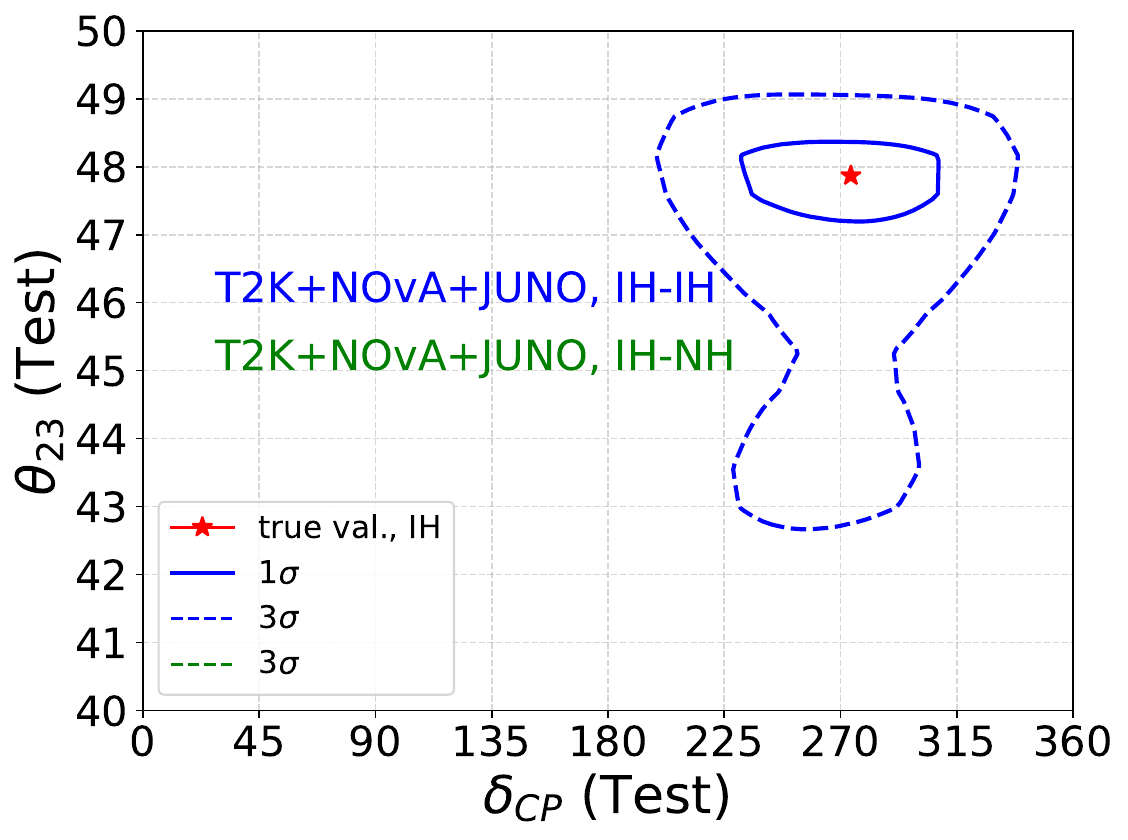}
        \caption{}
    \end{subfigure}

    \caption{Allowed regions in the test $\delta_{\rm CP}-\theta_{23}$ plane for the true values of $\theta_{23} = 47.9\degree$ and $\delta_{\rm CP} = 274\degree ({\rm or~} -86\degree )$ considering IH as true mass hierarchy. These true values and best-fit values of other oscillation parameters are adopted from NuFIT 6.0 (2024). In all the above plots, blue colour corresponds to the case where IH is assumed in both true and test (i.e. IH-IH), and green colour represents the IH-NH case. The absence of green colour contours means that there are no allowed regions for the IH-NH case within the considered significance limit.  }
    \label{fig:testth23_testdcp_IHtrue}
\end{figure}

Figures~\ref{fig:testth23_testdcp_NHtrue} and \ref{fig:testth23_testdcp_IHtrue} show the corresponding two--dimensional allowed regions in the $\sin^2\theta_{23}$--$\dcp$ plane for NH and IH as the true hierarchy. When NH is true, the global best--fit $\dcp=212^\circ$ corresponds to a favourable hierarchy--$\dcp$ combination. In this case, \nova can exclude the wrong hierarchy at $3\sigma$, while T2K cannot exclude it even at $1\sigma$. Both experiments allow the wrong octant at $3\sigma$. The combined NO$\nu$A+T2K data exclude the wrong hierarchy at $3\sigma$ and rule out the wrong half--plane of $\dcp$ at $3\sigma$, though the wrong octant persists. The addition of JUNO does not make any qualitative change.

For IH true ($\dcp=274^\circ$), an unfavourable hierarchy--$\dcp$ combination, neither \nova nor T2K can exclude the wrong hierarchy at $1\sigma$. T2K excludes most of the wrong half--plane of $\dcp$ due to its strong $\dcp$ sensitivity, while \nova cannot do so for NH as the test hierarchy. Both experiments allow the wrong octant at $3\sigma$. The combined NO$\nu$A+T2K data exclude the wrong hierarchy only at the $1\sigma$ level, and do not exclude the wrong octant at $3\sigma$. When JUNO is added, the hierarchy sensitivity increases dramatically: the wrong hierarchy is excluded at $3\sigma$.

There is also an improvement in the $\dcp$ precision for NH true, and around test $\sin^2\theta_{23}=0.5$ for IH true. This improvement in $\dcp$ precision originates from the synergy between $\theta_{23}$, $\dcp$, and $|\Delta_{31}|$ ($|\Delta_{32}|$) for true NH (IH): JUNO has superior standalone sensitivity to $|\dl|$ ($|\Delta_{32}|$) for NH (IH), and this improves the $|\dl|$($|\Delta_{32}|$) precision of the combined NO$\nu$A+T2K dataset when added with JUNO data. The combined NO$\nu$A+T2K+JUNO dataset therefore restricts test $|\dl|$ ($|\Delta_{32}|$) for true NH (IH) to a very narrow range at $3\sigma$, which in turn mildly sharpens the test $\dcp$ precision. 

\begin{figure}[h]
\hspace*{-1.5cm}
\centering \includegraphics[width=8cm,height=7.0cm]{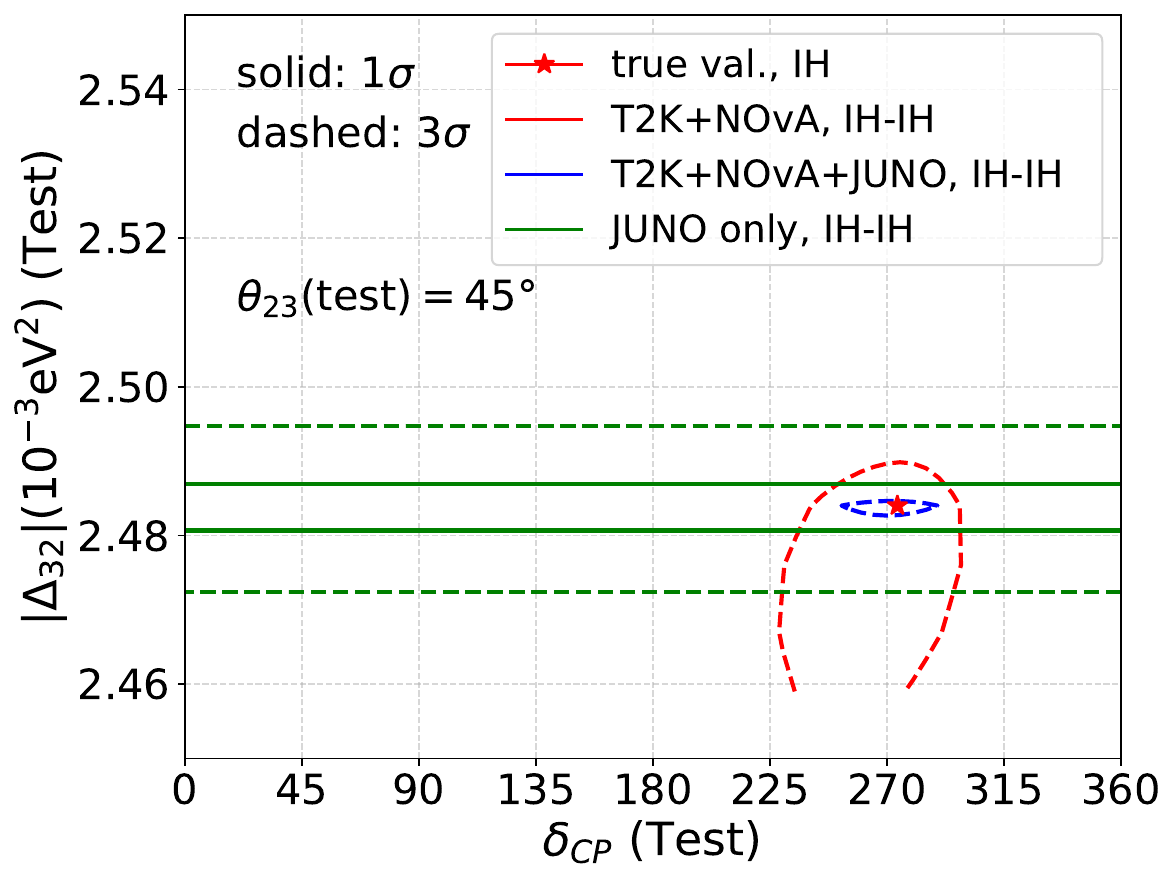}
\caption{Allowed regions in the test $|\Delta_{32}|-\delta_{\rm CP}$ plane for IH true and test while keeping the value of $\tz$ (test) fixed at $45^\circ$.}
 \label{fig:dm32_dcp_th23_plot}
\end{figure}

The improvement in the precision measurement of test $\dcp$ for IH has been highlighted in Fig.~\ref{fig:dm32_dcp_th23_plot}. In this plot, we have shown the allowed region on the test $|\Delta_{32}|-\dcp$ plane for test value of $\tz$ fixed at $45^\circ$ for true hierarchy being IH. We can see that at test $\tz=45^\circ$, the NO$\nu$A+T2K future dataset allows a large section of test $|\Delta_{32}|$ and $\dcp$ at $3\,\sigma$. However, after the addition of JUNO dataset, the precision measurement of $|\Delta_{32}|$ improves dramatically. This in turn restricts the allowed range of test $\dcp$ at $3\,\sigma$. We do not see any $1\,\sigma$ allowed region for NO$\nu$A+T2K and NO$\nu$A+T2K+JUNO, because test $\tz=45^\circ$ is excluded at $1\,\sigma$ by these two datasets for true IH, as shown in Fig.~\ref{fig:testth23_testdcp_IHtrue}. This improvement in test $\dcp$ due to the synergy of $|\dl|$ can be shown for NH true as well.

\section{Latest data from JUNO}
\label{juno_data}

The JUNO Collaboration has recently published its first physics results based on an exposure of 59.1 days~\cite{JUNO:2025gmd}. A total of 2379 inverse beta decay (IBD) candidate events were identified. From these data, JUNO obtained
\begin{align}
    \sin^2\theta_{12} &= 0.3092 \pm 0.0087, \\
    \ds &= (7.50 \pm 0.12)\times 10^{-5}\,{\rm eV}^2,
\end{align}
achieving a precision already superior to the current global-fit results summarised in Ref.~\cite{Esteban_2024}.

\begin{figure}[htbp]
    \centering

    \begin{subfigure}[b]{0.45\textwidth}
        \centering
        \includegraphics[width=\textwidth, height=6cm]{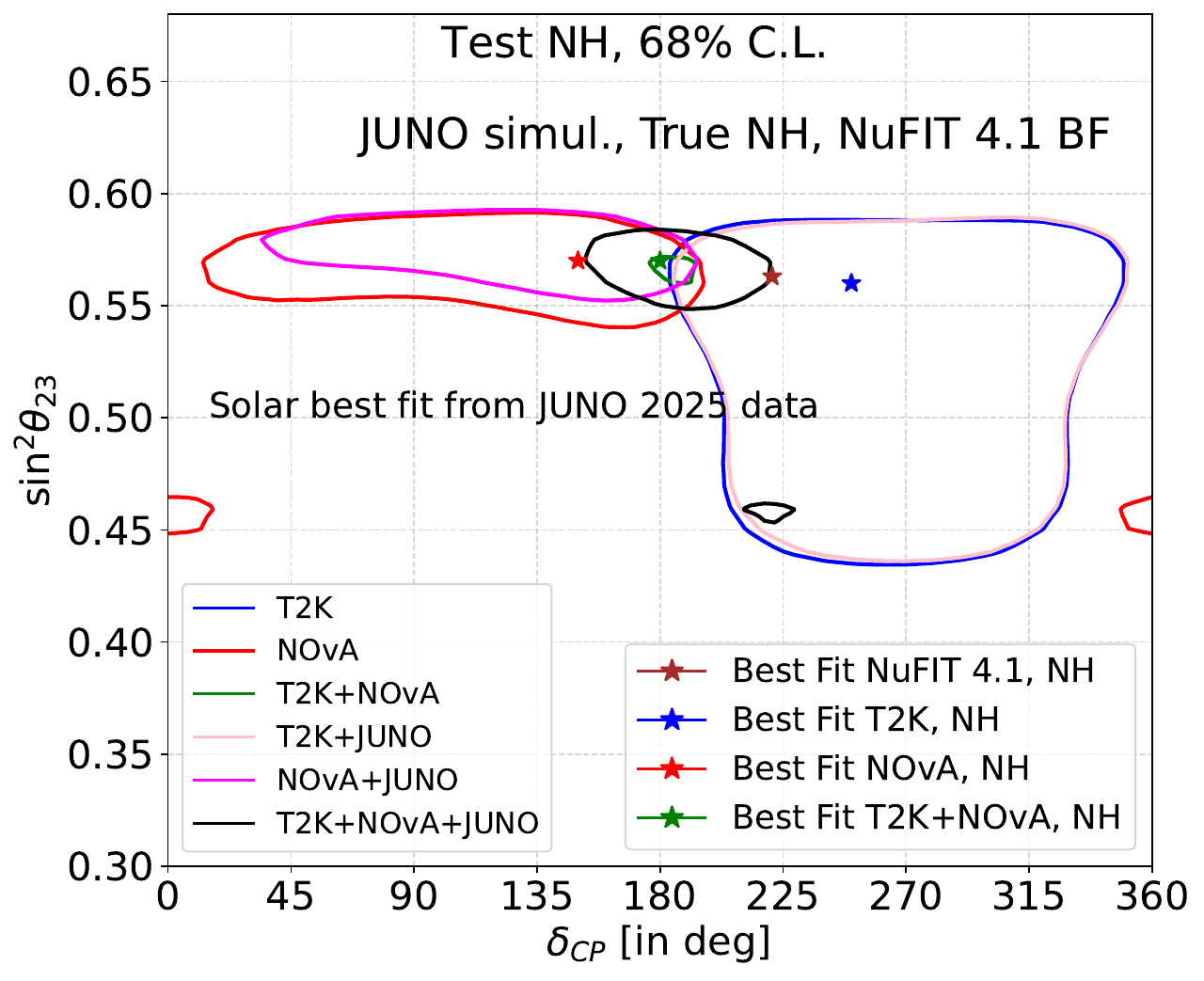}
        \caption{}
    \end{subfigure}
    \hfill
    \begin{subfigure}[b]{0.45\textwidth}
        \centering
        \includegraphics[width=\textwidth, height=6cm]{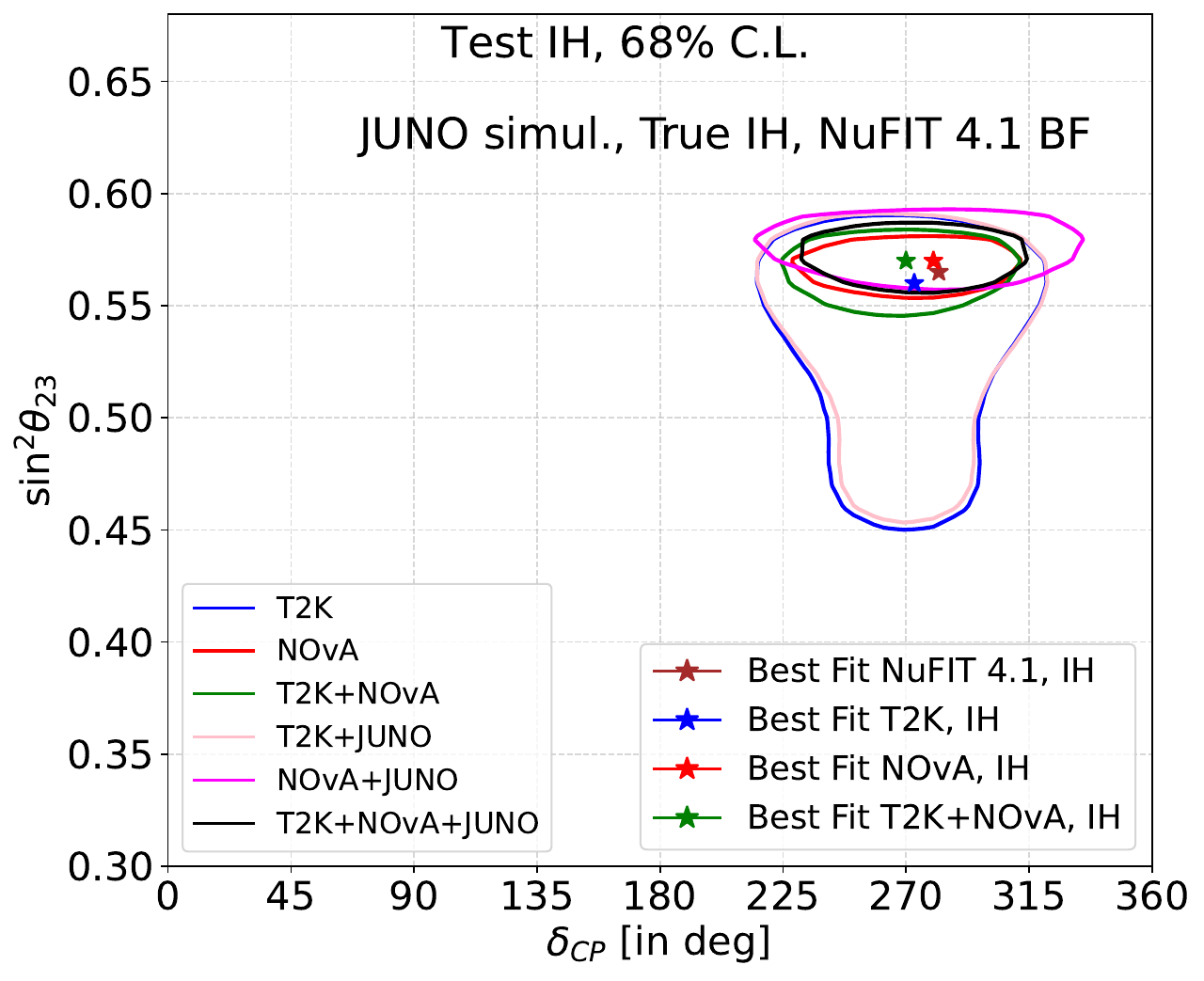}
        \caption{}
    \end{subfigure}

    \vspace{0.5cm} 

    \begin{subfigure}[b]{0.45\textwidth}
        \centering
        \includegraphics[width=\textwidth, height=6cm]{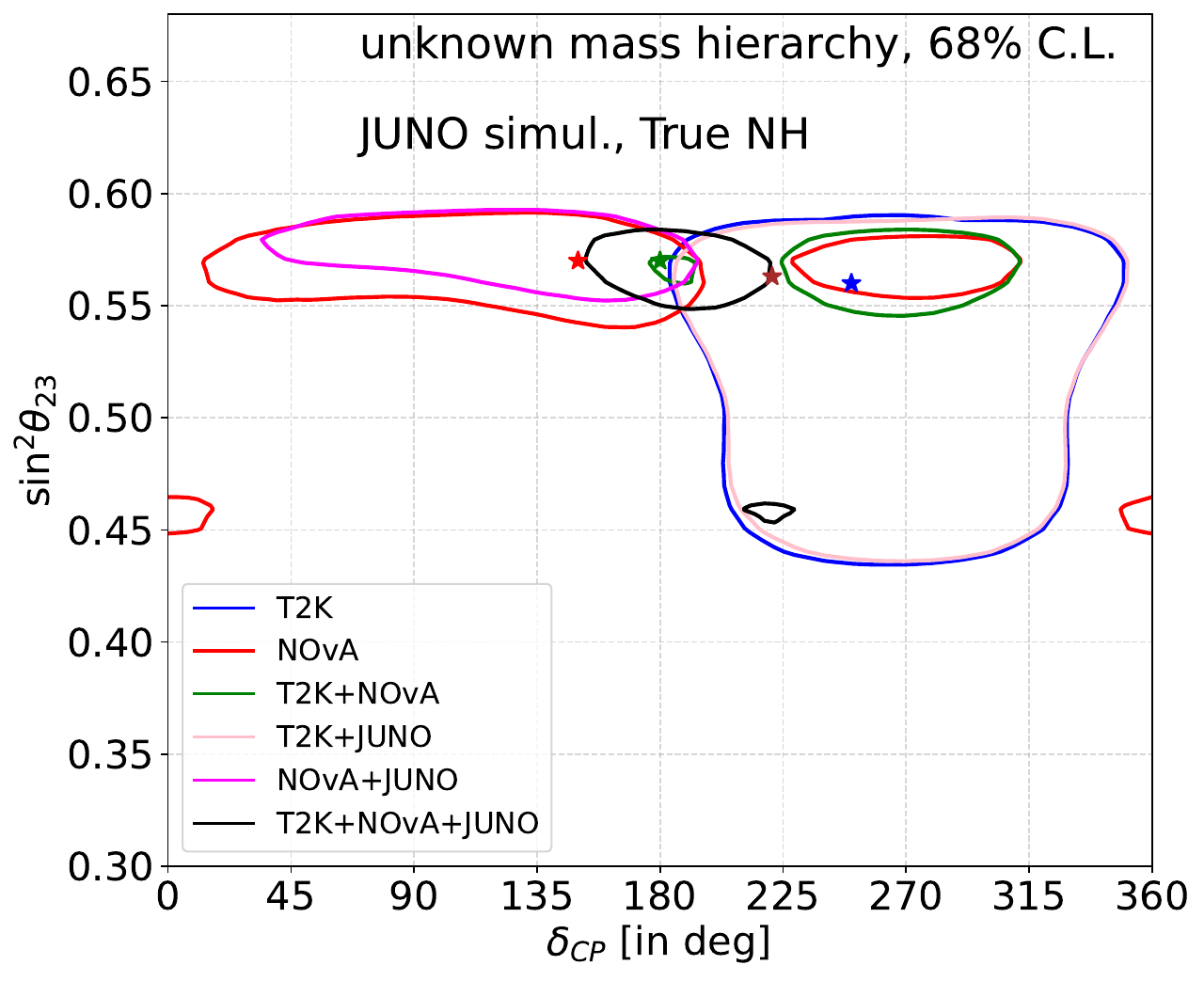}
        \caption{}
    \end{subfigure}
    \hfill
    \begin{subfigure}[b]{0.45\textwidth}
        \centering
        \includegraphics[width=\textwidth, height=6cm]{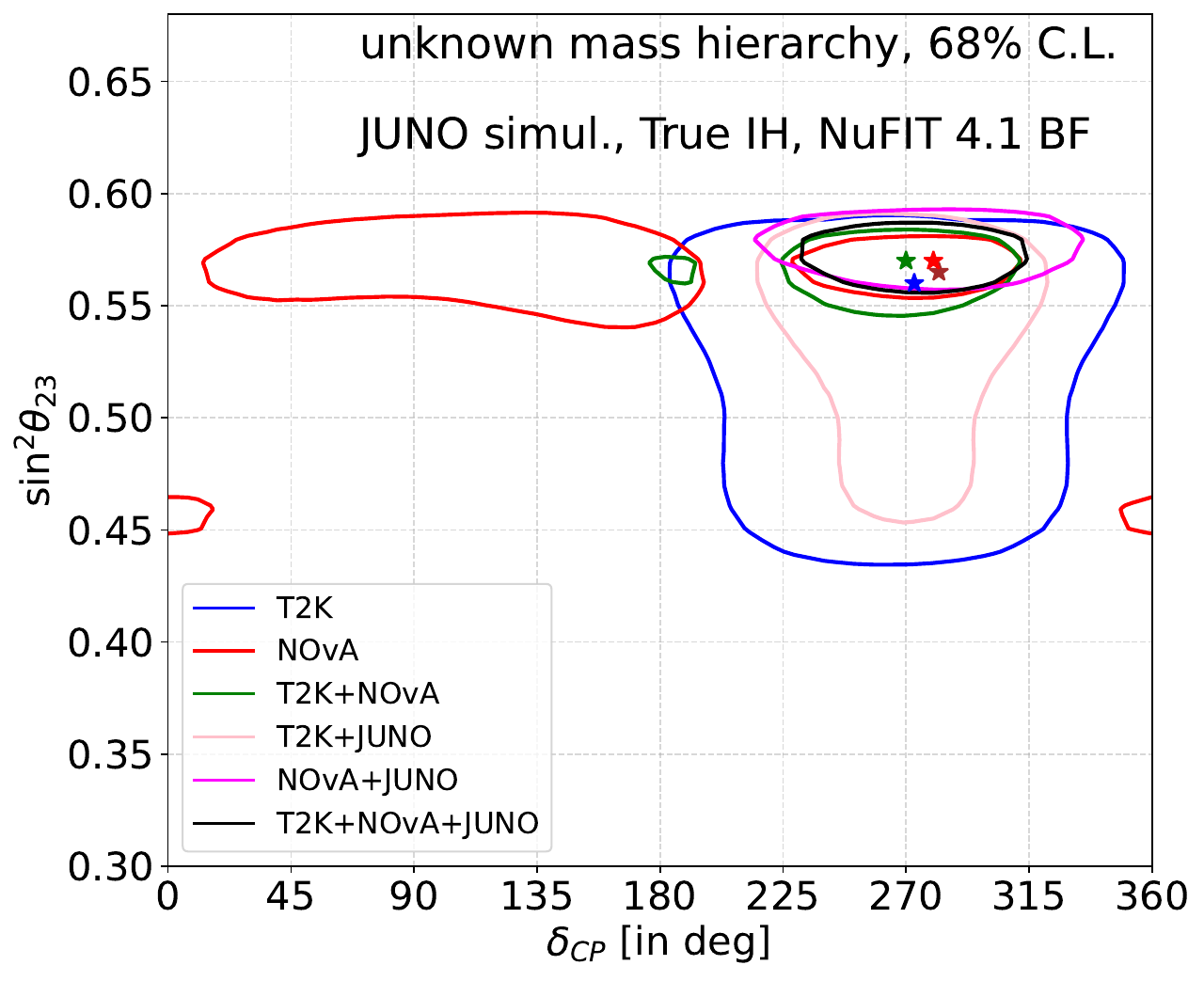}
        \caption{}
    \end{subfigure}

    \caption{Allowed regions in the $\sin^2{\theta_{23}}-\delta_{\rm CP}$ plane including T2K, \nova data and JUNO simulation. The true events of JUNO have been simulated with the best-fit values from the latest solar parameters measured by JUNO in 2025  and remaining parameters from NuFIT 4.1 ~\cite{Esteban:2018azc} as the true values of oscillation parameters. The top (bottom) panels present the cases for known (unknown) mass hierarchy and the left (right) panels present the cases when the true hierarchy for JUNO is NH (IH).}
    \label{fig:s23_dcp_juno25_nufit2019}
\end{figure}

In Fig.~\ref{fig:s23_dcp_juno25_nufit2019}, we illustrate the impact of incorporating future JUNO simulations—now updated using these newly measured values of $\sin^2\theta_{12}$ and $\ds$—into the present \nova and T2K datasets. The simulated JUNO true event rates are generated with $\sin^2\theta_{12}$ and $\ds$ fixed at JUNO’s measured best-fit values, while the remaining oscillation parameters are set to their NuFIT~4.1 global best-fit values~\cite{Esteban:2018azc}. 

For the test event rates, we vary $\sin^2\theta_{12}$ and $\ds$ within the $3\sigma$ ranges allowed by the latest JUNO measurements, and marginalise over the corresponding NuFIT~4.1 $3\sigma$ ranges for the remaining parameters. As seen in Fig.~\ref{fig:s23_dcp_juno25_nufit2019}, the inclusion of JUNO (with these updated solar-parameter uncertainties) leads to a noticeable improvement in the precision of the combined NO$\nu$A+T2K constraints, particularly in $\sin^2\theta_{23}$ and $\dcp$.

\section{Conclusions}
\label{conclusion}

The present \nova data exhibit a pronounced hierarchy--$\dcp$ degeneracy: the combinations
NH with $\dcp$ in the UHP and IH with $\dcp$ in the LHP
form two nearly degenerate solutions. In contrast, the current T2K data do not display such
a degeneracy. Although T2K cannot distinguish between NH and IH, it consistently prefers
$\dcp \simeq 270^\circ$ for both orderings. As a result, the \nova and T2K allowed regions are
in tension at the $2\sigma$ level if NH is assumed, whereas they are mutually consistent
under IH. The origin of this tension has been discussed in detail in Ref.~\cite{Rahaman:2021zzm}.

Because of the hierarchy--$\dcp$ degeneracy, the present \nova data have limited $\dcp$
sensitivity and cannot exclude any $\dcp$ value at $2\sigma$. T2K, on the other hand, has
substantially better $\dcp$ sensitivity and excludes the entire UHP at more than $2\sigma$,
with $\dcp = 90\degree$ disfavoured at over $3\sigma$. 

The first JUNO data release already provides improved precision on
$\sin^2\theta_{12}$ and $\ds$. When these updated constraints are included in the
simulation of future JUNO data, the combined NO$\nu$A+T2K fit shows an enhanced
precision in $\sin^2\theta_{23}$ and $\dcp$, as seen in Figs.~\ref{fig:s23_dcp_nufit2019} and \ref{fig:s23_dcp_juno25_nufit2019}.

Future JUNO data will play a decisive role in addressing the existing NO$\nu$A--T2K tension.
JUNO has strong hierarchy sensitivity and essentially no $\dcp$ sensitivity. When combined
with NO$\nu$A, JUNO removes the wrong-ordering branch of the \nova allowed region, thereby
lifting the hierarchy--$\dcp$ degeneracy. It also improves the hierarchy sensitivity of T2K.
However, because JUNO does not constrain $\dcp$, the CP-violation sensitivity of T2K and
\nova individually remains largely unchanged. Consequently:
\begin{itemize}
    \item If NH is the true hierarchy, the NO$\nu$A--T2K tension in $\dcp$ persists even after
    incorporating JUNO data.
    \item If IH is the true hierarchy, JUNO selects the correct ordering, and the \nova and T2K
    results become fully mutually consistent, as shown in Figs.~\ref{fig:s23_dcp_nufit2019}-\ref{fig:s23_dcp_t2k_nova}.
\end{itemize}

By eliminating the hierarchy--$\dcp$ degeneracy, future JUNO data also enhance the
$\dcp$ sensitivity of NO$\nu$A: for NH (IH) as the true hierarchy, the combined NO$\nu$A+JUNO
data can exclude the LHP (UHP) at more than $2\sigma$ ($3\sigma$).
The $\dcp$ sensitivity of T2K remains essentially unaffected by JUNO. The addition of future JUNO data improves
the precision of $|\dl|$ ($|\Delta_{32}|$) for NH (IH). For NO$\nu$A, this improved
measurement propagates into modestly better constraints on $\sin^2\theta_{23}$ and $\dcp$ at $1\sigma$.

Looking ahead, combining future \nova and T2K data with JUNO will further improve the
CP-violation discovery potential assuming the mass hierarchy is unknown, particularly for the
unfavourable combinations NH--UHP and IH--LHP. In these cases, JUNO’s hierarchy
measurement removes the wrong-ordering minima, restoring the CP violation discovery sensitivity to the level
expected when the hierarchy is known. Combining future datasets of JUNO with the accelerator neutrino datasets will also improve the $|\dl|$ ($|\Delta_{32}|$) sensitivity for NH (IH) true, which in turn will improve the $\dcp$ sensitivity moderately.

\medskip

In conclusion, the joint analysis of NO$\nu$A, T2K, and future JUNO data provides the most
robust strategy for resolving the hierarchy--$\dcp$ degeneracy present in current long-baseline
experiments. JUNO’s precise hierarchy measurement strengthens both the hierarchy and
$\dcp$ sensitivities of NO$\nu$A, improves the hierarchy reach of T2K, and may ultimately helps in being decisive about the present NO$\nu$A--T2K tension. JUNO's strong precision of $|\dl|$ also leads to stronger constraint from \nova on $\sin^2\tz$ and $
\dcp$ at $1\sigma$ for NH. In future too combining JUNO data will improve the hierarchy and CP violation discovery sensitivity of the accelerator neutrino dataset, and may improve the $\dcp$ sensitivity modestly as well.

\section*{Acknowledgements}
We thank Suprabh Prakash and Joachim Kopp for useful discussions. SG acknowledges the 
J.C.~Bose Fellowship (JCB/2020/000011) of the Anusandhan National Research Foundation,
Government of India, and the Department of Space, Government of India and Department of Space, Government of India.   We also acknowledge 
the International Centre for Theoretical Sciences (ICTS) for supporting the programme 
\emph{Understanding the Universe Through Neutrinos} (code: ICTS/Neus2024/04), where this 
work was initiated. AG would like to thank Kausik Das for helping with the computational facility. All numerical simulations in this work were carried out on the ``High-Performance Computing Facility" at the Saha Institute of Nuclear Physics, Kolkata, India.

\bibliographystyle{JHEP}
\bibliography{reference.bib,decoherence_ref}
\newpage 
\appendix
\section{Robustness of pseudo experiment}
\begin{figure}[htbp]
\centering
\includegraphics[width=0.8\textwidth,height=0.4\textheight]{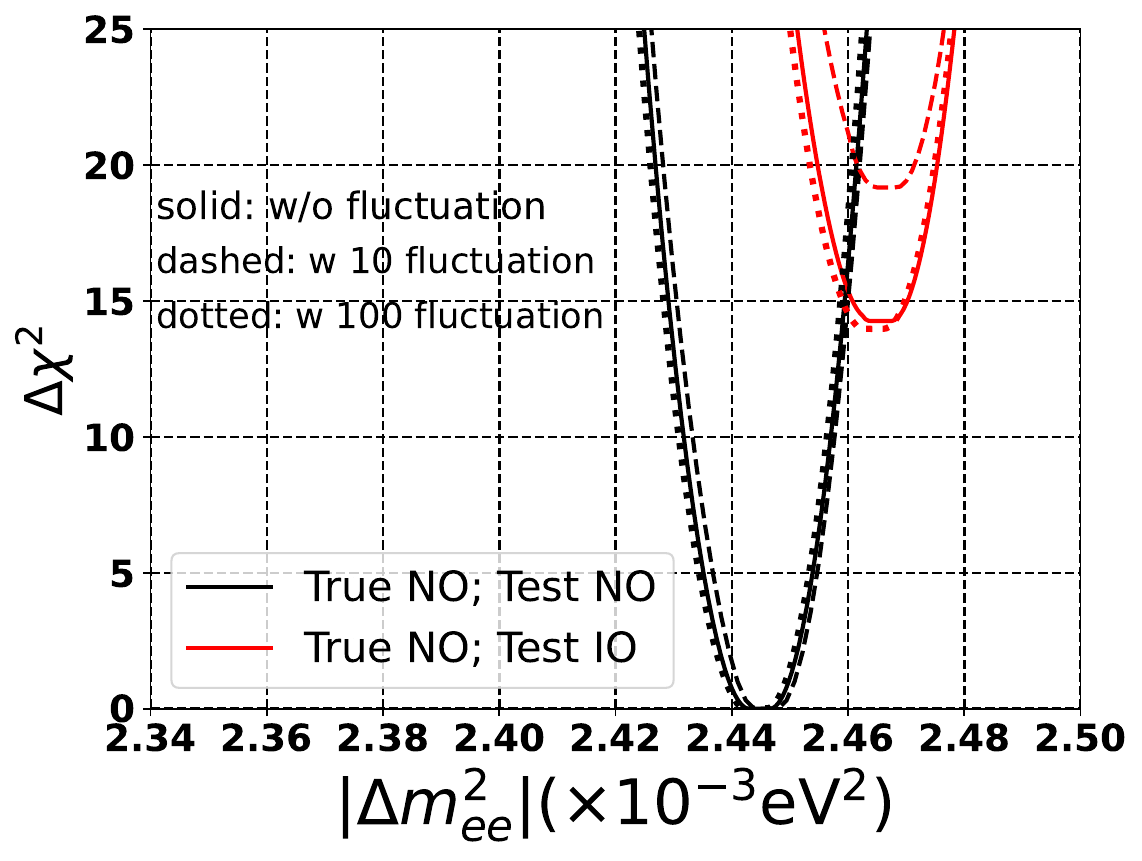}

	\caption{The comparison of MH sensitivity with and without fluctuations.}
	\label{fig:fluctuation_compare}
\end{figure}

To test the robustness of the averaging of JUNO pseudo experiments, we have provided a comparison of the reproduction of physics results, provided by the JUNO collaboration in Figures 2-8 of their TDR \cite{JUNO:2015zny}. We have reproduced JUNO TDR Fig 2-8 for no fluctuation, 10 pseudo experiments and 100 pseudo experiments in Fig.~\ref{fig:fluctuation_compare}. As we can see, with 100 pseudo-experiments, it is possible to reproduce the result of JUNO TDR. However, with only 10 pseudo-experiments, we cannot match the result for true NH and test IH.

Furthermore, we have also checked the variation of $\Delta \chi^2$ for true NH and test IH at the benchmark $|\Delta m^{2}_{ee}|=2.464\times 10^{-3}\, {\rm eV}^2$ as a function of numbers of pseudo experiments and found that 100 pseudo experiments are sufficient for convergence.




\section{Effect of JUNO simulations on \nova and T2K data: hierarchy and $\dcp-$octant structure with different true parameter values for JUNO simulations}\label{app:A}
In this section, we are presenting the results discussed in the subsection \ref{subsec: fig5_6_7} for JUNO simulated with true parameter values fixed at T2K best-fit points, \nova best-fit points and T2K+\nova best-fit points.

\begin{figure}[htbp]
    \centering

    \begin{subfigure}[b]{0.45\textwidth}
        \centering
        \includegraphics[width=\textwidth, height=6cm]{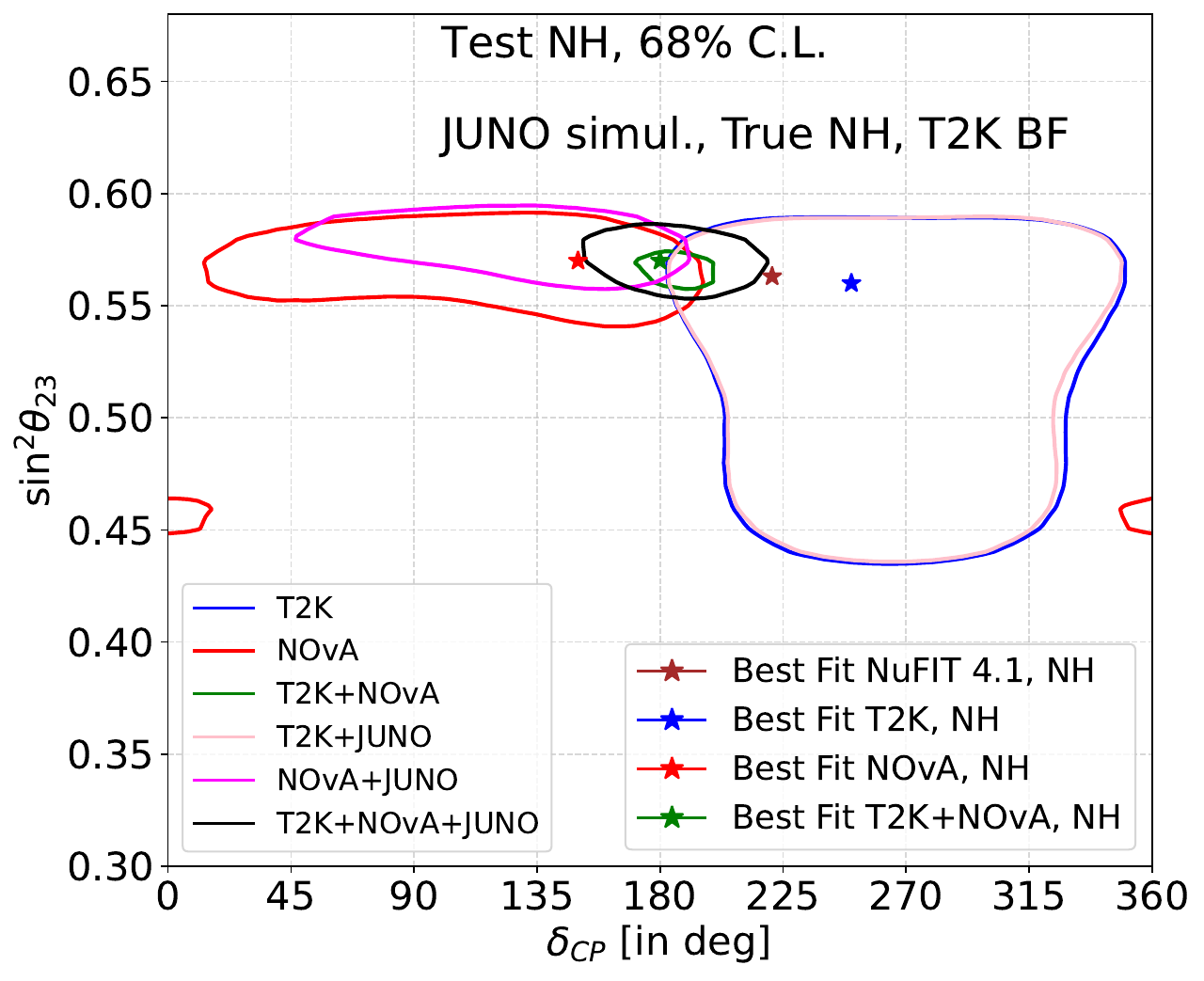}
        \caption{}
    \end{subfigure}
    \hfill
    \begin{subfigure}[b]{0.45\textwidth}
        \centering
        \includegraphics[width=\textwidth, height=6cm]{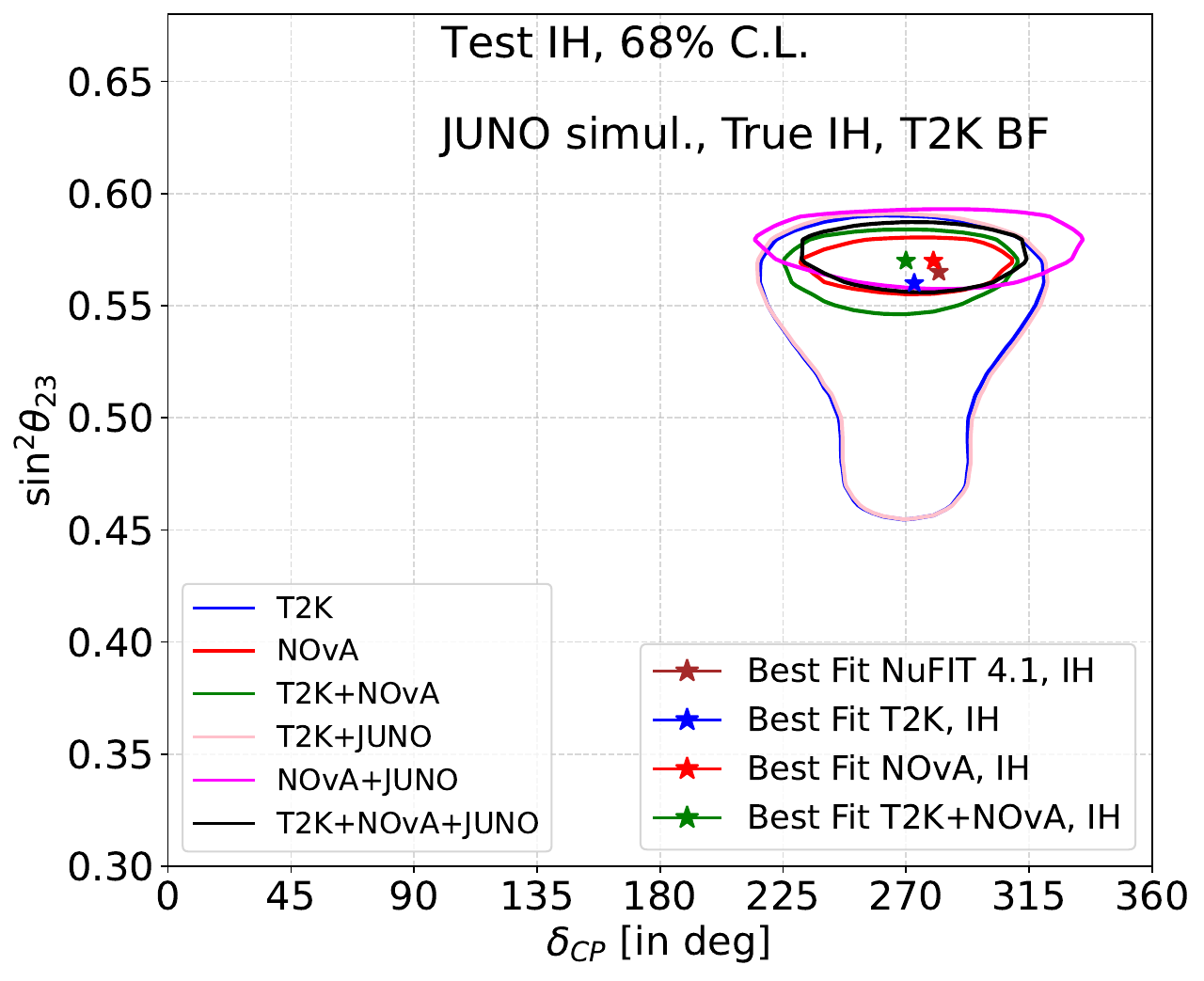}
        \caption{}
    \end{subfigure}

    \vspace{0.5cm} 

    \begin{subfigure}[b]{0.45\textwidth}
        \centering
        \includegraphics[width=\textwidth, height=6cm]{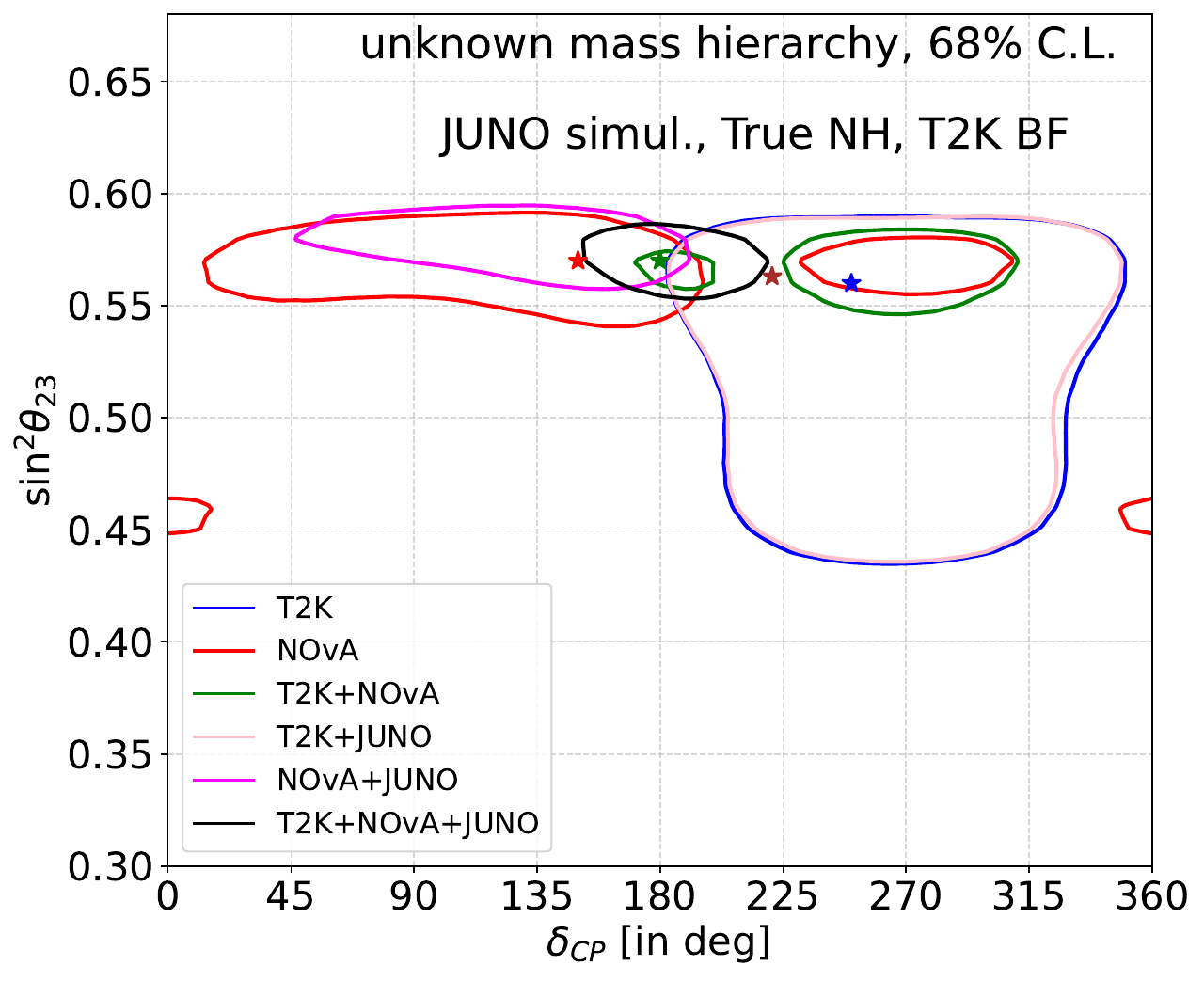}
        \caption{}
    \end{subfigure}
    \hfill
    \begin{subfigure}[b]{0.45\textwidth}
        \centering
        \includegraphics[width=\textwidth, height=6cm]{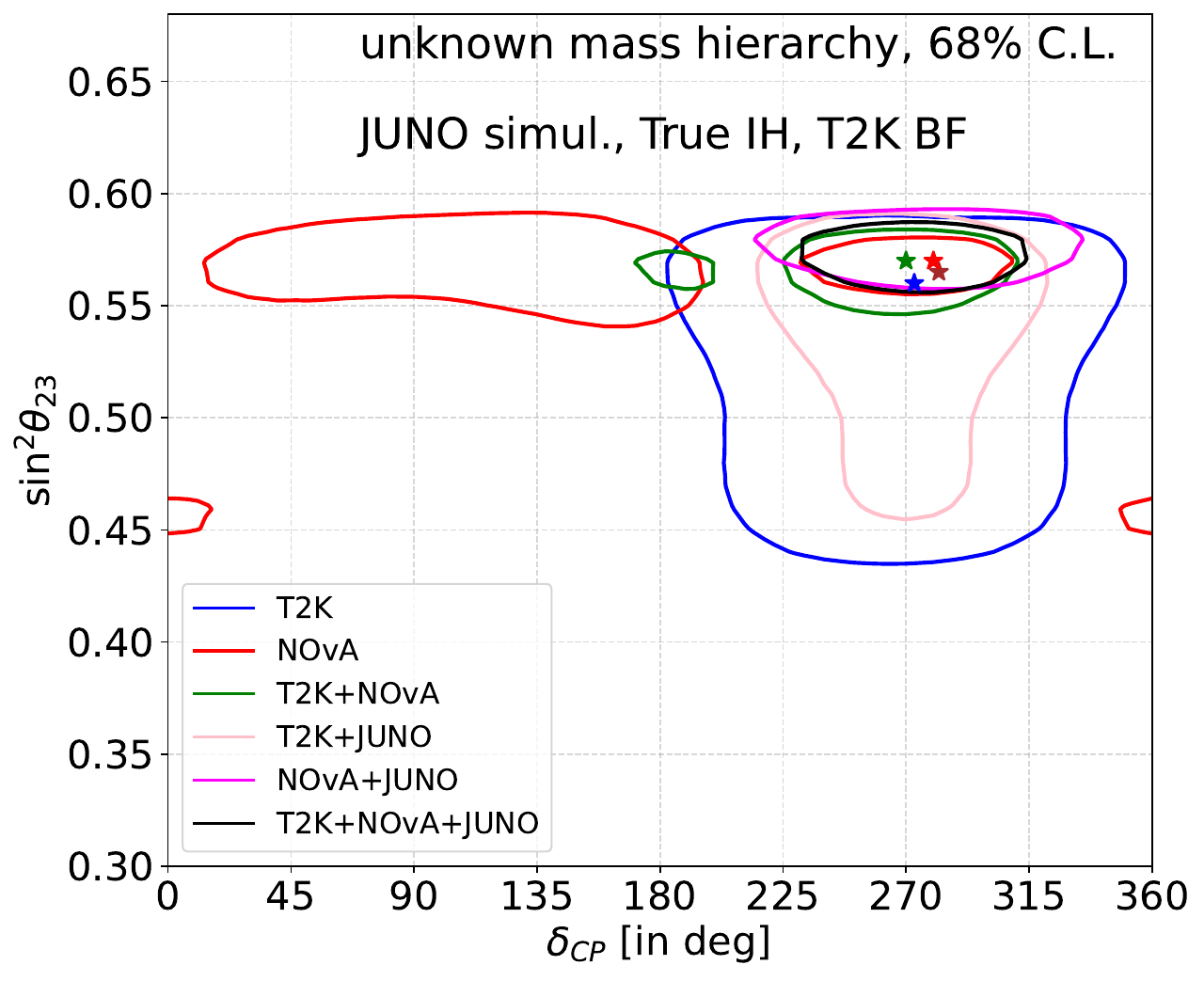}
        \caption{}
    \end{subfigure}

    \caption{Allowed regions in the $\sin^2{\theta_{23}}-\delta_{\rm CP}$ plane including T2K and \nova data, and JUNO simulation. The true events of JUNO have been simulated with the best-fit values from the T2K results \cite{T2K:2021xwb} as the true values of oscillation parameters. The top (bottom) panels present the cases for known (unknown) mass hierarchy and the left (right) panels present the cases when the true hierarchy for JUNO is NH (IH).}
    \label{fig:s23_dcp_t2k}
\end{figure}

\begin{figure}[htbp]
    \centering

    \begin{subfigure}[b]{0.45\textwidth}
        \centering
        \includegraphics[width=\textwidth, height=6cm]{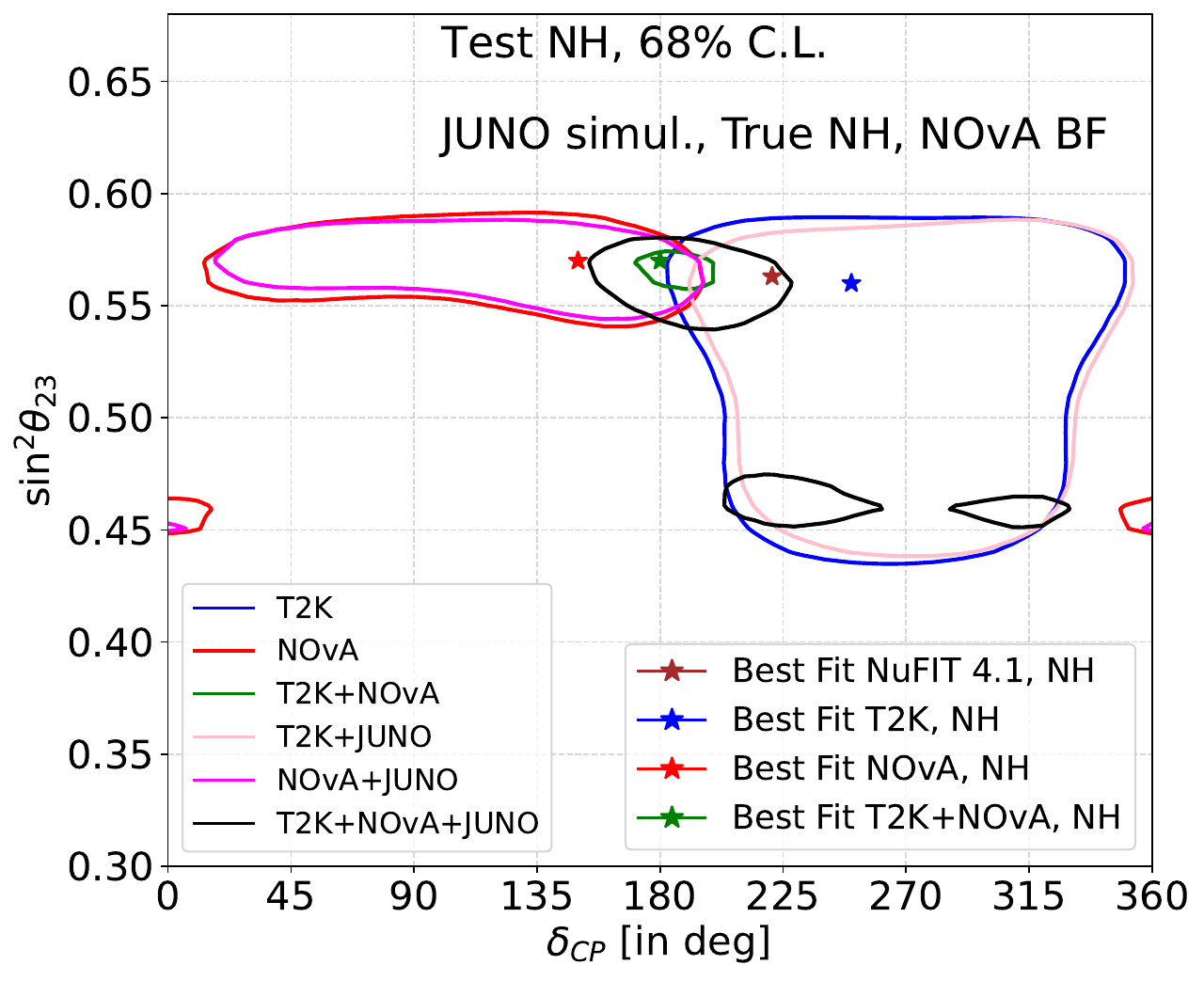}
        \caption{}
    \end{subfigure}
    \hfill
    \begin{subfigure}[b]{0.45\textwidth}
        \centering
        \includegraphics[width=\textwidth, height=6cm]{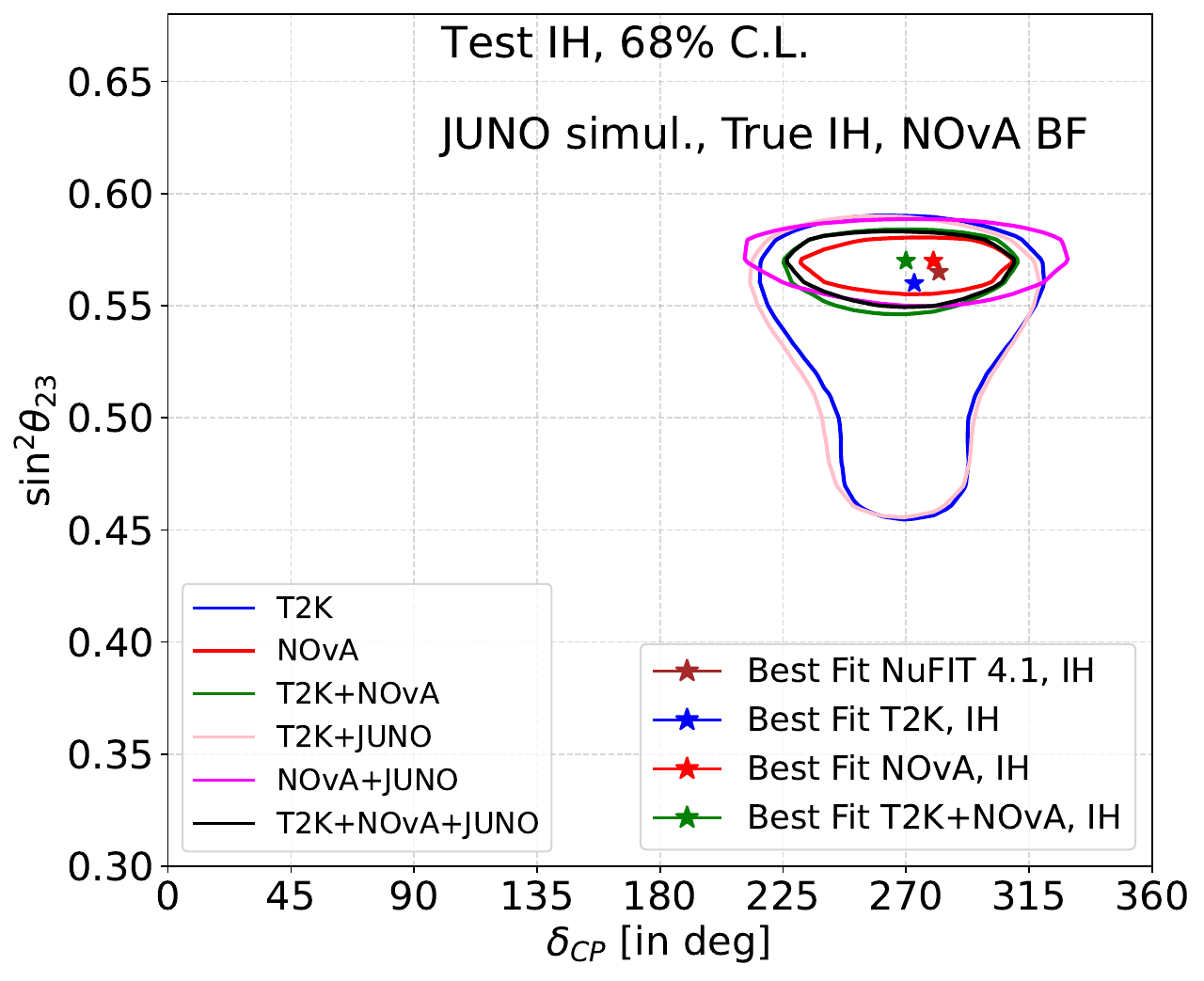}
        \caption{}
    \end{subfigure}

    \vspace{0.5cm} 

    \begin{subfigure}[b]{0.45\textwidth}
        \centering
        \includegraphics[width=\textwidth, height=6cm]{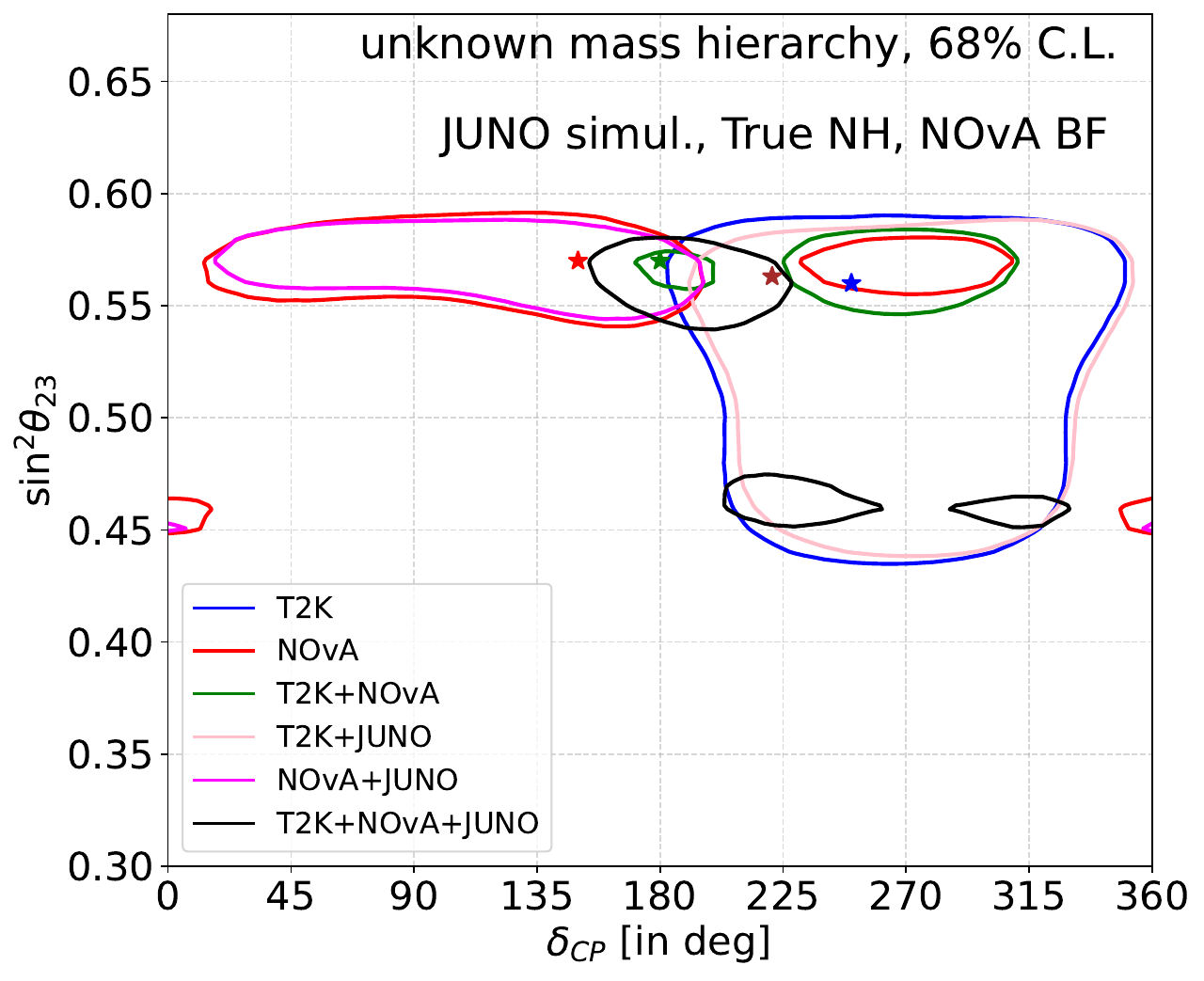}
        \caption{}
    \end{subfigure}
    \hfill
    \begin{subfigure}[b]{0.45\textwidth}
        \centering
        \includegraphics[width=\textwidth, height=6cm]{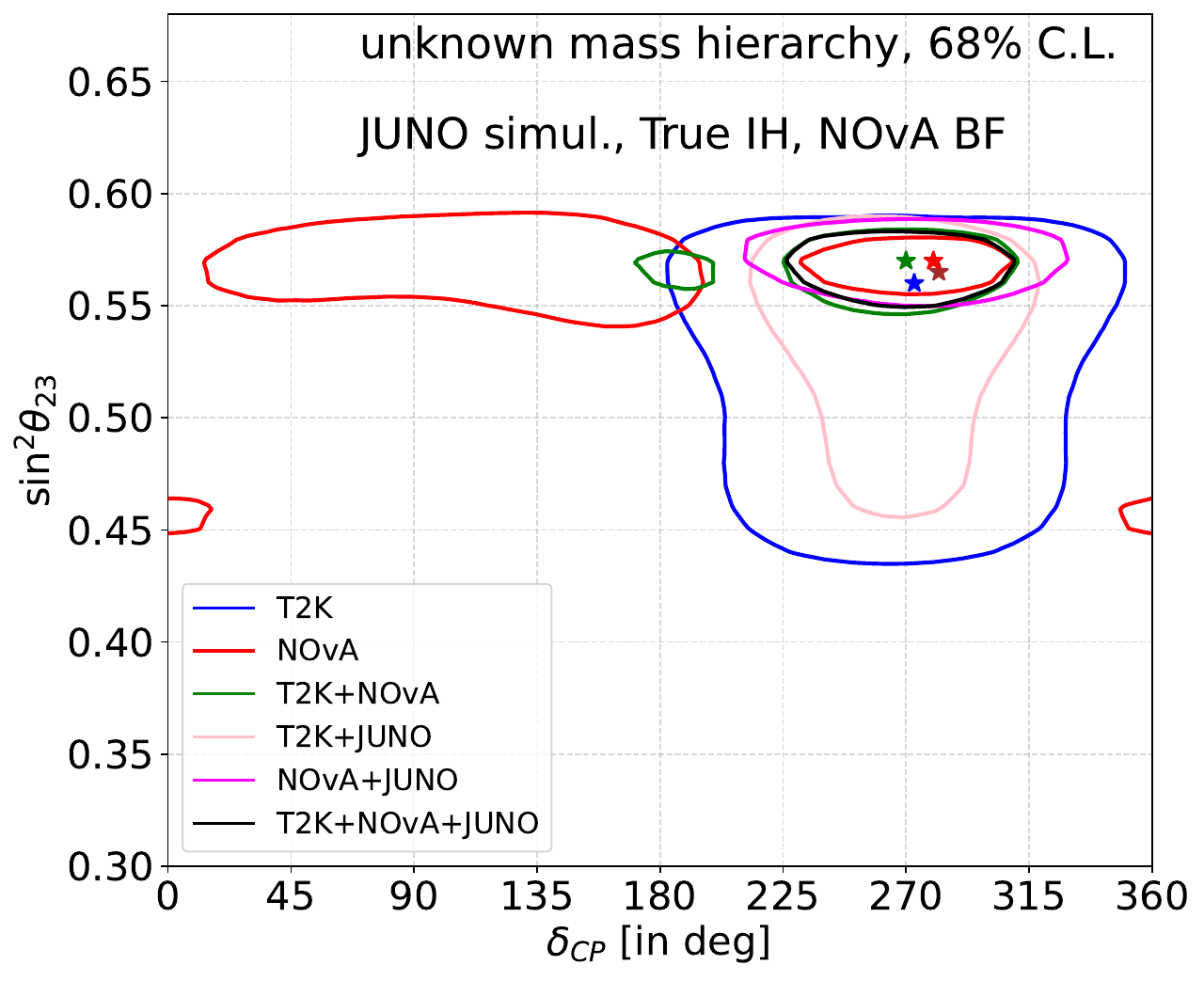}
        \caption{}
    \end{subfigure}

    \caption{Allowed regions in the $\sin^2{\theta_{23}}-\delta_{\rm CP}$ plane including T2K and \nova data, and JUNO simulation with. The true events of JUNO have been simulated with the best-fit values from \nova analysis~\cite{Wolcott:2024} as the true values of oscillation parameters. The top (bottom) panels present the cases for known (unknown) mass hierarchy and the left (right) panels present the cases when the true hierarchy for JUNO is NH (IH).}
    \label{fig:s23_dcp_nova}
\end{figure}
\begin{figure}[htbp]
    \centering

    \begin{subfigure}[b]{0.45\textwidth}
        \centering
        \includegraphics[width=\textwidth, height=6cm]{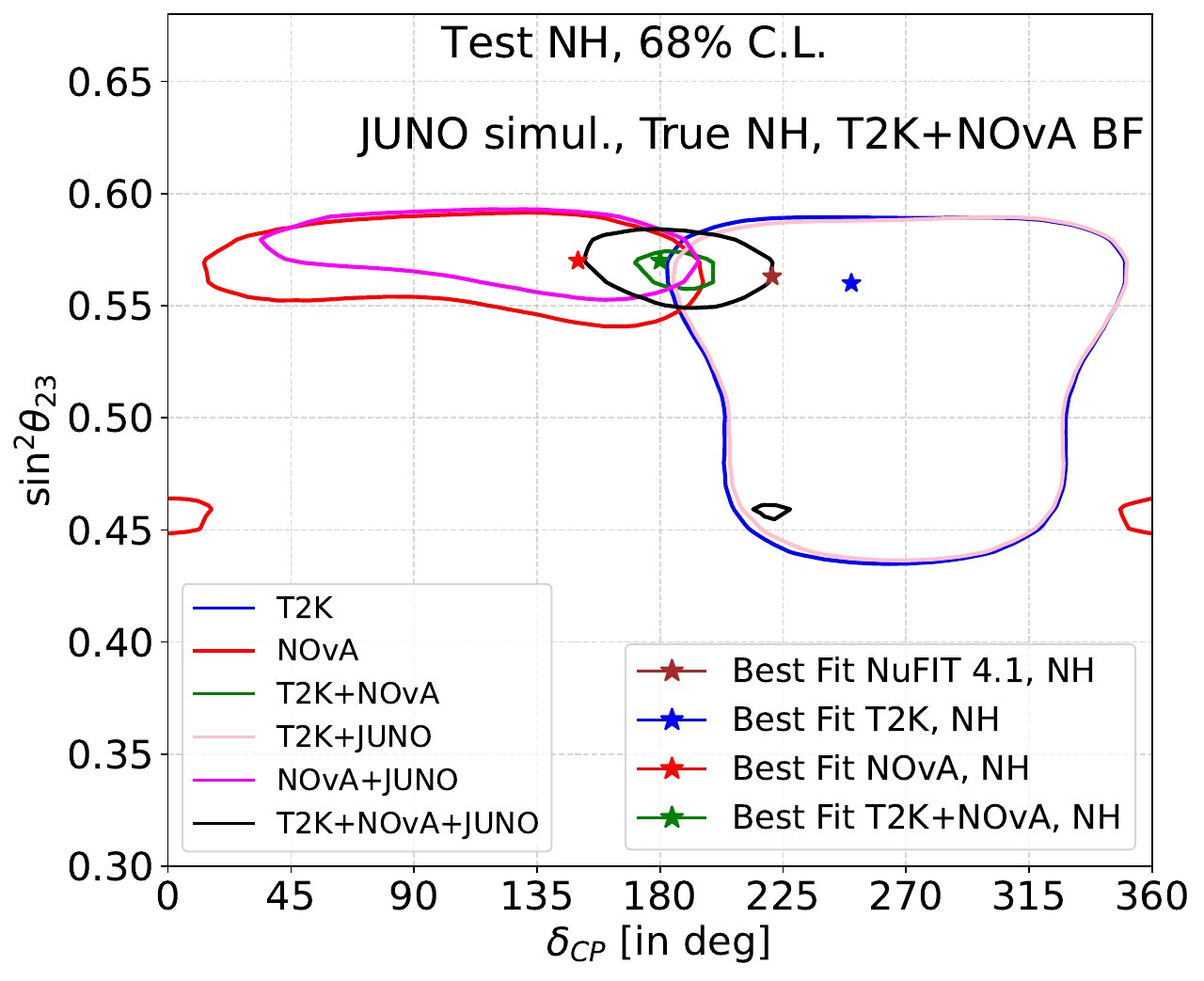}
        \caption{}
    \end{subfigure}
    \hfill
    \begin{subfigure}[b]{0.45\textwidth}
        \centering
        \includegraphics[width=\textwidth, height=6cm]{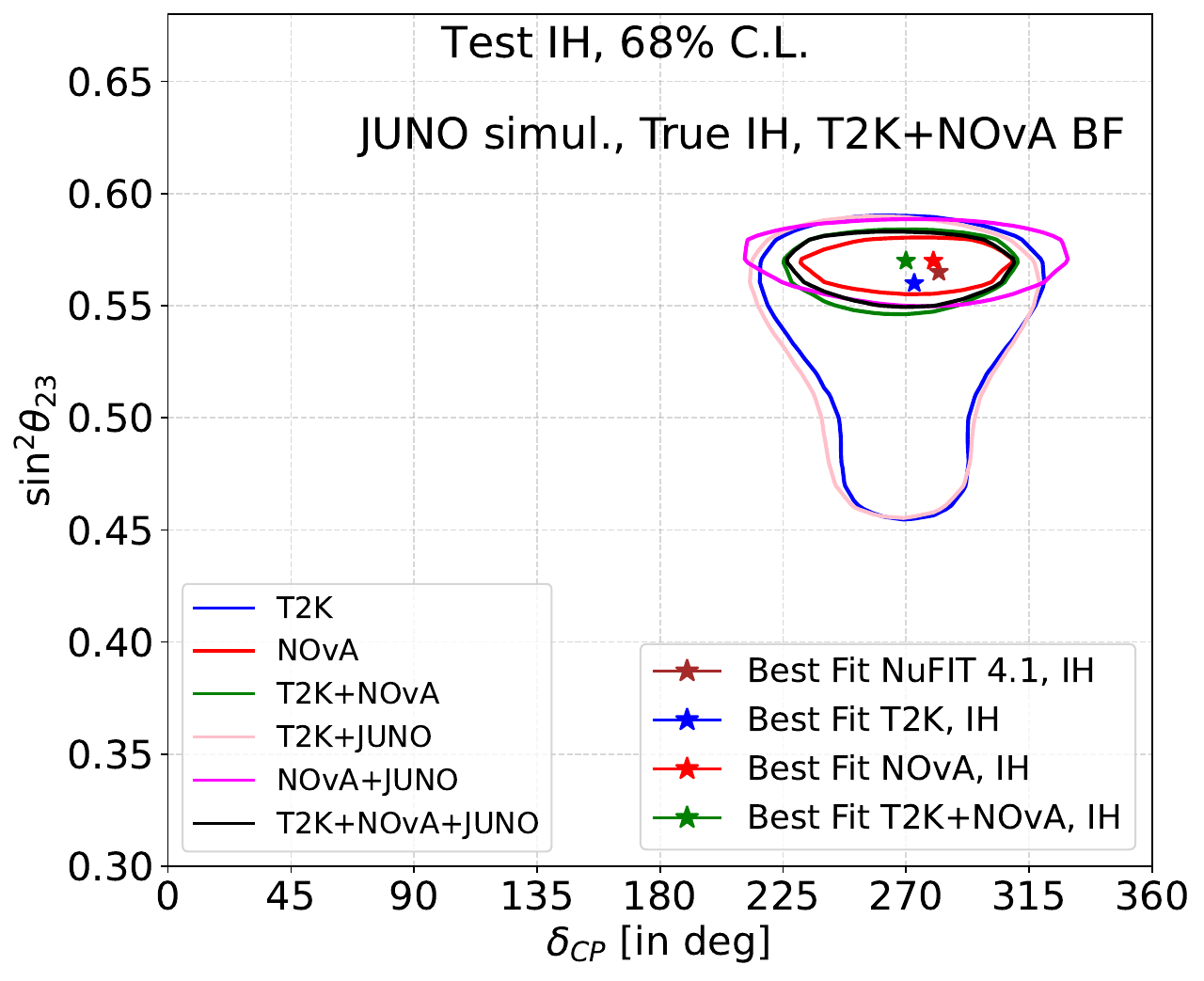}
        \caption{}
    \end{subfigure}

    \vspace{0.5cm} 

    \begin{subfigure}[b]{0.45\textwidth}
        \centering
        \includegraphics[width=\textwidth, height=6cm]{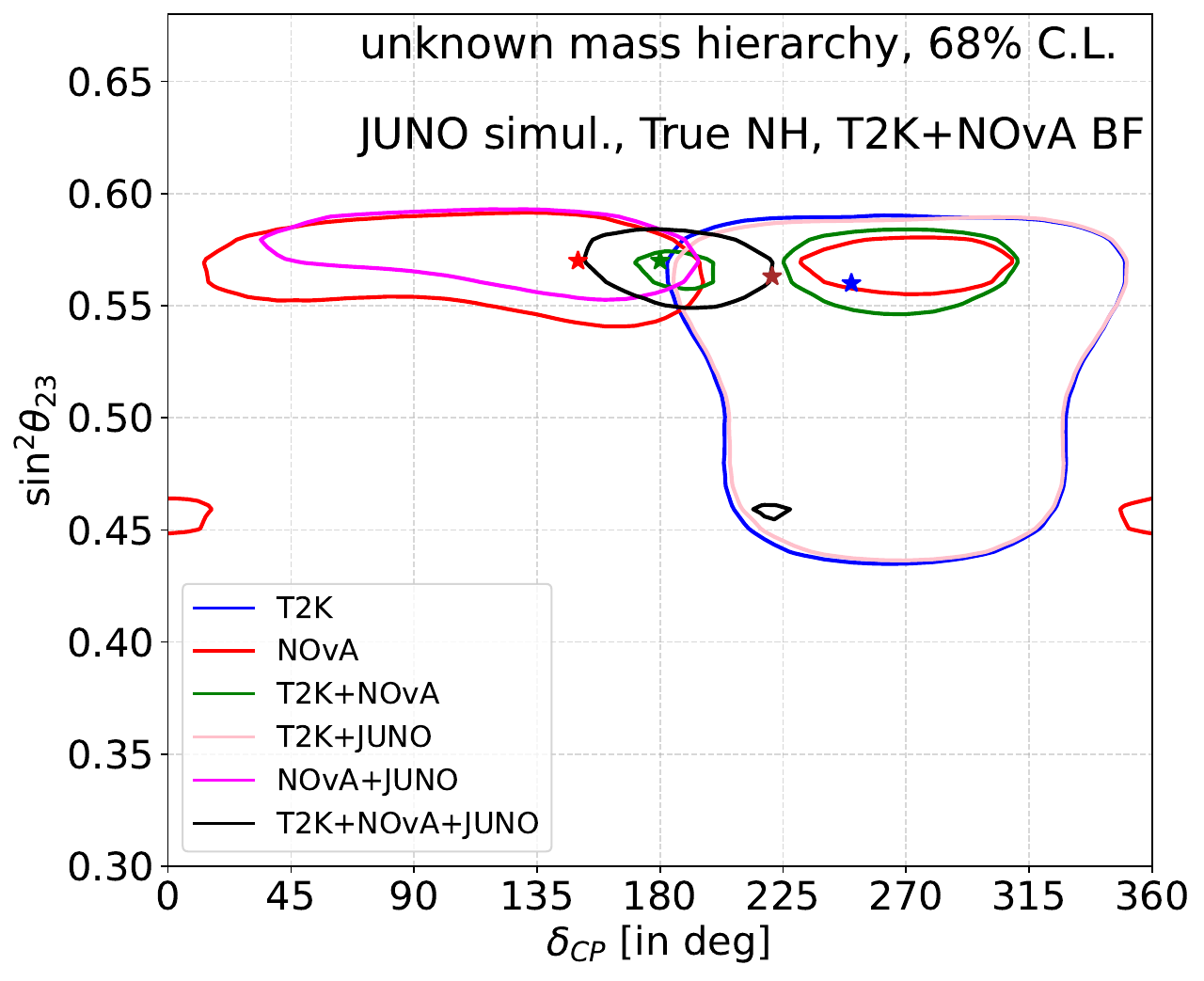}
        \caption{}
    \end{subfigure}
    \hfill
    \begin{subfigure}[b]{0.45\textwidth}
        \centering
        \includegraphics[width=\textwidth, height=6cm]{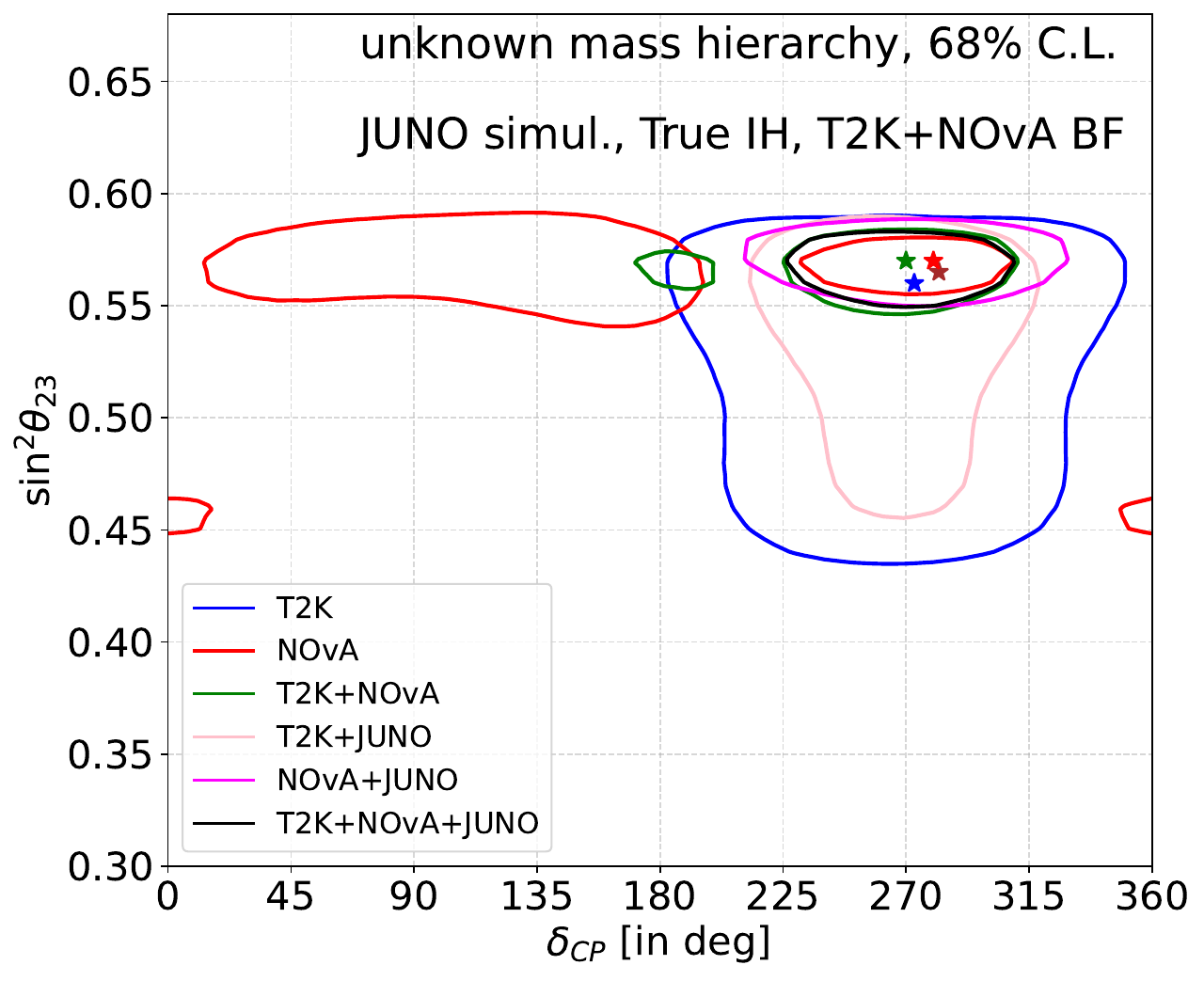}
        \caption{}
    \end{subfigure}

    \caption{Allowed regions in the $\sin^2{\theta_{23}}-\delta_{\rm CP}$ plane including T2K and \nova data, and JUNO simulation with NO$\nu$A+T2K bast-fit at the true parameter values. The top (bottom) panels present the cases for known (unknown) mass hierarchy and the left (right) panels present the cases when the true hierarchy for JUNO is NH (IH).}
    \label{fig:s23_dcp_t2k_nova}
\end{figure}
\newpage
\section{Effects of JUNO simulations on T2K and \nova data: CP sensitivity with different true parameter values for JUNO simulations}\label{app:B}
In this section, we display the results for CP sensitivity plots where JUNO data is simulated with the true parameter values fixed at T2K best-fit points, \nova best-fit points and T2K+\nova best-fit points.
\begin{figure}[htbp]
    \centering
    \begin{subfigure}[b]{0.45\textwidth}
        \centering
        \includegraphics[width=\textwidth, height=6cm]{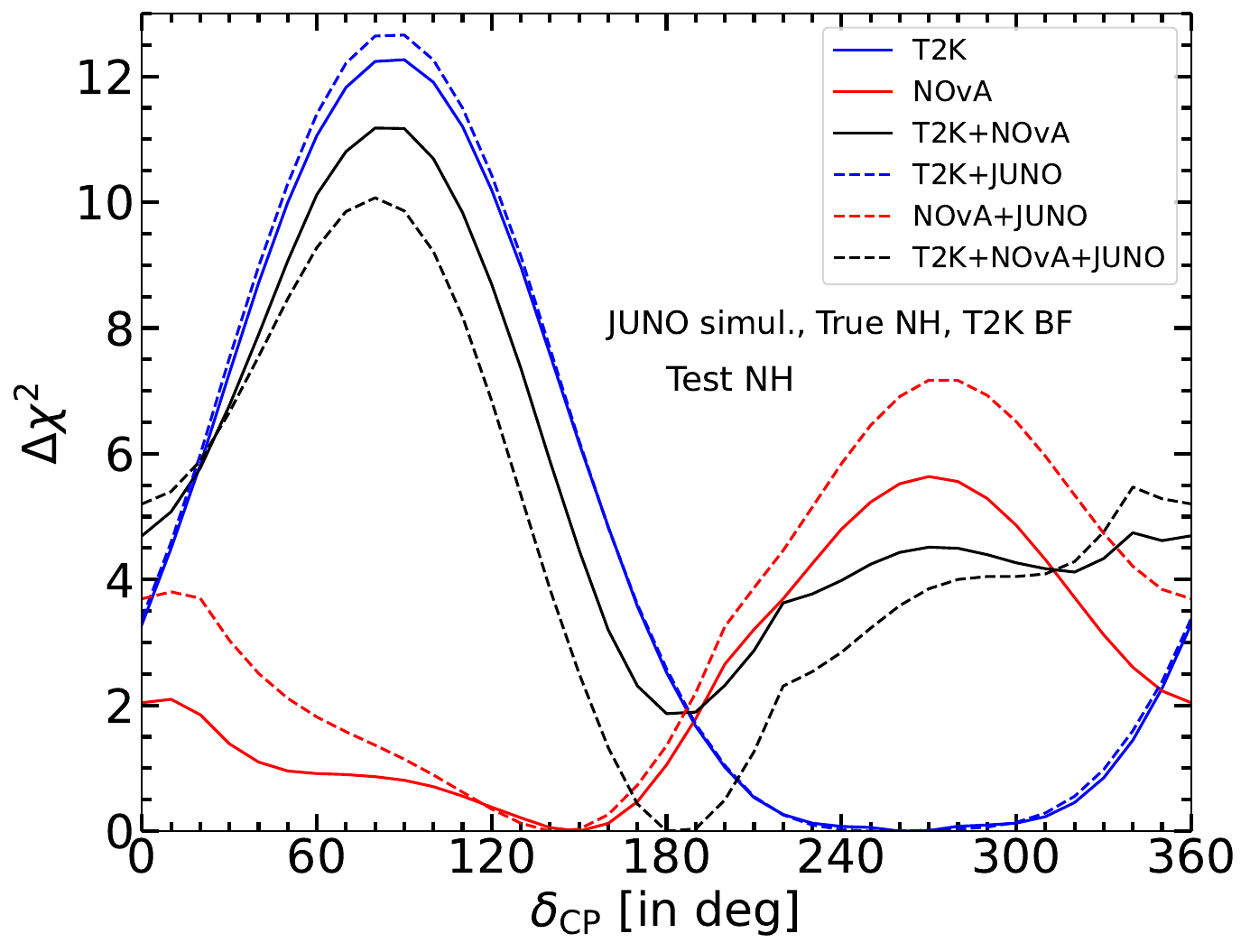}
        \caption{}
    \end{subfigure}
    \hfill
    \begin{subfigure}[b]{0.45\textwidth}
        \centering
        \includegraphics[width=\textwidth, height=6cm]{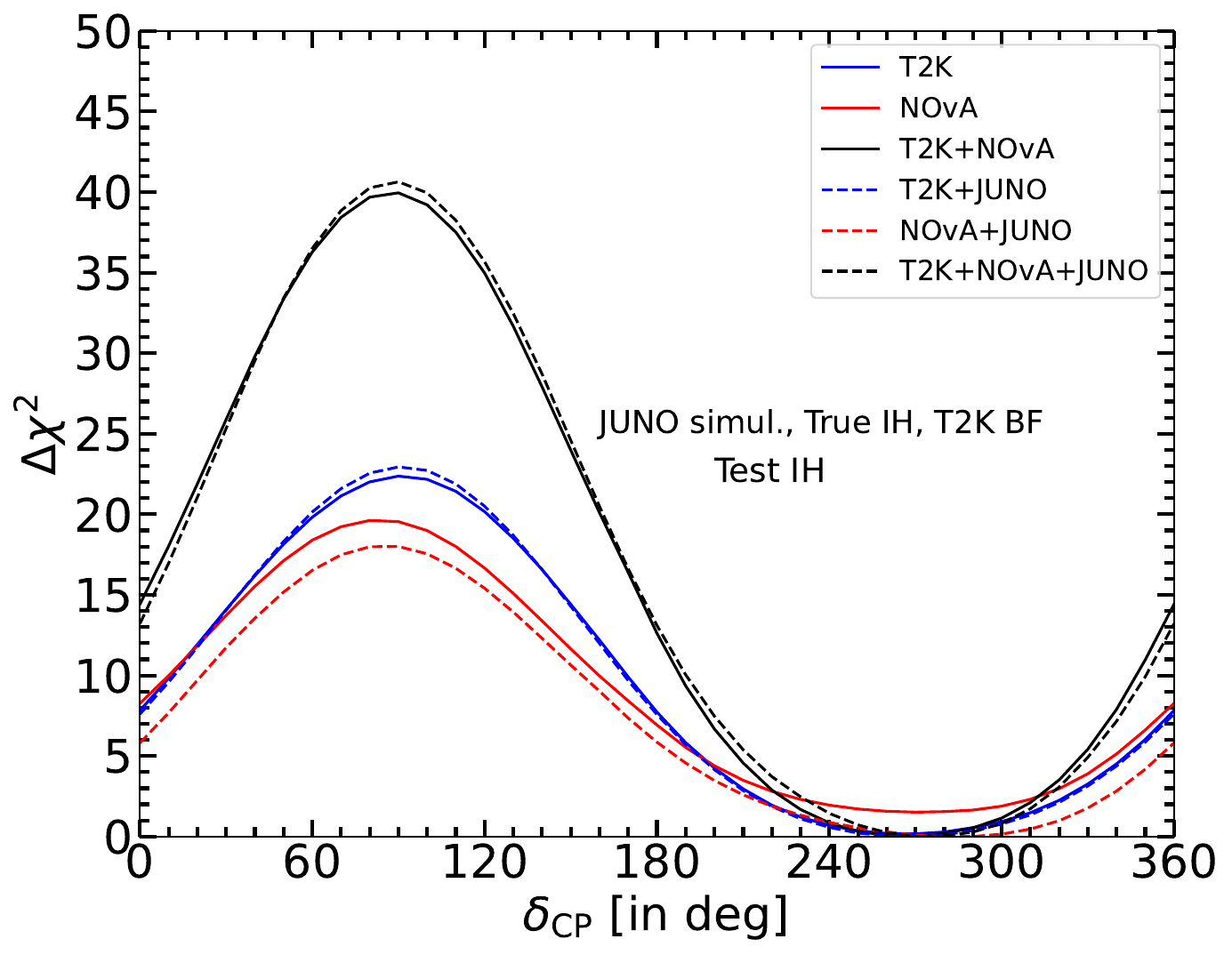}
        \caption{}
    \end{subfigure}

    \vspace{0.5cm} 

    \begin{subfigure}[b]{0.45\textwidth}
        \centering
        \includegraphics[width=\textwidth, height=6cm]{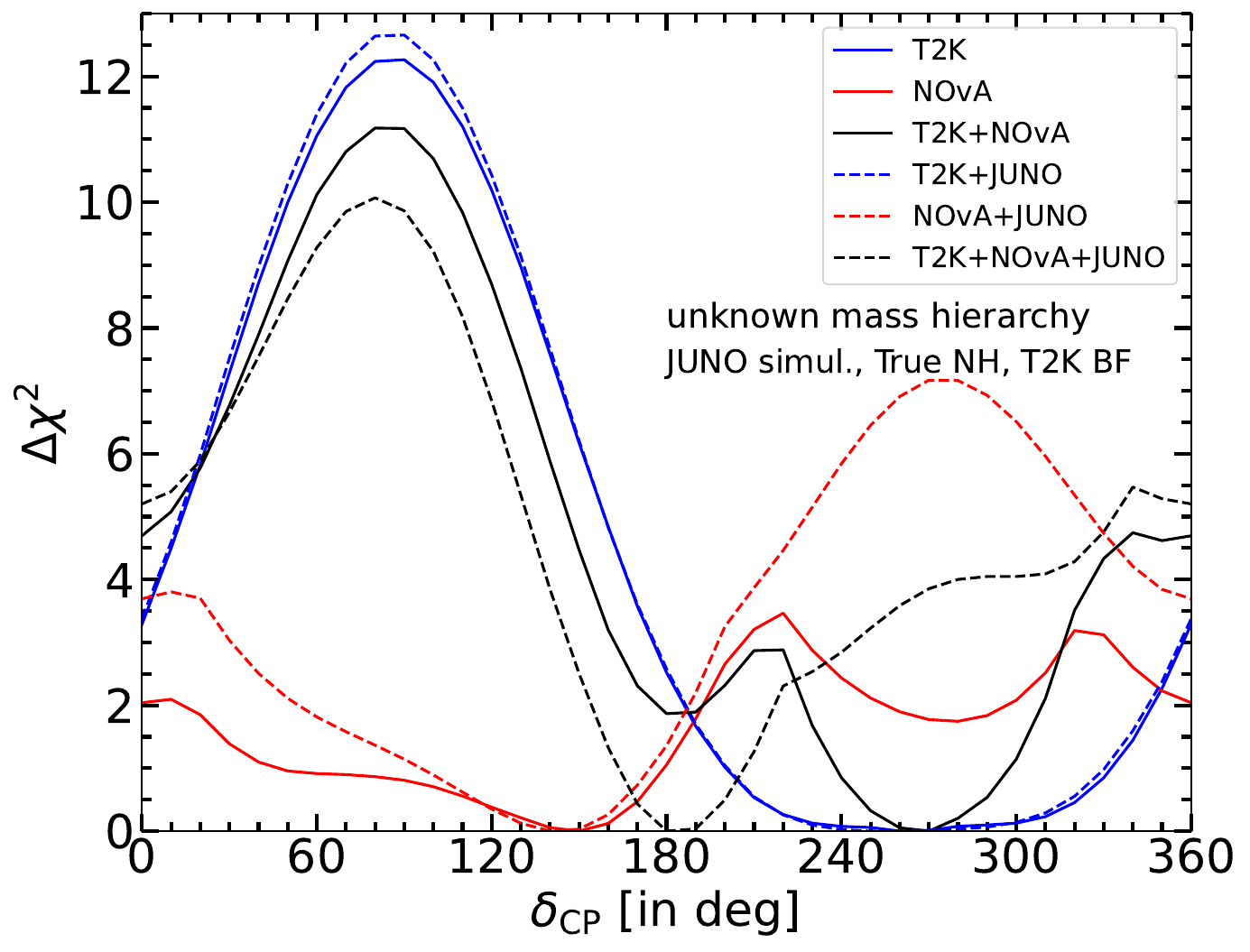}
        \caption{}
    \end{subfigure}
    \hfill
    \begin{subfigure}[b]{0.45\textwidth}
        \centering
        \includegraphics[width=\textwidth, height=6cm]{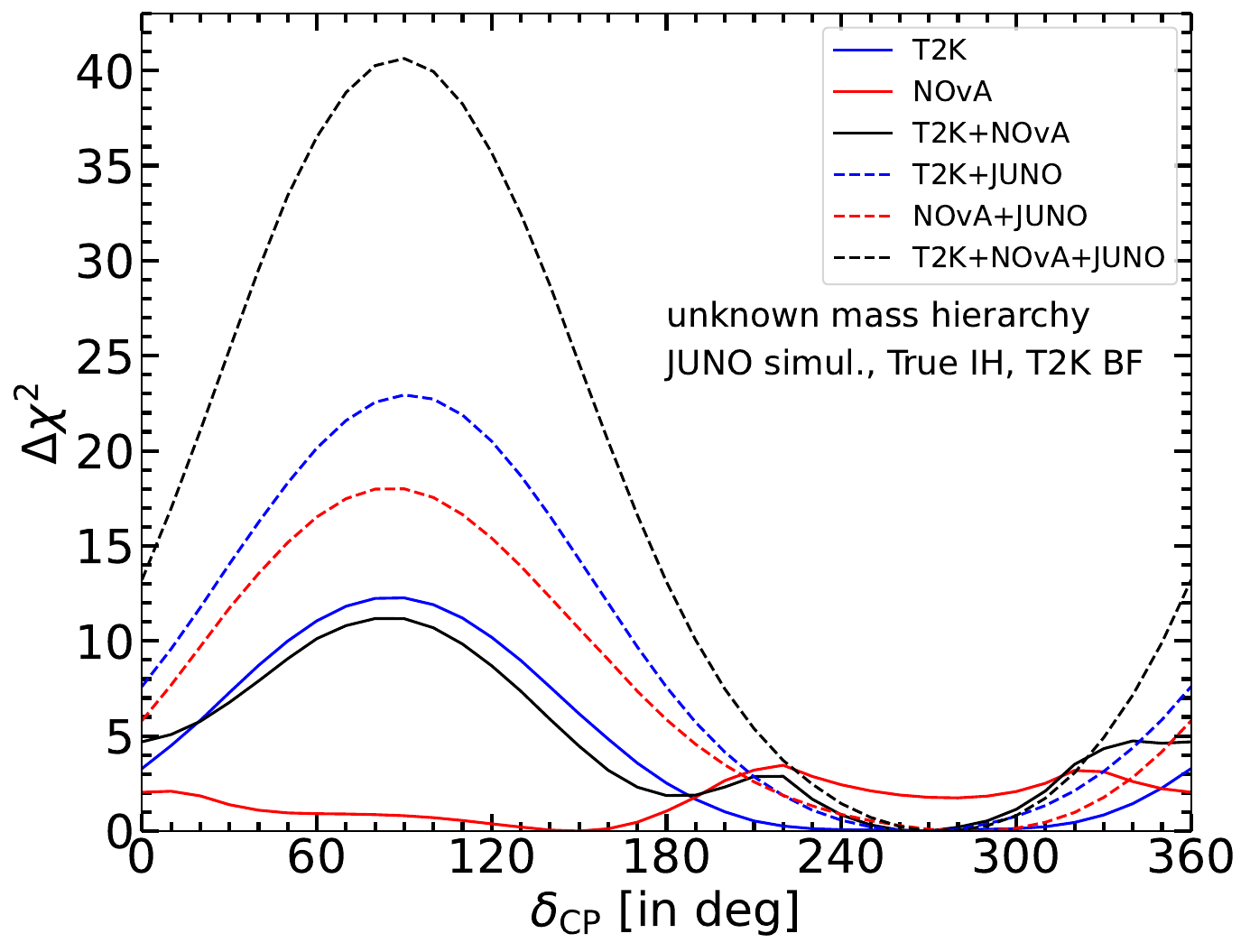}
        \caption{}
    \end{subfigure}

    \caption{CPV sensitivity plots for T2K and \nova data, and JUNO simulation with T2K best-fit at the true parameter values. The top (bottom) panels present the cases for known (unknown) mass hierarchy and the left (right) panels present the cases when the true hierarchy for JUNO is NH (IH).}
    \label{fig:res_t2k_bestfit}
\end{figure}

\begin{figure}[htbp]
    \centering
    \begin{subfigure}[b]{0.45\textwidth}
        \centering
        \includegraphics[width=\textwidth, height=6cm]{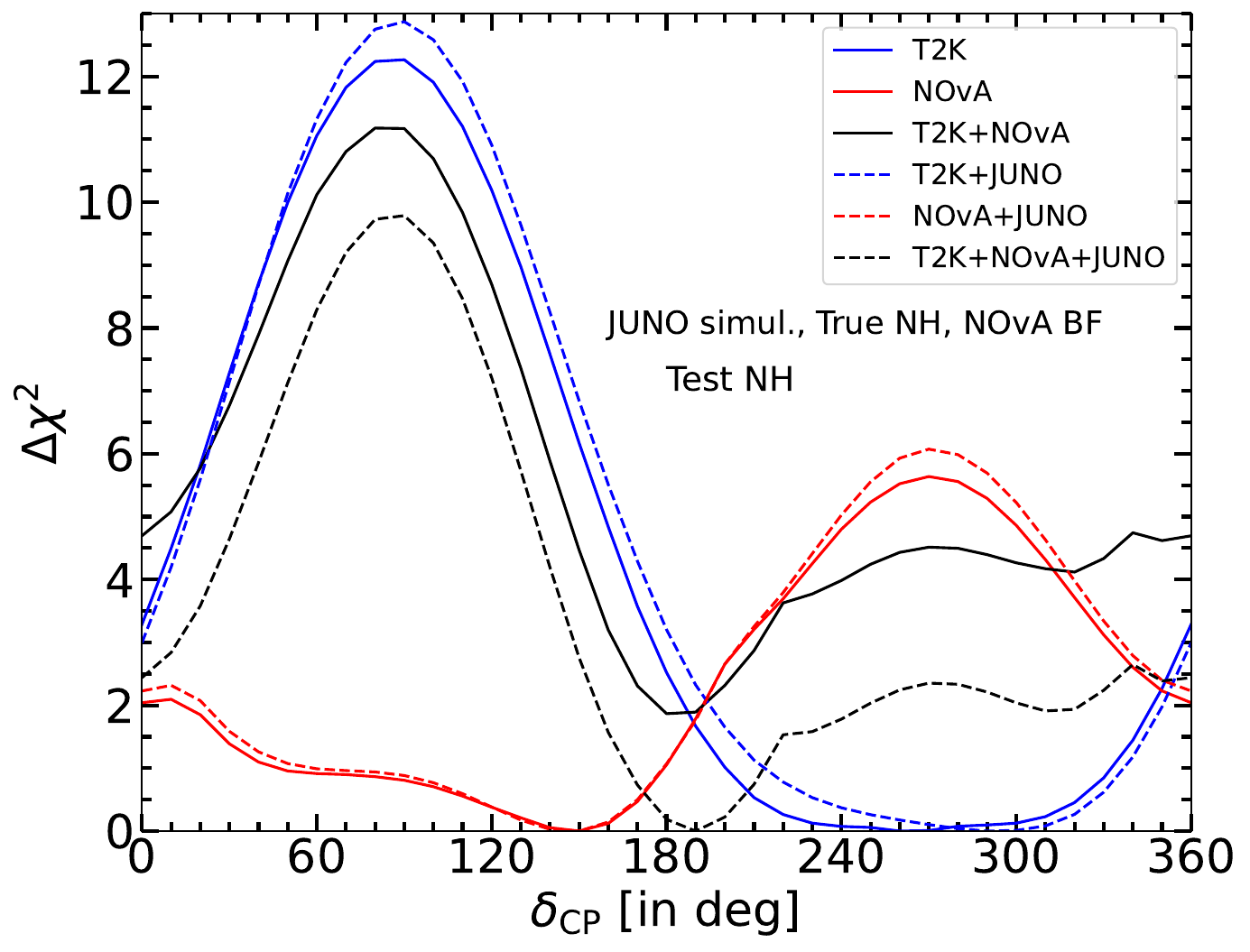}
        \caption{}
    \end{subfigure} 
    \hfill
    \begin{subfigure}[b]{0.45\textwidth}
        \center ing
        \includegraphics[width=\textwidth, height=6cm]{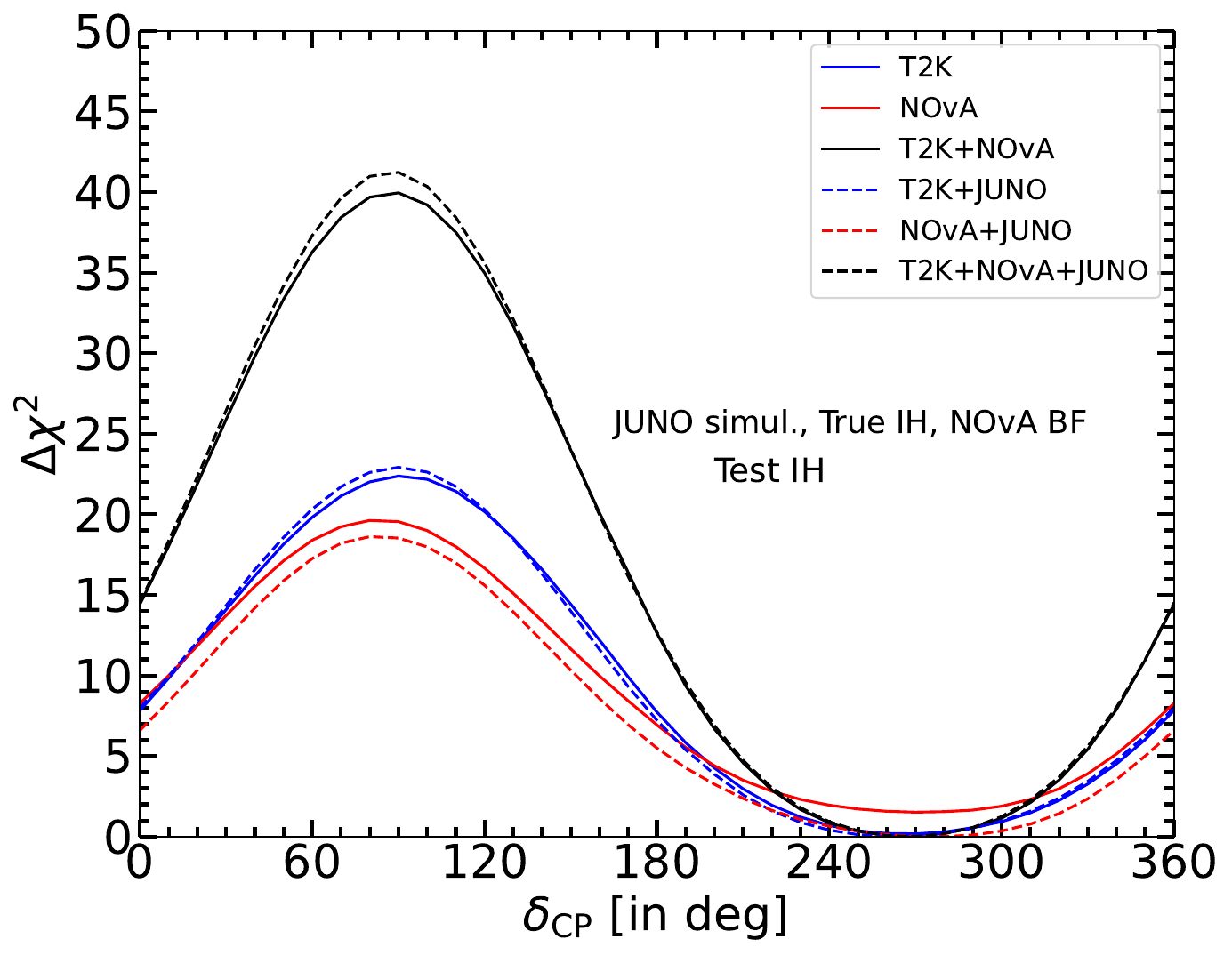}
        \caption{}
    \end{subfigure}

    \vspace{0.5cm} 

    \begin{subfigure}[b]{0.45\textwidth}
        \centering
        \includegraphics[width=\textwidth, height=6cm]{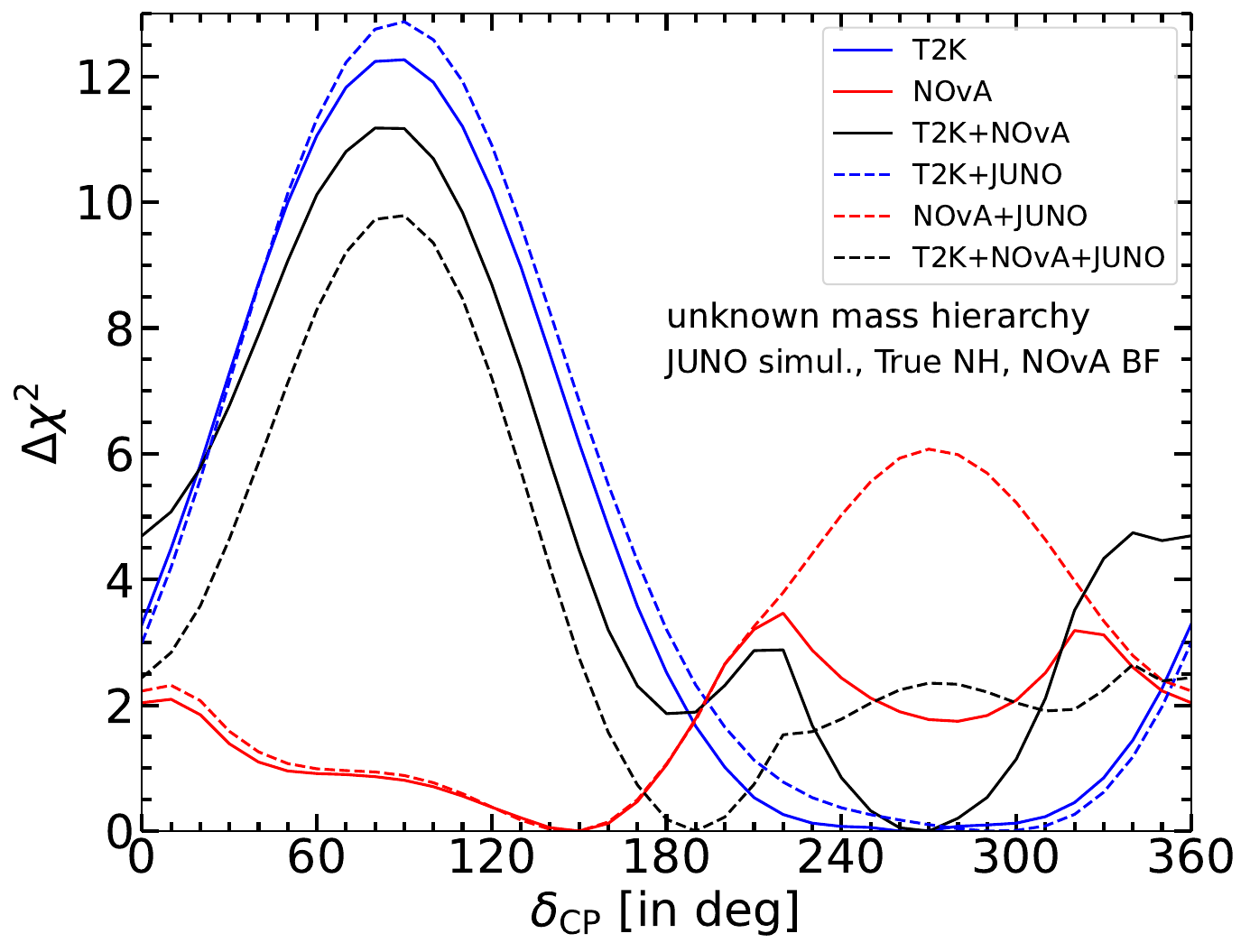}
        \caption{}
    \end{subfigure}
    \hfill
    \begin{subfigure}[b]{0.45\textwidth}
        \centering
        \includegraphics[width=\textwidth, height=6cm]{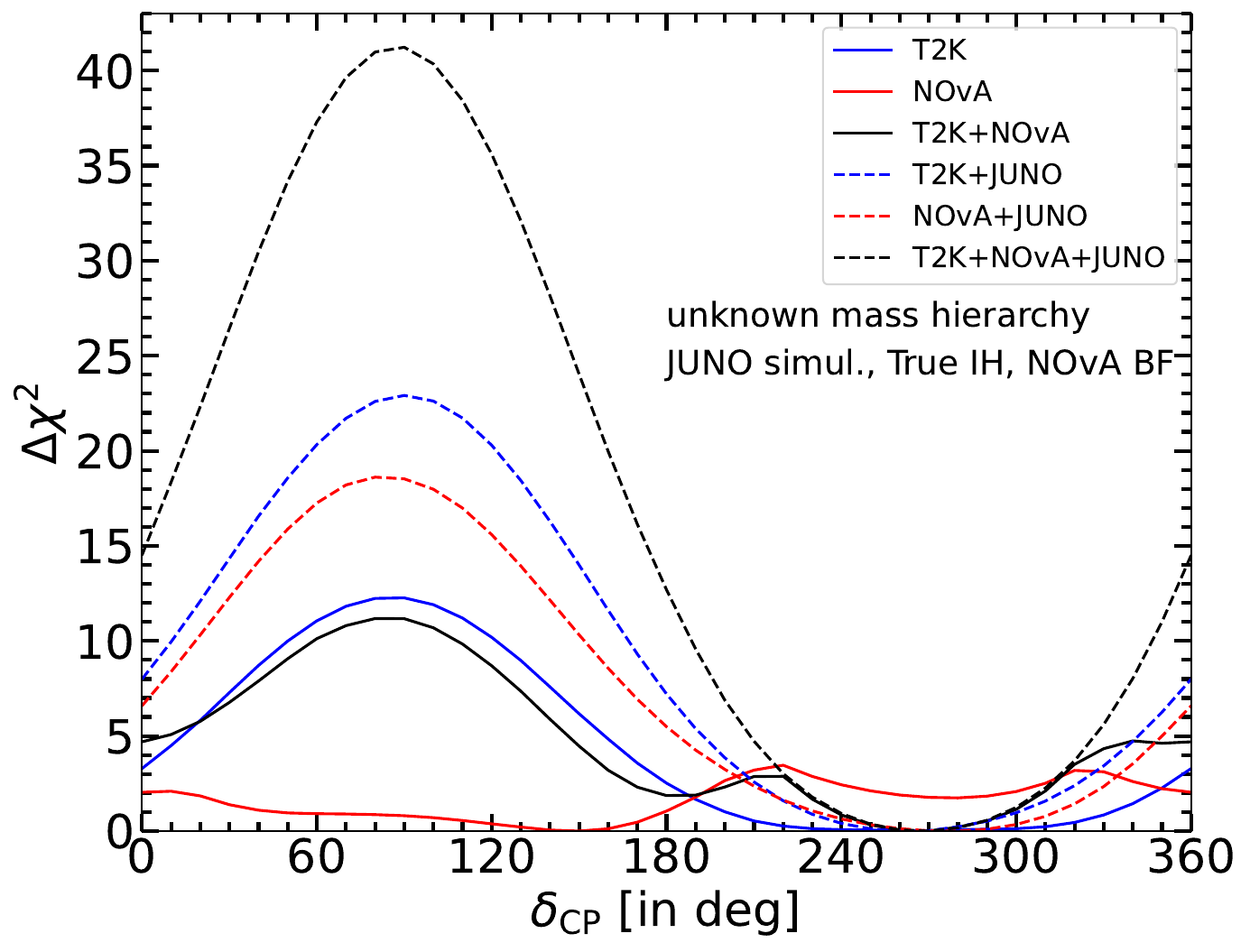}
        \caption{}
    \end{subfigure}

    \caption{CPV sensitivity plots for T2K and \nova data, and JUNO simulation with \nova best-fit at the true parameter values. The top (bottom) panels present the cases for known (unknown) mass hierarchy and the left (right) panels present the cases when the true hierarchy for JUNO is NH (IH).}
    \label{fig:res_nova_bestfit}
\end{figure}

\begin{figure}[htbp]
    \centering
    \begin{subfigure}[b]{0.45\textwidth}
        \centering
        \includegraphics[width=\textwidth, height=6cm]{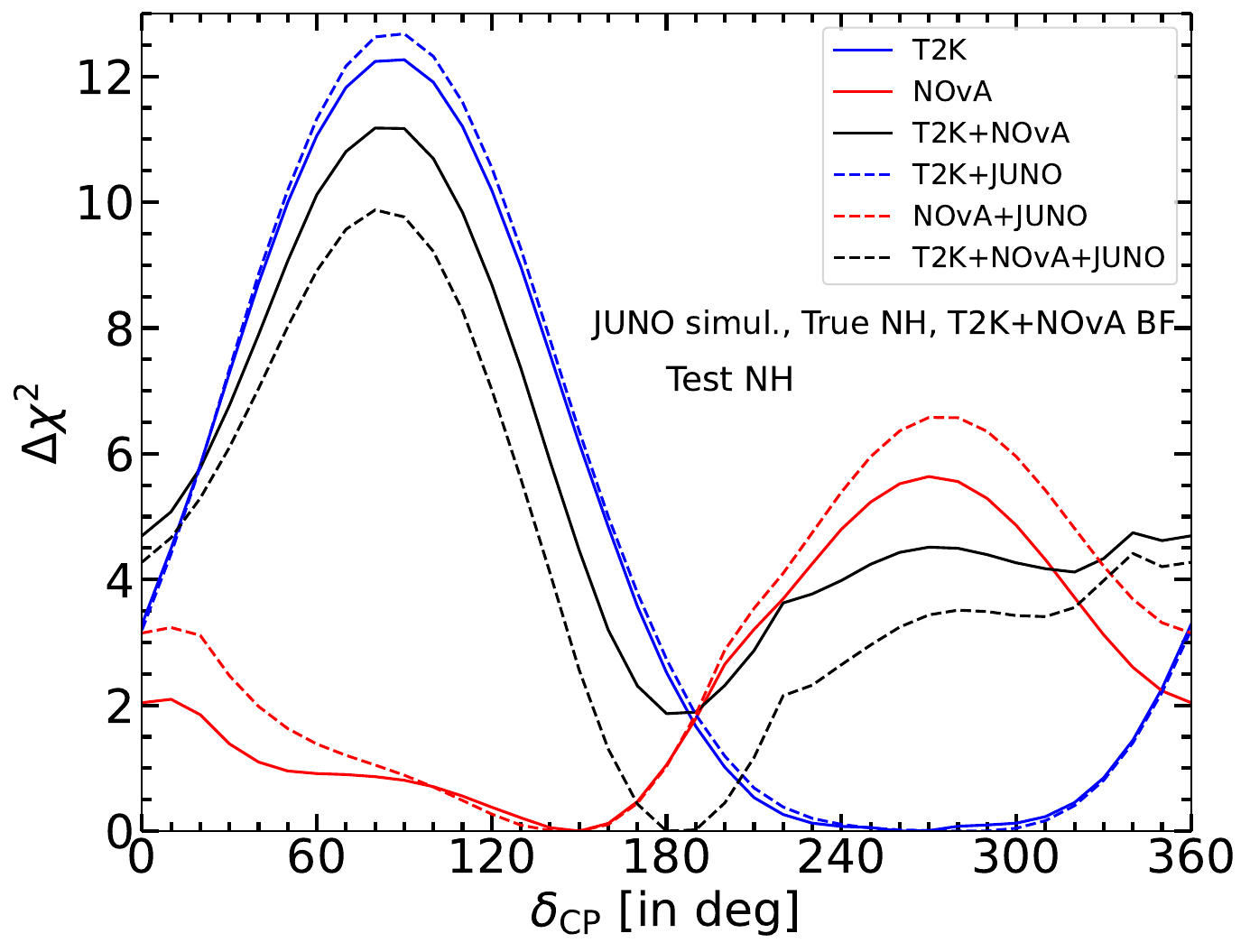}
        \caption{}
    \end{subfigure}
    \hfill
    \begin{subfigure}[b]{0.45\textwidth}
        \centering
        \includegraphics[width=\textwidth, height=6cm]{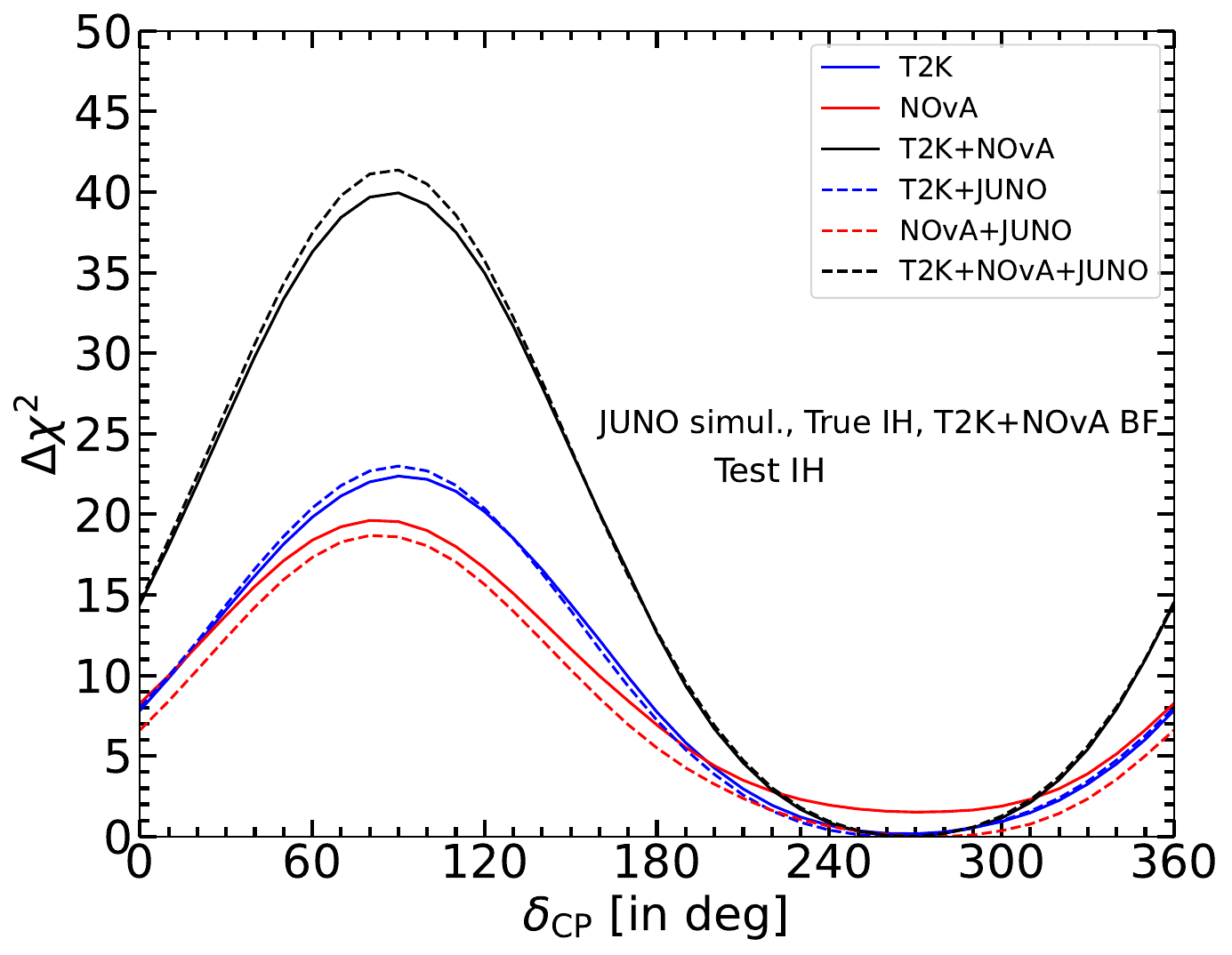}
        \caption{}
    \end{subfigure}

    \vspace{0.5cm} 

    \begin{subfigure}[b]{0.45\textwidth}
        \centering
        \includegraphics[width=\textwidth, height=6cm]{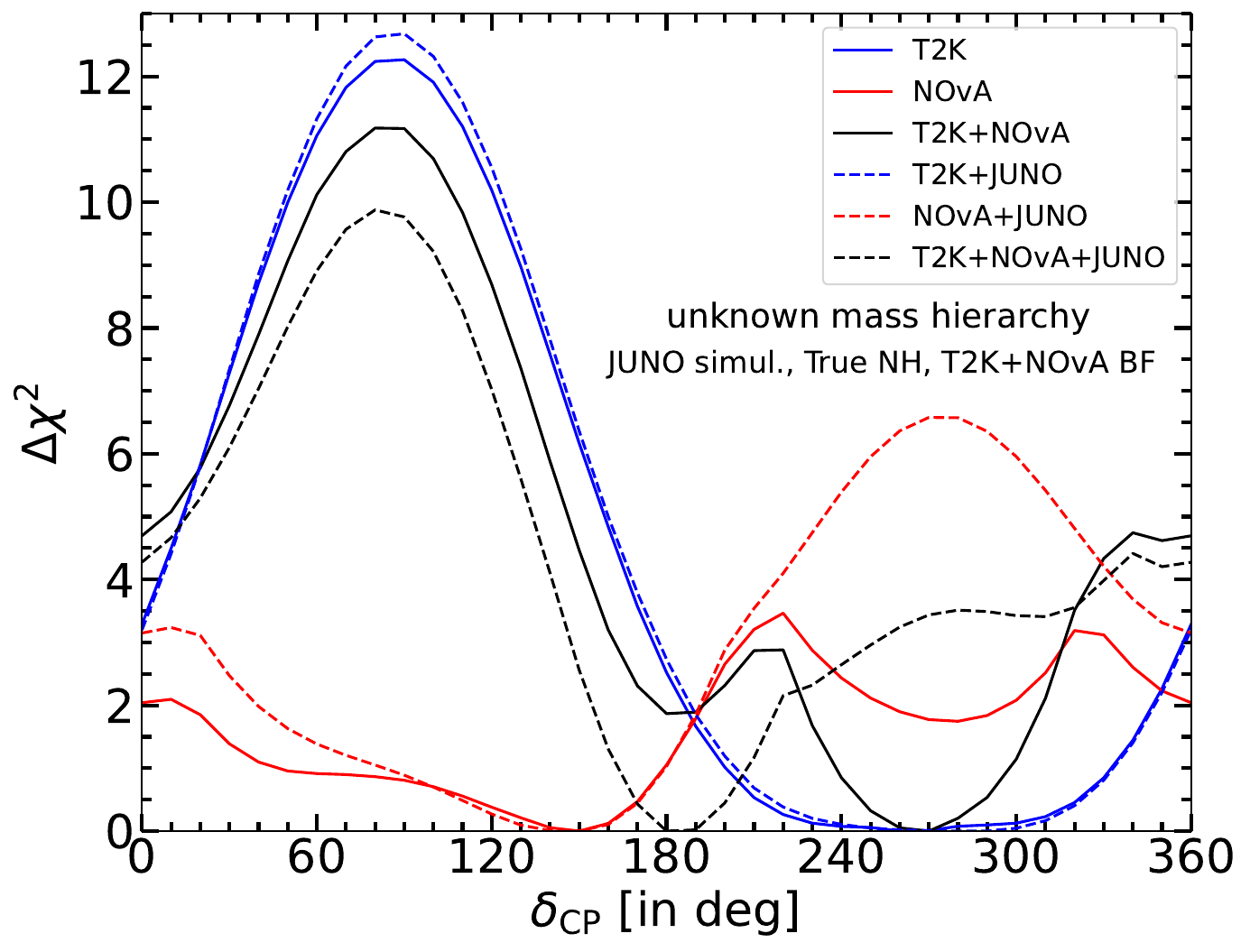}
        \caption{}
    \end{subfigure}
    \hfill
    \begin{subfigure}[b]{0.45\textwidth}
        \centering
        \includegraphics[width=\textwidth, height=6cm]{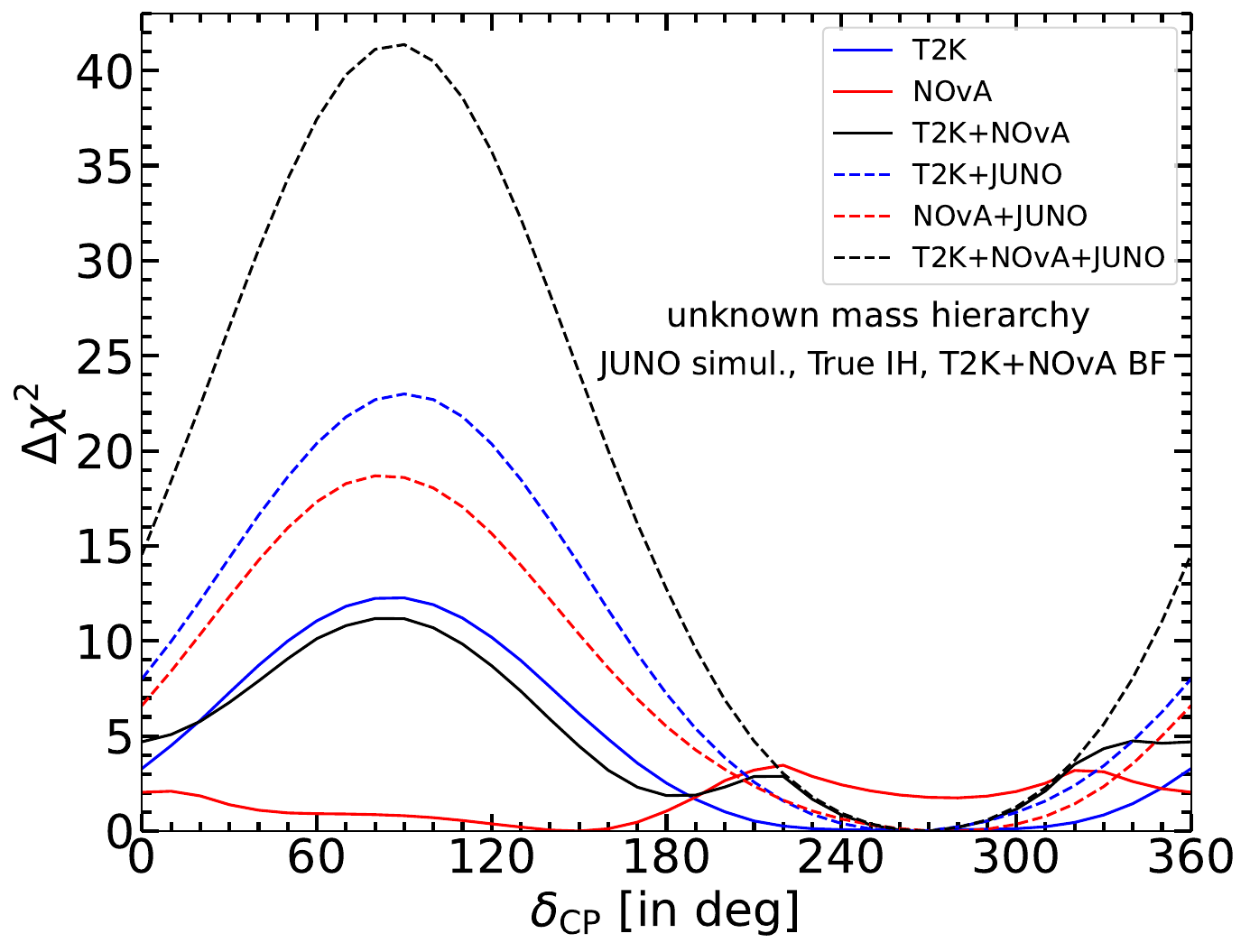}
        \caption{}
    \end{subfigure}

    \caption{CPV sensitivity plots for T2K and \nova data, and JUNO simulation with NO$\nu$A+T2K best-fit at the true parameter values. The top (bottom) panels present the cases for known (unknown) mass hierarchy, and the left (right) panels present the cases when the true hierarchy for JUNO is NH (IH).}
    \label{fig:res_t2k_nova_bestfit}
\end{figure}
\end{document}